\documentclass[twocolumn,showpacs,amsfonts,aps,prc,floatfix,%
superscriptaddress]{revtex4} 

\usepackage{amsmath}
\usepackage{bm}
\usepackage{graphicx}

\newcommand{\be}[1]{\begin{equation}\label{#1}}
\newcommand{\ee}{\end{equation}}

\newcommand{\ie}{\emph{i.e. }}

\newcommand{\eq}{{\,=\,}}

\newcommand{\du}{\nabla^{\left\langle\mu\right.}u^{\left.\nu\right\rangle}}

\def\La{\langle}
\def\Ra{\rangle}

\usepackage{color}

\begin{document}


\title{Causal viscous hydrodynamics in 2+1 dimensions for 
relativistic heavy-ion collisions}
\date{\today}

\author{Huichao Song}
\email[Correspond to\ ]{song@mps.ohio-state.edu}
\affiliation{Department of Physics, The Ohio State University, 
Columbus, OH 43210, USA}
\author{Ulrich Heinz}
\affiliation{Department of Physics, The Ohio State University, 
Columbus, OH 43210, USA}
\affiliation{CERN, Physics Department, Theory Division, CH-1211 Geneva 23, 
             Switzerland}

\begin{abstract}
We explore the effects of shear viscosity on the hydrodynamic evolution and
final hadron spectra of Cu+Cu collisions at ultrarelativistic collision
energies, using the newly developed (2+1)-dimensional viscous hydrodynamic
code VISH2+1. Based on the causal Israel-Stewart formalism, this code 
describes the transverse evolution of longitudinally boost-invariant systems 
without azimuthal symmetry around the beam direction. Shear viscosity is 
shown to decelerate the longitudinal and accelerate the transverse 
hydrodynamic expansion. For fixed initial conditions, this leads to a 
longer quark-gluon plasma (QGP) lifetime, larger radial flow in the final 
state, and flatter transverse momentum spectra for the emitted hadrons 
compared to ideal fluid dynamic simulations. We find that the elliptic 
flow coefficient $v_2$ is particularly sensitive to shear viscosity: even 
the lowest value allowed by the AdS/CFT conjecture $\eta/s{\,\geq\,}1/4\pi$ 
suppresses $v_2$ enough to have significant consequences for the 
phenomenology of heavy-ion collisions at the Relativistic Heavy Ion 
Collider. A comparison between our numerical results and earlier analytic 
estimates of viscous effects within a blast-wave model parametrization of 
the expanding fireball at freeze-out reveals that the full dynamical 
theory leads to much tighter constraints for the specific shear 
viscosity $\eta/s$, thereby supporting the notion that the quark-gluon 
plasma created at RHIC exhibits almost ``perfect fluidity''.  
\end{abstract}

\maketitle

\section{Introduction}\label{sec1}

Hydrodynamics is an efficient tool to describe the expansion of the
fireballs generated in relativistic heavy-ion collisions. As a 
macroscopic description that provides the 4-dimensional space-time 
evolution of the energy-momentum tensor $T^{\mu\nu}(x)$ it is much 
less demanding than microscopic descriptions based on kinetic theory 
that evolve the (on-shell) distribution function $f(x,p)$ in 7-dimensional 
phase-space.

Ideal fluid dynamics is even more efficient since it reduces the number 
of independent fields needed to describe the symmetric energy-momentum 
tensor from 10 to 5: the local energy density $e(x)$, pressure $p(x)$ and 
the normalized flow 4-velocity $u^\mu(x)$ (which has 3 independent 
components). The equation of state (EOS) $p(e)$ provides a further constraint 
which closes the set of four equations $\partial_\mu T^{\mu\nu}(x)=0$. 

Ideal fluid dynamics is based on the strong assumption that the
fluid is in local thermal equilibrium and evolves isentropically. While
local momentum isotropy in the comoving frame is sufficient for a unique
decomposition of the energy-momentum tensor in terms of $e$, $p$ and 
$u^\mu$, it does not in general guarantee a unique relationship $p(e)$. 
Generically, the equation of state $p(e)$ (a key ingredient for closing 
the set of hydrodynamic equations) becomes unique only after entropy 
maximization, \ie after a locally thermalized state, with Maxwellian 
(or Bose-Einstein and Fermi-Dirac) momentum distributions in the comoving 
frame, has been reached. For this assumption to be valid, the miscroscopic 
collision time scale must be much shorter than the macroscopic evolution 
time scale. Since the fireballs created in relativistic heavy-ion collisions 
are small and expand very rapidly, applicability of the hydrodynamic 
approach has long been doubted.

It came therefore as a surprise to many that the bulk of the matter 
produced in Au+Au collisions at the Relativistic Heavy Ion Collider (RHIC)
was found to behave like an almost ideal fluid. Specifically, ideal 
fluid dynamic models correctly reproduce the hadron transverse momentum 
spectra in central and semi-peripheral collisions, including their 
anisotropy in non-central collisions given by the elliptic flow 
coefficient $v_2(p_T)$ and its dependence on the hadron rest mass,
for transverse momenta up to about 1.5--2\,GeV/$c$ \cite{Rev-hydro}
which covers more than 99\% of the emitted particles. This observation 
has led to the conclusion that the quark-gluon plasma (QGP) created in 
RHIC collisions thermalizes very fast and must therefore be strongly
(non-perturbatively) interacting \cite{Heinz:2001xi}, giving rise to 
the notion that the QGP is a strongly coupled plasma 
\cite{Gyulassy:2004vg,Gyulassy:2004zy,Shuryak:2004cy} that behaves like
an almost perfect fluid \cite{fn0}. 

At RHIC energies, the almost perfect ideal fluid dynamical description 
of experimental data gradually breaks down in more peripheral collisions, 
at high transverse momenta, and at forward and backward rapidities; at 
lower energies it lacks quantitative accuracy even in the most central 
collisions at midrapidity \cite{hydro-Heinz05}. Most of these deviations 
from ideal fluid dynamical predictions can be understood as the result of 
strong viscous effects during the late hadronic stage of the fireball 
expansion \cite{Hirano:2005xf} after the QGP has hadronized. As the initial 
energy density of the fireball decreases, the dissipative dynamics of the 
hadronic stage takes on increasing importance, concealing the perfect 
fluidity of any quark-gluon plasma possibly created at the beginning of the 
collision. However, as also pointed out in \cite{Hirano:2005xf}, persisting 
uncertainties about the initial conditions in heavy-ion collisions leave 
room for a small amount of viscosity even during the early QGP stage. 
Furthermore, the observed deviations of the elliptic flow parameter $v_2(p_T)$
at large $p_T$ even in the largest collision systems at the highest available
collision energies are consistent with viscous effects during the early
epoch of the fireball \cite{Molnar:2001ux,Teaney:2003kp}. During this
epoch, the matter is so dense and strongly interacting that a microscopic
description based on classical kinetic theory of on-shell partons 
\cite{Molnar:2001ux} may be questionable. We therefore develop here a 
dissipative generalization of the macroscopic hydrodynamic approach,
viscous relativistic fluid dynamics.

The need for such a framework is further highlighted by the recent insight
that, due to quantum mechanical uncertainty \cite{Gyulassy85}, no classical
fluid can have exactly vanishing viscosity (as is assumed in the ideal
hydrodynamic approach). Even in the limit of infinitely strong coupling,
the QGP must hence maintain a nonzero viscosity. Recent calculations of
the shear viscosity to entropy ratio (the ``specific shear viscosity''
$\eta/s$) in a variety of conformal gauge field theories which share 
some properties with QCD, using the AdS/CFT correspondence, suggest a 
lower limit of $\frac{\eta}{s}\eq\frac{1}{4\pi}$ 
\cite{Ads-CFT,son,Janik:2006ft}. This is much smaller than the value 
obtained from weak coupling calculations in QCD \cite{QCD-Viscosity}
(although close to a recent first numerical result from lattice QCD
\cite{Meyer}) and more than an order of magnitude below the lowest 
measured values in standard fluids \cite{son}. Some alternative ideas
how small effective viscosities could be generated by anomalous effects 
(chaoticity) in anisotropically expanding plasmas \cite{anom_viscosity} 
or by negative eddy viscosity in 2-dimensional turbulent flows 
\cite{Romatschke:2007eb} have also been proposed.

Initial attempts to formulate relativistic dissipative fluid dynamics
as a relativistic generalization of the Navier-Stokes equation 
\cite{Eckart,Landau} ran into difficulties because the resulting 
equations allowed for acausal signal propagation, and their solutions
developed instabilities. These difficulties are avoided in the 
``2nd order formalism'' developed 30 years ago by Israel and Stewart
\cite{Israel:1976tn} which expands the entropy current to 2nd order in the
dissipative flows and replaces the instantaneous identification of the
dissipative flows with their driving forces multiplied by some transport
coefficient (as is done in Navier-Stokes theory) by a kinetic equation 
that evolves the dissipative flows rapidly but smoothly towards their 
Navier-Stokes limit. This procedure eliminates causality and stability 
problems at the expense of numerical complexity: the dissipative flows 
become independent dynamical variables whose kinetic equations of motion 
are coupled and must be solved simultaneously with the hydrodynamic 
evolution equations. This leads effectively to more than a doubling of 
the number of coupled partial differential equations to be solved 
\cite{Heinz:2005bw}.

Only recently computers have become powerful enough to allow efficient 
solution of the Israel-Stewart equations. The last 5 years have seen the
development of numerical codes which solve these equations (or slight 
variations thereof \cite{Israel:1976tn,Heinz:2005bw,Muronga:2001zk,%
Teaney:2004qa,Baier:2006um}) numerically, for systems with boost-invariant 
longitudinal expansion and transverse expansion in zero 
\cite{Muronga:2001zk,Baier:2006um}, one 
\cite{Teaney:2004qa,Muronga:2004sf,Chaudhuri:2005ea,Baier:2006gy}
and two dimensions \cite{Chaudhuri:2007zm,Romatschke:2007mq,Song:2007fn,%
Dusling:2007gi} (see also Ref.~\cite{Chaudhuri:2006jd} for a numerical 
study of the relativistic Navier-Stokes equation in 2+1 dimensions). The 
process of verification and validation of these numerical codes is still 
ongoing: While different initial conditions and evolution parameters used 
by the different groups of authors render a direct comparison of their results 
difficult, it seems unlikely that accounting for these differences will 
bring all the presently available numerical results in line with each 
other. 

We here present results obtained with an independently developed 
(2+1)-dimensional causal viscous hydrodynamic code, VISH2+1 
\cite{fn1}. While a short account of some of our main findings has
already been published \cite{Song:2007fn}, we here report many more
details, including extensive tests of the numerical accuracy of the 
code: We checked that (i) in the limit of vanishing viscosity, it accurately 
reproduces results obtained with the (2+1)-d ideal fluid code AZHYDRO 
\cite{AZHYDRO}; (ii) for homogeneous transverse density distributions (\ie 
in the absence of transverse density gradients and transverse flow) and 
vanishing relaxation time it accurately reproduces the known analytic 
solution of the relativistic Navier-Stokes equation for boost-invariant 
1-dimensional longitudinal expansion \cite{Danielewicz:1984ww}; 
(iii) for very short kinetic relaxation times our Israel-Stewart code 
accurately reproduces results from a separately coded (2+1)-d relativistic 
Navier-Stokes code, under restrictive conditions where the latter produces 
numerically stable solutions; and (iv) for simple analytically parametrized 
anisotropic velocity profiles the numerical code correctly computes the 
velocity shear tensor that drives the viscous hydrodynamic effects. 

In its present early state, and given the existing open questions about
the mutual compatibility of various numerical results reported in the 
recent literature, we believe that it is premature to attempt a detailed 
comparison of VISH2+1 with experimental data, in order to empirically 
constrain the specific shear viscosity of the QGP. Instead, we concentrate 
in this paper on describing and trying to understand the robustness of a
variety of fluid dynamical effects generated by shear viscosity in a 
relativistic QGP fluid. We report here only results for Cu+Cu collisions, 
with initial entropy densities exceeding significantly those that can be 
reached in such collisions at RHIC. The reasons for doing so are purely 
technical: Initially our numerical grid was not large enough to accomodate 
Au+Au collision fireballs with sufficient resolution, and while this 
restriction has been lifted in the meantime, a large body of instructive 
numerical results had already been accumulated which would have been quite 
expensive to recreate for the Au+Au system. The unrealistic choice of 
initial conditions was driven by the wish to allow for a sufficiently long 
lifetime of the QGP phase even in peripheral Cu+Cu collisions such that 
shear viscous effects on elliptic flow are still dominated by the 
quark-gluon plasma stage. The main goals of the present paper are: 
(i) to quantitatively establish shear viscous effects on the evolution of 
the energy and entropy density, of the flow profile, source eccentricity, 
and total momentum anisotropy, on the final hadron spectra, and on the 
elliptic flow in central and non-central heavy-ion collisions, under the 
influence of different equations of state; and (ii) to explore in detail 
and explain physically how these effects arise, trying to extract general 
rules and generic features which should also apply for other collision 
systems and collision energies. We note that recent calculations for 
Au+Au collisions \cite{next} have shown that viscous effects are somewhat 
bigger in the smaller Cu+Cu studied here than in the larger Au+Au system
for which the largest body of experimental data exists. The reader must 
therefore apply caution when trying to compare (in his or her mind's 
eye) our results with the well-known RHIC Au+Au data.

The paper is organized as follows: Section \ref{sec2} gives a brief review 
of the Israel-Stewart formalism for causal relativistic hydrodynamics 
for dissipative fluids, lists the specific form of these equations for
the (2+1)-dimensional evolution of non-central collision fireballs with
boost-invariant longitudinal expansion which are solved by VISH2+1, and 
details the initial consditions and the equation of state (EOS) employed
in our calculations. In Section \ref{sec3} we report results for central
($b\eq0$) Cu+Cu collisions. Section \ref{sec4} gives results for non-central 
collisions, including a detailed analysis of the driving forces behind the
strong shear viscous effects on elliptic flow observed by us. In Section 
\ref{sec5} we explore the influence of different initializations and 
different relaxation times for the viscous shear pressure tensor on the 
hydrodynamic evolution and establish the limits of applicability for
viscous hydrodynamics in the calculation of hadron spectra. Some 
technical details and the numerical tests performed to verify the 
accuracy of the computer code are discussed in Appendices 
\ref{appa}-\ref{appd}, and in Appendix~\ref{appe} we compare
our hydrodynamic results with analytical estimates of shear viscous 
effects by Teaney \cite{Teaney:2003kp} that were based on Navier-Stokes 
theory and a blast-wave model parametrization of the fireball.

\section{\bf Israel-Stewart theory of causal viscous hydrodynamics}
\label{sec2}

In this section, we review briefly the 2nd order Israel-Stewart formalism 
for viscous relativistic hydrodynamics and the specific set of equations
solved by VISH2+1 for anisotropic transverse expansion in longitudinally
boost-invariant fireballs. Details of the derivation can be found in 
\cite{Heinz:2005bw}, with a small correction pointed out by Baier {\it 
et al.} in \cite{Baier:2006um}. For simplicity, and in view of the intended
application to RHIC collisions whose reaction fireballs have almost
vanishing net baryon density, the discussion will be restricted to viscous 
effects, neglecting heat conduction and working in the Landau velocity 
frame \cite{Landau}. 

\subsection{Basics of Israel-Stewart theory}
\label{sec2a}

The general hydrodynamic equations arise from the local conservation of
energy and momentum, 
\begin{eqnarray}
  \partial_\mu T^{\mu \nu}(x)&=&0,
\end{eqnarray}
where the energy-momentum tensor is decomposed in the form
\begin{eqnarray}
\label{tmunu}
  T^{\mu\nu}&=& eu^{\mu}u^{\nu} - (p{+}\Pi)\Delta^{\mu \nu}
               + \pi^{\mu \nu}.
\end{eqnarray}
Here $e$ and $p$ are the local energy density and thermal equilibrium 
pressure, and $u^{\mu}$ is the (timelike and normalized, $u^\mu u_\mu\eq1$) 
4-velocity of the energy flow. $\Pi$ is the bulk viscous pressure; it 
combines with the thermal pressure $p$ to the total bulk pressure. In 
Eq.~(\ref{tmunu}) it is multiplied by the projector 
$\Delta^{\mu\nu}{\eq}g^{\mu\nu}-u^{\mu}u^{\nu}$ transverse to the flow 
velocity, \ie in the local fluid rest frame the bulk pressure is 
diagonal and purely spacelike, $(p+\Pi)\delta_{ij}$. $\pi^{\mu \nu}$ 
is the traceless shear viscous pressure tensor, also transverse to the 
4-velocity ($\pi^{\mu \nu}u_{\nu}\eq0$) and thus purely spatial in the 
local fluid rest frame.

For ideal fluids, $\Pi$ and $\pi^{\mu\nu}$ vanish, and the only dynamical
fields are $e(x)$, $p(x)$ and $u^\mu(x)$, with $e$ and $p$ related by 
the equation of state $p(e)$. In dissipative fluids without heat
conduction, $\Pi$ and the 5 independent components of $\pi^{\mu\nu}$ enter
as additional dynamical variables which require their own evolution
equations. In relativistic Navier-Stokes theory, these evolution equations 
degenerate to instantaneous constituent equations,
\begin{eqnarray}
\label{pimunu}
  \Pi = -\zeta\,\nabla{\cdot}u,\quad
  \pi^{\mu \nu} = 2 \eta\,\du,
\end{eqnarray}
which express the {\em dissipative flows} $\Pi$ and $\pi^{\mu\nu}$ 
directly in terms of their driving forces, the local expansion rate 
$\theta{\,\equiv\,}\nabla{\cdot}u$ and velocity shear tensor 
$\sigma^{\mu\nu}{\,\equiv\,}\du$, multiplied by phenomenological transport 
coefficients $\zeta,\,\eta{\,\geq\,}0$ (the bulk and shear viscosity,
respectively). Here $\nabla^\nu{\,\equiv\,}\Delta^{\mu\nu}\partial_\nu$
is the gradient in the local fluid rest frame, and 
$\du{\,\equiv\,}\frac{1}{2}(\nabla^\mu u^\nu{+}\nabla^\nu u^\mu)
- \frac{1}{3}(\nabla{\cdot}u)\Delta^{\mu\nu}$, showing that, 
like $\pi^{\mu\nu}$, the velocity shear tensor is traceless and transverse 
to $u^\mu$. The instantantaneous identification (\ref{pimunu}) leads to
causality problems through instantaneous signal propagation, so that
this straightforward relativistic generalization of the Navier-Stokes 
formalism turns out not to be a viable relativistic theory.

The Israel-Stewart approach \cite{Israel:1976tn} avoids these problems by 
replacing the instantaneous identifications (\ref{pimunu}) with the kinetic
evolution equations
\begin{eqnarray}
\label{Pi-transport}
  D{\Pi}&=&-\frac{1}{\tau_{\Pi}}\big(\Pi+\zeta \nabla{\cdot}u\big),
\\
  D\pi^{\mu \nu} &=&-\frac{1}{\tau_{\pi}}\big(\pi^{\mu\nu}-2\eta\du\big)
\nonumber\\
\label{pi-transport}
  && -\bigl(u^\mu\pi^{\nu\alpha} + u^\nu\pi^{\mu\alpha}\bigr)
      Du_\alpha, 
\end{eqnarray}
where $D=u^{\mu}\partial_{\mu}$ is the time derivative in the local fluid
rest frame, and the last term in the bottom equation ensures that the
kinetic evolution preserves the tracelessness and transversality of 
$\pi^{\mu\nu}$ \cite{fn2}. $\tau_{\Pi}$ and $\tau_{\pi}$ are
relaxation times and related to the 2nd order expansion coefficients
in the entropy current \cite{Israel:1976tn,Muronga:2001zk}. The fact
that in the Israel-Stewart approach the dissipative flows $\Pi$ and
$\pi^{\mu \nu}$ no longer respond to the corresponding thermodynamic 
forces $\nabla\cdot u$ and $\du$ instantaneously, but on finite albeit
short kinetic time scales, restores causality of the theory 
\cite{Israel:1976tn}.

We should caution that the form of the kinetic evolution equations
(\ref{Pi-transport},\ref{pi-transport}) is not generally agreed upon,
due to unresolved ambiguities in their derivation \cite{Israel:1976tn,%
Heinz:2005bw,Muronga:2001zk,Teaney:2004qa,Baier:2006um,Muronga:2004sf,%
Chaudhuri:2007zm,Romatschke:2007mq,Song:2007fn,Dusling:2007gi}. We will
here use the form given in Eqs.~(\ref{Pi-transport},\ref{pi-transport})
and comment on differences with other work when we discuss our results.

In the following calculations we further simplify the problem by
neglecting bulk viscosity. Bulk viscosity vanishes classically
for a system of massless partons, and quantum corrections arising
from the trace anomaly of the energy-momentum tensor are expected to 
be small, rendering bulk viscous effects much less important than 
those from shear viscosity. This expectation has been confirmed by 
recent lattice calculations \cite{Meyer:2007ic,Meyer:2007dy} which 
yield very small bulk viscosity in the QGP phase. The same calculations 
show, however, a rapid rise of the bulk viscosity near the quark-hadron 
phase transition \cite{Meyer:2007dy}, consistent with earlier predictions
\cite{Paech:2006st,Kharzeev:2007wb}. In the hadronic phase it is again
expected to be small \cite{Paech:2006st}. We leave the discussion of
possible dynamical effects of bulk viscosity near the quark-hadron
phase transition for a future study. Bulk viscous pressure effects can 
be easily restored by substituting $p{\,\to\,}p{+}\Pi$ everywhere below
and adding the kinetic evolution equation (\ref{Pi-transport}) for $\Pi$.

\subsection{Viscous hydrodynamics in 2+1 dimensions}
\label{sec2b}

In the present paper we eliminate one of the three spatial dimensions by
restricting our discussion to longitudinally boost-invariant systems.
These are conveniently described in curvilinear $x^m\eq(\tau,x,y,\eta)$ 
coordinates, where $\tau\eq\sqrt{t^2{-}z^2}$ is the longitudinal proper time, 
$\eta\eq\frac{1}{2}\ln\bigl(\frac{t{+}z}{t{-}z}\bigr)$ is the space-time
rapidity, and $(x,y)$ are the usual Cartestsian coordinates in the plane 
transverse to the beam direction $z$. In this coordinate system, the 
transport equations for the full energy momentum tensor $T^{\mu\nu}$ are 
written as~\cite{Heinz:2005bw}
\begin{eqnarray}
&&\partial_\tau \widetilde{T}^{\tau\tau}
 +\partial_x (v_x\widetilde{T}^{\tau\tau})
 +\partial_y (v_y\widetilde{T}^{\tau\tau}) = {\cal S}^{\tau\tau},
\nonumber \\
&&\partial_\tau \widetilde{T}^{\tau x}
 +\partial_x (v_x\widetilde{T}^{\tau x})
 +\partial_y (v_y\widetilde{T}^{\tau x}) = {\cal S}^{\tau x},
\label{transport-T}\\
&&\partial_\tau \widetilde{T}^{\tau y}
 +\partial_x (v_x\widetilde{T}^{\tau y}) 
 +\partial_y (v_y\widetilde{T}^{\tau y}) = {\cal S}^{\tau y}.
\nonumber
\end{eqnarray}
Here $\widetilde{T}^{m n}\equiv\tau(T_0^{mn}{+}\pi^{mn})$, 
$T_0^{mn}\eq{e}u^mu^n{-}p\Delta^{mn}$ being the ideal fluid contribution, 
$u^m\eq(u^\tau, u^x, u^y, 0)=\gamma_\perp(1,v_x,v_y,0)$ is the flow profile
(with $\gamma_\perp\eq\frac{1}{\sqrt{1{-}v_x^2{-}v_y^2}}$), 
and $g^{mn}\eq\mathrm{diag}(1,-1,-1,-1/\tau^2)$ is the metric tensor for 
our coordinate system. The source terms ${\cal S}^{mn}$ on the right hand 
side of Eqs.~(\ref{transport-T}) are given explicitly as
\begin{eqnarray}
  {\cal S}^{\tau\tau} &=& -p -\tau^2\pi^{\eta\eta} 
                   -\tau\partial_x(p v_x{+}\pi^{x\tau}{-}v_x\pi^{\tau\tau})
\nonumber\\
               && \qquad\qquad\quad\ 
                   -\,\tau\partial_y(p v_y{+}\pi^{y\tau}{-}v_y\pi^{\tau\tau})
\nonumber\\
\label{S00}
         &\approx& -p -\tau^2\pi^{\eta\eta} -\tau\partial_x(p v_x)
                                            -\tau\partial_y(p v_y),
\\
  {\cal S}^{\tau x}   &=& -\tau\partial_x(p{+}\pi^{xx}{-}v_x\pi^{\tau x})
                   -\tau\partial_y(\pi^{xy}{-}v_y\pi^{\tau x})
\nonumber\\
\label{S01}
         &\approx& -\tau\partial_x(p{+}\pi^{xx}),
\\
  {\cal S}^{\tau y}   &=& -\tau\partial_x(\pi^{xy}{-}v_x\pi^{\tau y})
                   -\tau\partial_y(p{+}\pi^{yy}{-}v_y\pi^{\tau y})
\nonumber\\
\label{S02}
         &\approx& -\tau\partial_y(p{+}\pi^{yy}).
\end{eqnarray}
We will see later (see the right panel of Fig.~\ref{AverPi} below) that 
the indicated approximations of these source terms isolate the dominant 
drivers of the evolution and provide a sufficiently accurate quantitative 
understanding of its dynamics.

The transport equations for the shear viscous pressure tensor are
\begin{eqnarray}
\label{transport-pi}
  &&(\partial_\tau + v_x \partial_x + v_y \partial_y)\tilde\pi^{mn} =
  -\frac{1}{\gamma_\perp\tau_\pi}(\tilde{\pi}^{mn}{-}2\eta\tilde{\sigma}^{mn})
\nonumber\\
  &&\qquad
  -\bigl(u^m\tilde{\pi}^n_{\ j}{+}u^n\tilde{\pi}^m_{\ j}\bigr)
   (\partial_\tau + v_x \partial_x + v_y \partial_y)u^j.
\end{eqnarray}
The expressions for $\tilde{\sigma}^{mn}$ and $\tilde{\pi}^{mn}$ are found
in Eqs.~(\ref{tilde-pi},\ref{tilde-sigma}); they differ from
$\pi^{mn}$ in Eqs.~(\ref{S00}-\ref{S02}) and $\sigma^{mn}$ given 
in Ref. ~\cite{Heinz:2005bw} by a Jacobian $\tau^2$ factor in the 
$(\eta\eta)$-component: $\tilde{\pi}^{\eta\eta}\eq\tau^2\pi^{\eta\eta}$,
$\tilde{\sigma}^{\eta\eta}\eq\tau^2\sigma^{\eta\eta}$. This factor arises 
from the curved metric where the local time derivative $D\eq{u^m}d_m$ must 
be evaluated using covariant derivatives $d_m$ \cite{fn3}. Since $u^\eta\eq0$,
no such extra Jacobian terms arise in the derivative $Du^j$ in the 
second line of Eq.~(\ref{transport-pi}).

The algorithm requires the propagation of $\pi^{\tau\tau}$, $\pi^{\tau x}$, 
$\pi^{\tau y}$, and $\pi^{\eta\eta}$ with Eq.~(\ref{pi-transport}) even 
though one of the first three is redundant (see \cite{Song:2007fn} and 
Appendix~\ref{appb}). In addition, we have chosen to evolve several more, 
formally redundant components of $\pi^{mn}$ using Eq.~(\ref{pi-transport}),
and to use them for testing the numerical accuracy of the code, by checking 
that the transversality conditions $u_m\pi^{mn}\eq0$ and the tracelessness
$\pi^m_m\eq0$ are preserved over time. We find them to be satisfied 
with an accuracy of better than $1-2\%$ everywhere except for the fireball
edge where the $\pi^{mn}$ are very small and the error on the transversality 
and tracelessness constraints can become as large as 5\%.

\subsection{Initial conditions}
\label{sec2c}

For the ideal part $T_0^{\tau\tau},\, T_0^{\tau x},\, T_0^{\tau x}$ of 
the energy momentum tensor we use the same initialization scheme as 
for ideal hydrodynamics. For simplicity and ease of comparison with
previous ideal fluid dynamical studies we here use a simple Glauber model 
initialization with zero initial transverse flow velocity where the initial 
energy density in the transverse plane is taken proportional to the wounded 
nucleon density \cite{Kolb:1999it}:
\begin{eqnarray}
&& e_0(x,y;b)=K n_\mathrm{WN}(x,y;b) 
\\
&& = K \biggl\{T_A\bigl(x{+}{\textstyle\frac{b}{2}},y)\biggl[1-
       \biggl(1-\frac{\sigma T_B\left(x{-}\frac{b}{2},y\right)}{B}\biggr)^B 
       \biggr] 
\nonumber\\ 
&& \quad\ \  +\, T_B\bigl(x{-}{\textstyle\frac{b}{2}},y\bigr) \biggl[1-
       \biggl(1-\frac{\sigma T_A\left(x{+}\frac{b}{2},y\right)}{A}\biggr)^A
       \biggr]\biggr\}.
\nonumber
\end{eqnarray}
Here $\sigma$ is the total inelastic nucleon-nucleon cross section 
for which we take $\sigma\eq40$\,mb. $T_{A,B}$ is the nuclear thickness 
function of the incoming nucleus A or B, $T_A(x,y)\eq\int^\infty_{-\infty}
dz \rho_A(x,y,z)$; $\rho_A(x,y,z)$ is the nuclear density given by a 
Woods-Saxon profile: $\rho_A(\bm{r})\eq\frac{\rho_0}{1+\exp[(r-R_A)/\xi]}$. 
For Cu+Cu collisions we take $R_\mathrm{Cu}\eq4.2$\,fm, $\xi\eq0.596$\,fm,
and $\rho_0\eq0.17$\,fm$^{-3}$. The proportionality constant $K$ does not
depend on collision centrality but on collision energy; it fixes the 
overall scale of the initial energy density and, via the associated 
entropy, the final hadron multiplicity to which it must be fitted as a
function of collision energy. We here fix it to give 
$e_0{\,\equiv\,}e(0,0;b{=}0)\eq30$\,GeV/fm$^3$ for the peak energy density 
in central Cu+Cu collisions, at an initial time $\tau_0$ for the hydrodynamic 
evolution that we set as $\tau_0\eq0.6$\,fm/$c$. As already mentioned in
the Introduction, this exceeds the value reached in Cu+Cu collisions at 
RHIC (it would be appropriate for central Au+Au collisions at 
$\sqrt{s}\eq200\,A$\,GeV \cite{Rev-hydro}). It ensures, however, 
a sufficiently long lifetime of the QGP phase in Cu+Cu collisions that
most of the final momentum anisotropy is generated during the QGP stage, 
thereby permitting a meaningful investigation of QGP viscosity on the 
elliptic flow generated in the collision.
 
Lacking a microscopic dynamical theory for the early pre-equilibrium 
stage, initializing the viscous pressure tensor $\pi^{mn}$ requires 
some guess-work. The effect of different choices for the initial 
$\pi^{mn}$ on viscous entropy production during boost-invariant viscous
hydrodynamic evolution without transverse expansion was recently studied 
in \cite{Dumitru:2007qr}. We will here explore two options: (i) we set
$\pi^{mn}_0\eq0$ initially \cite{Romatschke:2007mq}; or (ii) we assume 
that at time $\tau_0$ one has $\pi^{mn}_0\eq2\eta\sigma^{mn}_0$ where the 
shear tensor $\sigma^{mn}_0$ is calculated from the initial velocity 
profile $u^m\eq(1,0,0,0)$ \cite{Chaudhuri:2005ea,Chaudhuri:2007zm,%
Chaudhuri:2006jd,Dusling:2007gi}. The second option is the default choice 
for most of the results shown in this paper. It gives $\tau^2 
\pi^{\eta\eta}_0\eq{-2}\pi^{xx}_0\eq{-2}\pi^{yy}_0\eq{-}\frac{4\eta}{3
\tau_0}$, i.e. a negative contribution to the longitudinal pressure and 
a positive contribution to the transverse pressure.

We here present results only for one value of the specific shear viscosity,
$\frac{\eta}{s}=\frac{1}{4\pi}\simeq0.08$, corresponding to its conjectured 
lower limit \cite{Ads-CFT}. The kinetic relaxation time $\tau_\pi$ will
be taken as $\tau_\pi\eq\frac{3\eta}{sT}$ except were otherwise mentioned.
This value is half the one estimated from classical kinetic theory for
a Boltzmann gas of non-interacting massless partons 
\cite{Israel:1976tn,Baier:2006um} -- we did not use the twice larger 
classical value because it led to uncomfortably large viscous pressure
tensor components $\pi^{mn}$ at early times, caused by large excursions
from the Navier-Stokes limit. To study the sensitivity to 
different relaxation times and the approach to Navier-Stokes theory, we 
also did a few calculations with $\tau_\pi\eq\frac{1.5\eta}{sT}$ in 
Section \ref{sec5b}.

\subsection{EOS}\label{sec2d}

Through its dependence on the Equation of State (EOS), hydrodynamic flow
constitutes an important probe for the existence and properties of the 
quark-hadron phase transition which softens the EOS near $T_c$. To isolate
effects induced by this phase transition from generic features of viscous
fluid dynamics we have performed calculations with two different equations
of state, EOS~I and SM-EOS~Q. They are described in this subsection.

{\bf EOS~I} models a non-interacting gas of massless quarks and gluons, 
with $p\eq\frac{1}{3}e$. It has no phase transition. Where needed,
the temperature is extracted from the energy density via the relation
$e\eq\left(16+\frac{21}{2}N_f\right)\frac{\pi^2}{30}\frac{T^4}{(\hbar c)^3}$,
corresponding to a chemically equilibrated QGP with $N_f\eq2.5$ effective 
massless quark flavors.

%
\begin{figure}[htb]
\includegraphics[width=\linewidth,clip=]{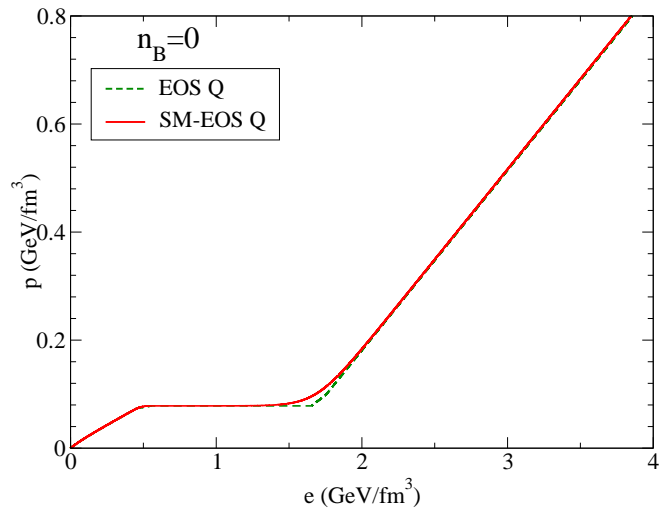}
\caption{The equations of state EOS~Q (dashed line) and SM-EOS~Q (``modified
EOS~Q'', solid line).}
\label{EOS}
\end{figure}
%

{\bf SM-EOS~Q} is a smoothed version of {\bf EOS~Q} \cite{Kolb:1999it} 
which connects a noninteracting QGP through a first order phase transition 
to a chemically equilibrated hadron resonance gas. In the QGP phase 
it is defined by the relation $p\eq\frac{1}{3}e{-}\frac{4}{3}B$ (i.e.
$c_s^2\eq\frac{\partial p}{\partial e}\eq\frac{1}{3}$). The vacuum energy 
(bag constant) $B^{1/4}\eq230$\,MeV is a parameter that is adjusted to 
yield a critical temperature $T_c=164$\,MeV. The hadron resonance gas 
below $T_c$ can be approximately characterized by the relation $p\eq0.15\,e$
(i.e. $c_s^2\eq0.15$) \cite{Kolb:1999it}. The two sides are matched 
through a Maxwell construction, yielding a relatively large latent heat 
$\Delta e_\mathrm{lat}\eq1.15$\,GeV/fm$^3$. For energy densities between
$e_\mathrm{H}\eq0.45$\,GeV/fm$^3$ and $e_\mathrm{Q}\eq1.6$\,GeV/fm$^3$ one 
has a mixed phase with constant pressure (i.e. $c_s^2\eq0$). The 
discontinuous jumps of $c_s^2$ from a value of 1/3 to 0 at $e_\mathrm{Q}$ 
and back from 0 to 0.15 at $e_\mathrm{H}$ generate propagating numerical 
errors in VISH2+1 which grow with time and cause problems. We avoid these 
by smoothing the function $c_s^2(e)$ with a Fermi distribution of width 
$\delta e\eq0.1$\,GeV/fm$^3$ centered at $e\eq{e_\mathrm{Q}}$
and another one of width $\delta e\eq0.02$\,GeV/fm$^3$ centered at 
$e\eq{e_\mathrm{H}}$. Both the original EOS~Q and our smoothed version 
SM-EOS~Q are shown in Figure~\ref{EOS}. A comparison of simulations using 
ideal hydrodynamics with EOS~Q and SM-EOS~Q is given in Appendix~\ref{appd1}.
It gives an idea of the magnitude of smoothing effects on the ideal 
fluid evolution of elliptic flow.

Another similarly smoothed EOS that matches a hadron resonance gas 
below $T_c$ with lattice QCD data above $T_c$ has also been constructed.
Results using this lattice based EOS will be reported elsewhere.

\subsection{Freeze-out procedure: Particle spectra and $v_2$}
\label{sec2e}

Final particle spectra are computed from the hydrodynamic output via a
modified Cooper-Frye procedure \cite{Cooper-Frye}. We here compute spectra
only for directly emitted particles and do not include feeddown from 
resonance decay after freeze-out. We first determine the freeze-out 
surface $\Sigma(x)$, by postulating (as common in hydrodynamic studies) 
that freeze-out from a thermalized fluid to free-streaming, non-interacting 
particles happens suddenly when the temperature drops below a critical 
value. As in the ideal fluid case with EOS~Q \cite{Kolb:1999it} we choose 
$T_\mathrm{dec}\eq130$\,MeV. The particle spectrum is then computed as an 
integral over this surface,
\begin{eqnarray}
\label{Cooper}
  E\frac{d^3N_i}{d^3p} &=& \frac{g_i}{(2\pi)^3}\int_\Sigma 
  p\cdot d^3\sigma(x)\, f_i(x,p)
\\
  &=& \frac{g_i}{(2\pi)^3}\int_\Sigma p\cdot d^3\sigma(x)
 \left[f_{\mathrm{eq},i}(x,p)+\delta f_i(x,p) \right],
\nonumber
\end{eqnarray}
where $g_i$ is the degeneracy factor for particle species $i$, 
$d^3\sigma^\mu(x)$ is the outward-pointing surface normal vector 
on the decoupling surface $\Sigma(x)$ at point $x$,
\begin{eqnarray}
\label{dsigma}
  p \cdot d^3\sigma(x) &=& \big[m_T\cosh(y{-}\eta)-\bm{p}_\perp\cdot
  \bm{\nabla}_\perp\tau_f(\bm{r})\big]
\nonumber\\
  &\times& 
  \tau_f(\bm{r})\, r dr\, d\phi\, d\eta
\end{eqnarray}
(with $\bm{r}\eq(x,y)\eq(r\cos\phi,r\sin\phi)$ denoting the transverse
position vector), and $f_i(x,p)$ is the local distribution function for 
particle species $i$, computed from hydrodynamic output. Equation 
(\ref{Cooper}) generalizes the usual Cooper-Frye prescription for ideal 
fluid dynamics \cite{Cooper-Frye} by accounting for the fact that in a 
viscous fluid the local distribution function is never exactly in
local equilibrium, but deviates from local equilibrium form by 
small terms proportional to the non-equilibrium viscous flows
\cite{Teaney:2003kp,Baier:2006um}. Both contributions can be extracted
from hydrodynamic output along the freeze-out surface. The equilibrium
contribution is
\begin{eqnarray}
\label{f0}
  f_{\mathrm{eq},i}(p,x) 
  = f_{\mathrm{eq},i}\Bigl(\frac{p{\cdot}u(x)}{T(x)}\Bigr)
  = \frac{1}{e^{p\cdot u(x)/T(x)}\pm 1},
\end{eqnarray}
where the exponent is computed from the temperature $T(x)$ and hydrodynamic
flow velocity $u^\mu\eq\gamma_\perp(\cosh\eta,v_\perp\cos\phi_v,
v_\perp\sin\phi_v,\sinh\eta)$ along the surface $\Sigma(x)$:
\begin{eqnarray}
\label{pdotu}
  p\cdot u(x)\eq\gamma_\bot[m_T\cosh(y{-}\eta) 
                          - p_T v_\perp \cos(\phi_p{-}\phi_v)].\ 
\end{eqnarray}
Here 
$m_T\eq\sqrt{p_T^2{+}m_i^2}$ is the particle's transverse mass. The 
viscous deviation from local equilibrium is given by 
\cite{Teaney:2003kp,Baier:2006um}
\begin{eqnarray}
\label{deltaf}
  \delta f_i(x,p) \!\!&=&\!\! 
  f_{\mathrm{eq},i}(p,x) \bigl(1{\mp}f_{\mathrm{eq},i}(p,x)\bigr)
  \frac{p^\mu p^\nu \pi_{\mu\nu}(x)}{2T^2(x)\left(e(x){+}p(x)\right)}
\nonumber\\
  \!\!&\approx&\!\! 
  f_{\mathrm{eq},i} \cdot \frac{1}{2} \frac{p^\mu p^\nu}{T^2}\,
                          \frac{\pi_{\mu\nu}}{e{+}p}.
\end{eqnarray}
The approximation in the second line is not used in our numerical results
but it holds (within the line thickness in all of our corresponding plots) 
since $(1{\mp}f_\mathrm{eq})$ deviates from 1 only when $p{\,\ll\,}T$ where 
the last factor is small. With it the spectrum (\ref{Cooper}) takes the 
instructive form
\begin{eqnarray}
\label{Cooper1}
  E\frac{d^3N_i}{d^3p} =
  \frac{g_i}{(2\pi)^3}\!\int_\Sigma\! p\cdot d^3\sigma(x)
  f_{\mathrm{eq},i}(x,p)\Bigl(1{+}{\textstyle{\frac{1}{2}}} 
                                \frac{p^\mu p^\nu}{T^2}\,
                                \frac{\pi_{\mu\nu}}{e{+}p}\Bigr).
\!\!\!\!\!\!\nonumber\\
\end{eqnarray}
The viscous correction is seen to be proportional to $\pi^{\mu\nu}(x)$ on 
the freeze-out surface (normalized by the equilibrium enthalpy $e{+}p$) 
and to increase quadratically with the particle's momentum (normalized by 
the temperature $T$). At large $p_T$, the viscous correction can exceed 
the equilibrium contribution, indicating a breakdown of viscous 
hydrodynamics. In that domain, particle spectra can not be reliably 
computed with viscous fluid dynamics. The limit of applicability
depends on the actual value of $\pi^{\mu\nu}/(e{+}p)$ and thus on the 
specific dynamical conditions encountered in the heavy-ion collision.
 
%
%
The viscous correction to the spectrum in Eq.~(\ref{Cooper1}) reads
explicitly 
\begin{eqnarray}
  p_\mu p_\nu \pi^{\mu\nu} &=&
  m_T^2\bigl(\cosh^2(y{-}\eta)\pi^{\tau\tau}+\sinh^2(y{-}\eta)
                 \tau^2\pi^{\eta\eta}\bigr)
\nonumber\\
  &-& 2m_T\cosh(y{-}\eta)\bigl(p_x\pi^{\tau x}+p_y\pi^{\tau y}\bigr)
\nonumber\\
  &+& \bigl(p_x^2\pi^{xx}+2p_xp_y\pi^{xy}+p_y^2\pi^{yy}\bigr).
\label{vis-correction}
\end{eqnarray}
%
We can use the expressions given in Appendix 2 of Ref.~\cite{Heinz:2005bw}
(in particular Eqs.~(A22) in that paper) to re-express this in terms
of the three independent components of $\pi^{mn}$ for which we choose 
\begin{equation}
\label{indep}
  \tilde{\pi}^{\eta\eta} = \tau^2\pi^{\eta\eta},\quad 
  \Sigma = \pi^{xx}{+}\pi^{yy},\quad 
  \Delta = \pi^{xx}{-}\pi^{yy}.
\end{equation}
This choice is motivated by our numerical finding (see Fig.~2 in 
\cite{Song:2007fn} and Sec.~\ref{sec4c}) that $\tilde{\pi}^{\eta\eta}$, 
$\pi^{xx}$ and $\pi^{yy}$ are about an order of magnitude larger than 
all other components of $\pi^{mn}$, and that in the azimuthally symmetric 
limit of central ($b\eq0$) heavy-ion collisions the azimuthal average of 
$\Delta$ vanishes (see Eq.~(\ref{C2})): $\langle\Delta\rangle_\phi\eq0$.
We find
\begin{widetext}
\begin{eqnarray}
  p_\mu p_\nu \pi^{\mu\nu} \!\!&=&\!\! \tilde{\pi}^{\eta\eta}
  \left[m_T^2\bigl(2\cosh^2(y{-}\eta){-}1\bigr)
      -2 \frac{p_T}{v_\perp} m_T\cosh(y{-}\eta)
         \frac{\sin(\phi_p{+}\phi_v)}{\sin(2\phi_v)}
       +\frac{p_T^2}{v_\perp^2}\frac{\sin(2\phi_p)}{\sin(2\phi_v)}\right]
\nonumber\\
 &+&\!\! \Sigma 
  \left[m_T^2\cosh^2(y{-}\eta)
       -2\frac{p_T}{v_\perp}m_T\cosh(y{-}\eta)
       \left(\frac{\sin(\phi_p{+}\phi_v)}{\sin(2\phi_v)} - \frac{v_\perp^2}{2}
             \frac{\sin(\phi_p{-}\phi_v)}{\tan(2\phi_v)}\right)
       +\frac{p_T^2}{2} 
       + \frac{p_T^2}{v_\perp^2}\left(1{-}\frac{v_\perp^2}{2}\right)
         \frac{\sin(2\phi_p)}{\sin(2\phi_v)}\right]
\!\!\!\!
\nonumber\\
  &+&\!\! \Delta
  \left[p_T m_T\cosh(y{-}\eta) v_\perp
        \frac{\sin(\phi_p{-}\phi_v)}{\sin(2\phi_v)}
       -\frac{p_T^2}{2}\frac{\sin(2(\phi_p{-}\phi_v))}{\sin(2\phi_v)}
  \right].
\label{vis-cor}
\end{eqnarray}
\end{widetext}
%

%
\begin{figure*}
\includegraphics[bb=1 32 710 557,width=.49\linewidth,clip=]{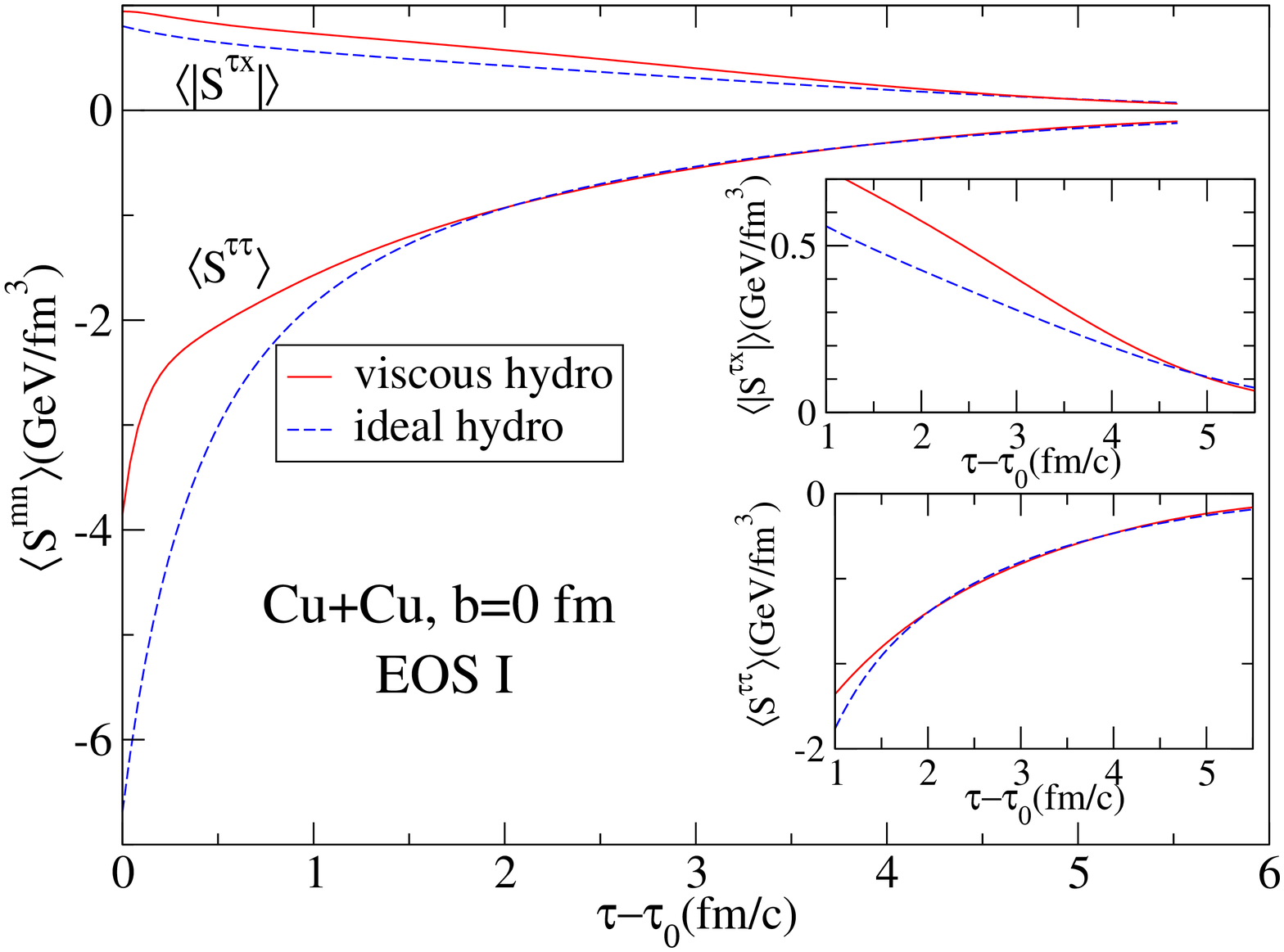}
\includegraphics[bb=1 32 710 557,width=.49\linewidth,clip=]{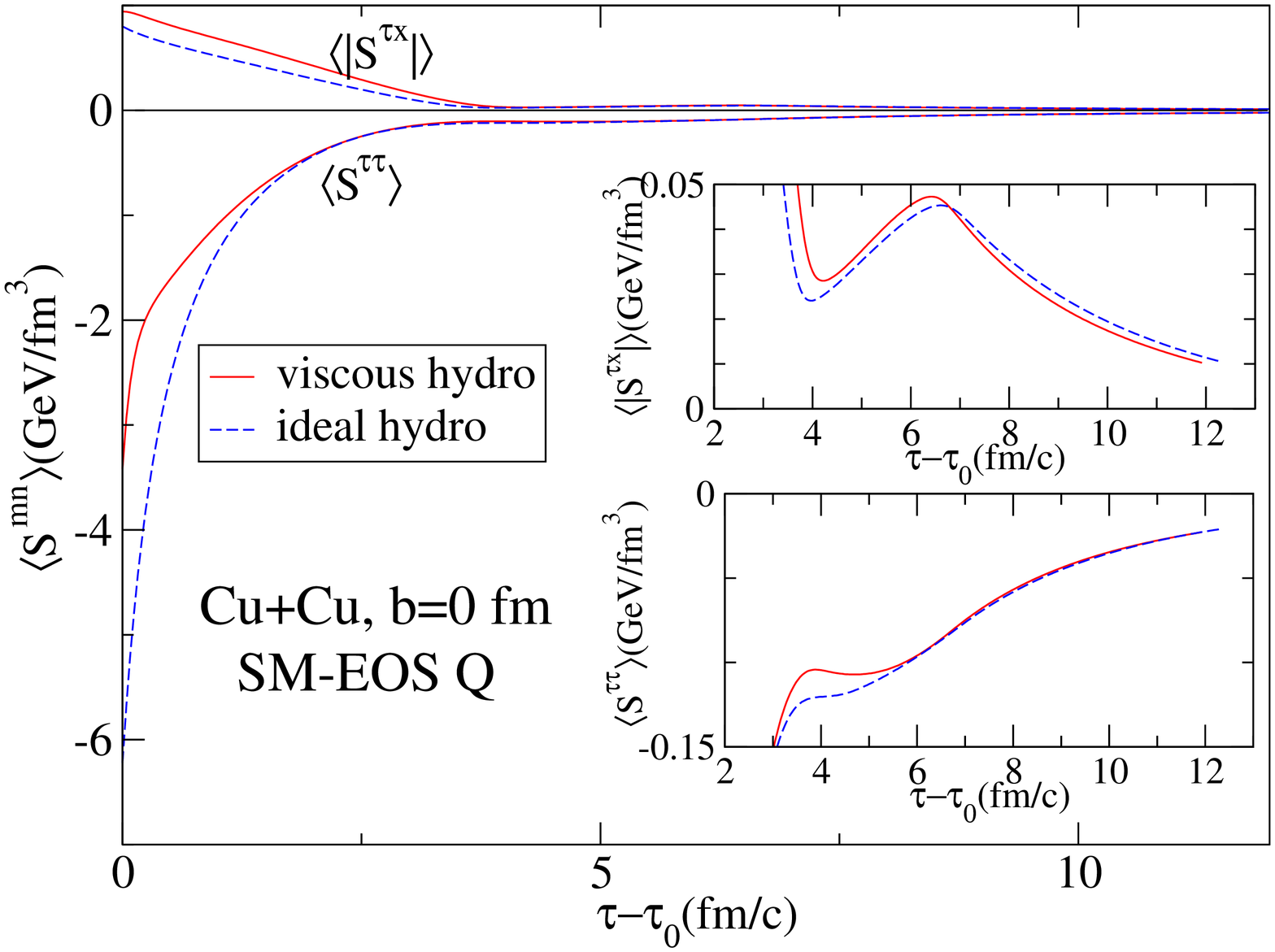}
\caption{(Color online) 
Time evolution of the hydrodynamic source terms (\ref{S00}-\ref{S02}),
averaged over the transverse plane, for central Cu+Cu collisions, 
calculated with EOS~I in the left panel and with SM-EOS~Q in the
right panel. The smaller insets blow up the vertical scale to show more 
detail. The dashed blue lines are for ideal hydrodynamics with 
$e_0\eq30$\,GeV/fm$^3$ and $\tau_0\eq0.6$\,fm/$c$. Solid red lines show 
results from viscous hydrodynamics with identical initial conditions and 
$\frac{\eta}{s}\eq\frac{1}{4\pi}{\,\approx\,}0.08$, $\tau_\pi\eq\frac{3\eta}
{sT}{\,\approx\,}0.24\left(\frac{200\,\mathrm{MeV}}{T}\right)$\,fm/$c$.
The positive source terms drive the transverse expansion while the negative
ones affect the longitudinal expansion.}
\label{source}
\end{figure*}
%


Due to longitudinal boost-invariance, the integration over space-time
rapidity $\eta$ in Eq.~(\ref{Cooper}) can be done analytically, 
resulting in a series of contributions involving modified Bessel 
functions \cite{concepts,Baier:2006gy}. 
VISH2+1 does not exploit this possibility and instead performs this
and all other integrations for the spectra numerically.

Once the spectrum (\ref{Cooper}) has been computed, a Fourier decomposition
with respect to the azimuthal angle $\phi_p$ yields the anisotropic flow
coefficients. For collisions between equal spherical nuclei followed by
longitudinally boost-invariant expansion of the collision fireball, only 
even-numbered coefficients contribute, the ``elliptic flow'' $v_2$ being
the largest and most important one:
\begin{eqnarray}
  && E\frac{d^3N_i}{d^3p}(b) = \frac{dN_i}{dy\, p_T dp_T\, d\phi_p}(b)
\\  
\nonumber
 &&\ = \frac{1}{2\pi}\frac{dN_i}{dy\, p_T dp_T}\big[1 + 2 v_2(p_T;b) 
     \cos(2\phi_p) + \dots \big].
\label{eq-V2}
\end{eqnarray}
In practice it is evaluated as the $\cos(2\phi_p)$-moment of the
final particle spectrum, 
\begin{eqnarray}
\label{cos2phi}
  v_2(p_T)=\langle\cos(2\phi_p)\rangle\equiv
  \frac{\int d\phi_p\,\cos(2\phi_p)\,\frac{dN}{dy\,p_T dp_T\,d\phi_p}}
                {\int d\phi_p\,\frac{dN}{dy\,p_T dp_T\,d\phi_p}}\,,\quad
\end{eqnarray}
where, according to Eq.~(\ref{Cooper}), the particle spectrum is a sum of
a local equilibrium and a non-equilibrium contribution (to be indicated 
symbolically as $N\eq{N}_\mathrm{eq}+\delta N$).

\section{Central collisions}
\label{sec3}
\subsection{Hydrodynamic evolution}
\label{sec3a}

%
\begin{figure*}
\includegraphics[bb=25 33 720 531,width=.49\linewidth,clip=]{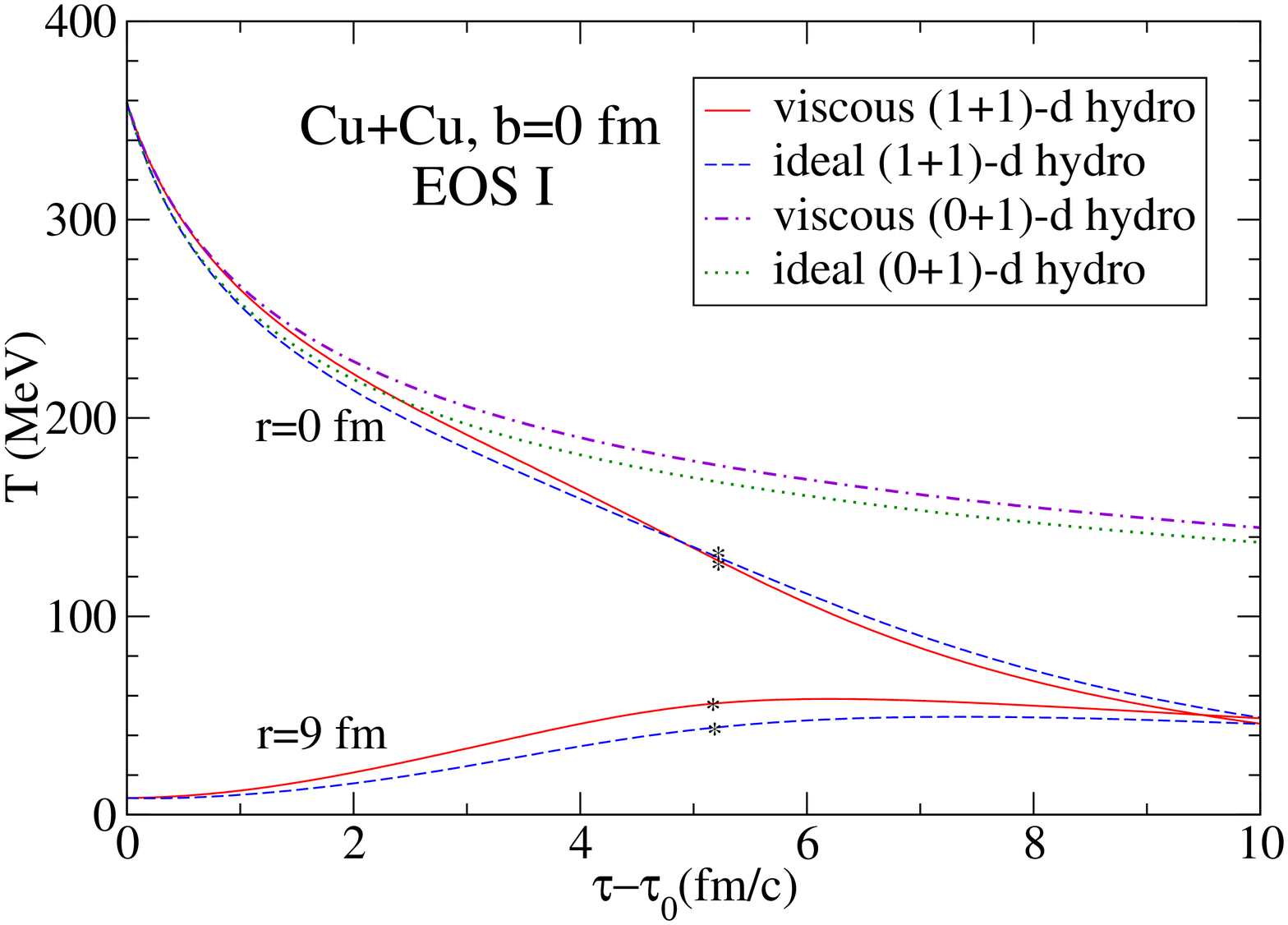}
\includegraphics[bb=25 33 720 522,width=.49\linewidth,clip=]{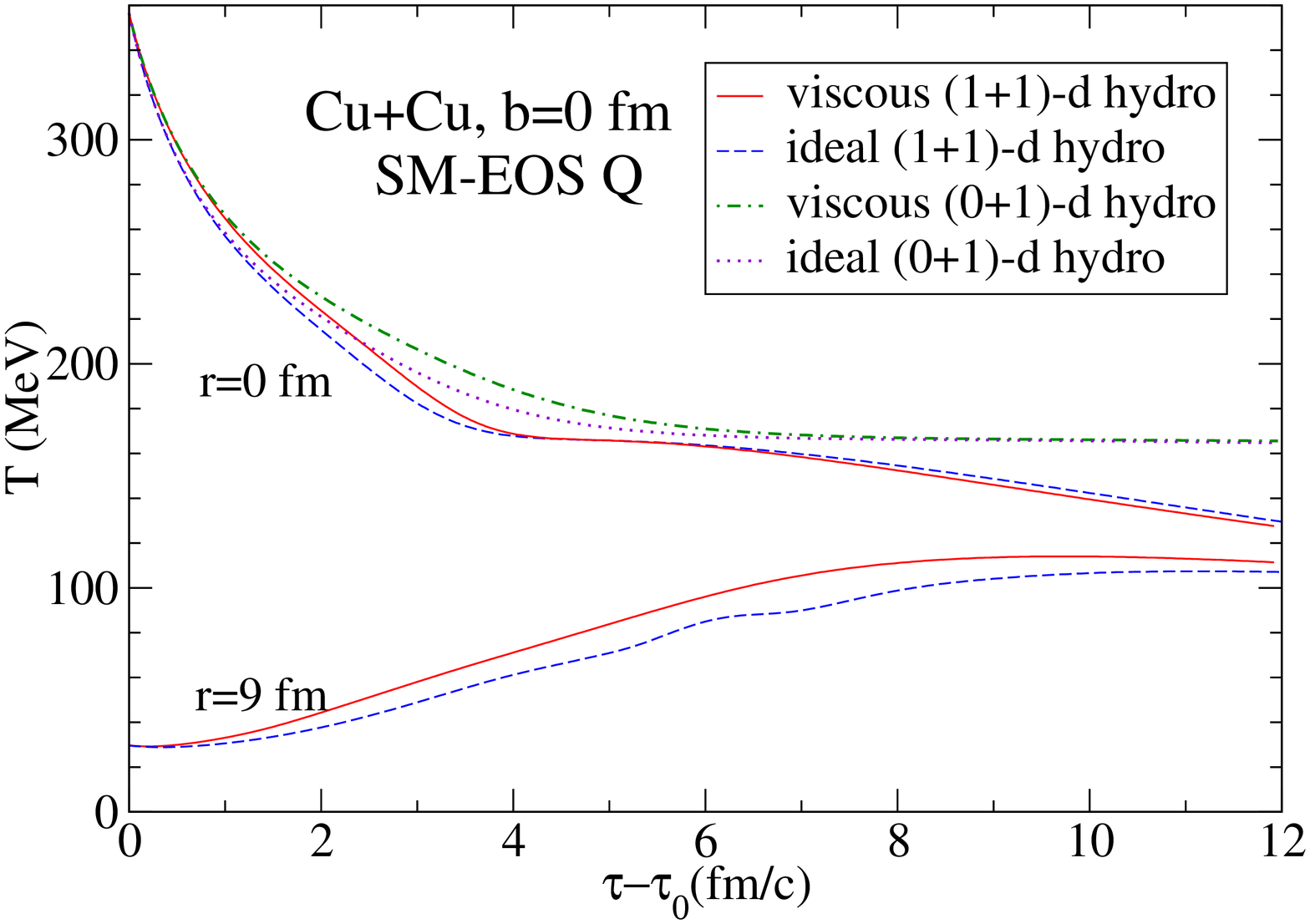}
\caption{(Color online) 
Time evolution of the local temperature in central Cu+Cu collisions, 
calculated with EOS~I (left) and SM-EOS~Q (right), for the center of
the fireball ($r\eq0$, upper set of curves) and a point near the edge
($r\eq9$\,fm, lower set of curves). Same parameters as in Fig.~\ref{source}.
See text for discussion.
}
\label{localT}
\end{figure*}
%
Even without transverse flow initially, the boost-invariant longitudinal
expansion leads to a non-vanishing initial stress tensor $\sigma^{mn}$
which generates non-zero target values for three components of the shear 
viscous pressure tensor: $\tau^2\pi^{\eta \eta}\eq\frac{-4\eta}{3\tau_0}$,
$\pi^{xx}\eq\pi^{yy}\eq\frac{2\eta}{3\tau_0}$. Inspection of the source
terms in Eqs.~(\ref{S00}-\ref{S02}) then reveals that the initially negative 
$\tau^2\pi^{\eta \eta}$ reduces the longitudinal pressure, thus reducing
the rate of cooling due to work done by the latter, while the initially 
positive values of $\pi^{xx}$ and $\pi^{yy}$ add to the transverse pressure
and accelerate the devolpment of transverse flow in $x$ and $y$ directions.
As the fireball evolves, the stress tensor $\sigma^{mn}$ receives additional 
contributions involving the transverse flow velocity and its derivatives
(see Eq.~(\ref{tilde-sigma})) which render an analytic discussion of
its effects on the shear viscous pressure impractical.

%
\begin{figure*}
\includegraphics[bb=30 42 710 524,width=.49\linewidth,clip=]{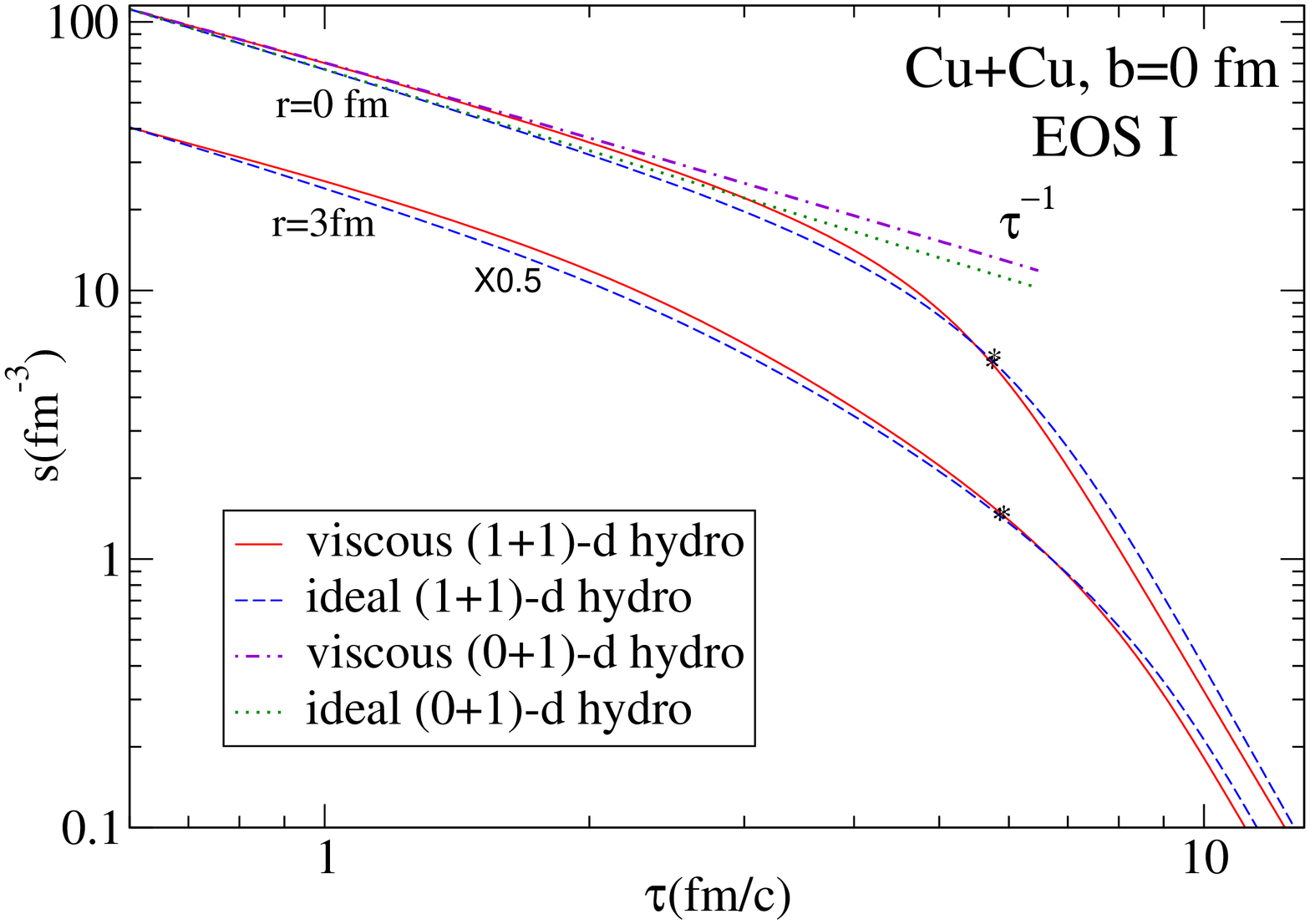}
\includegraphics[bb=30 42 710 522,width=.49\linewidth,clip=]{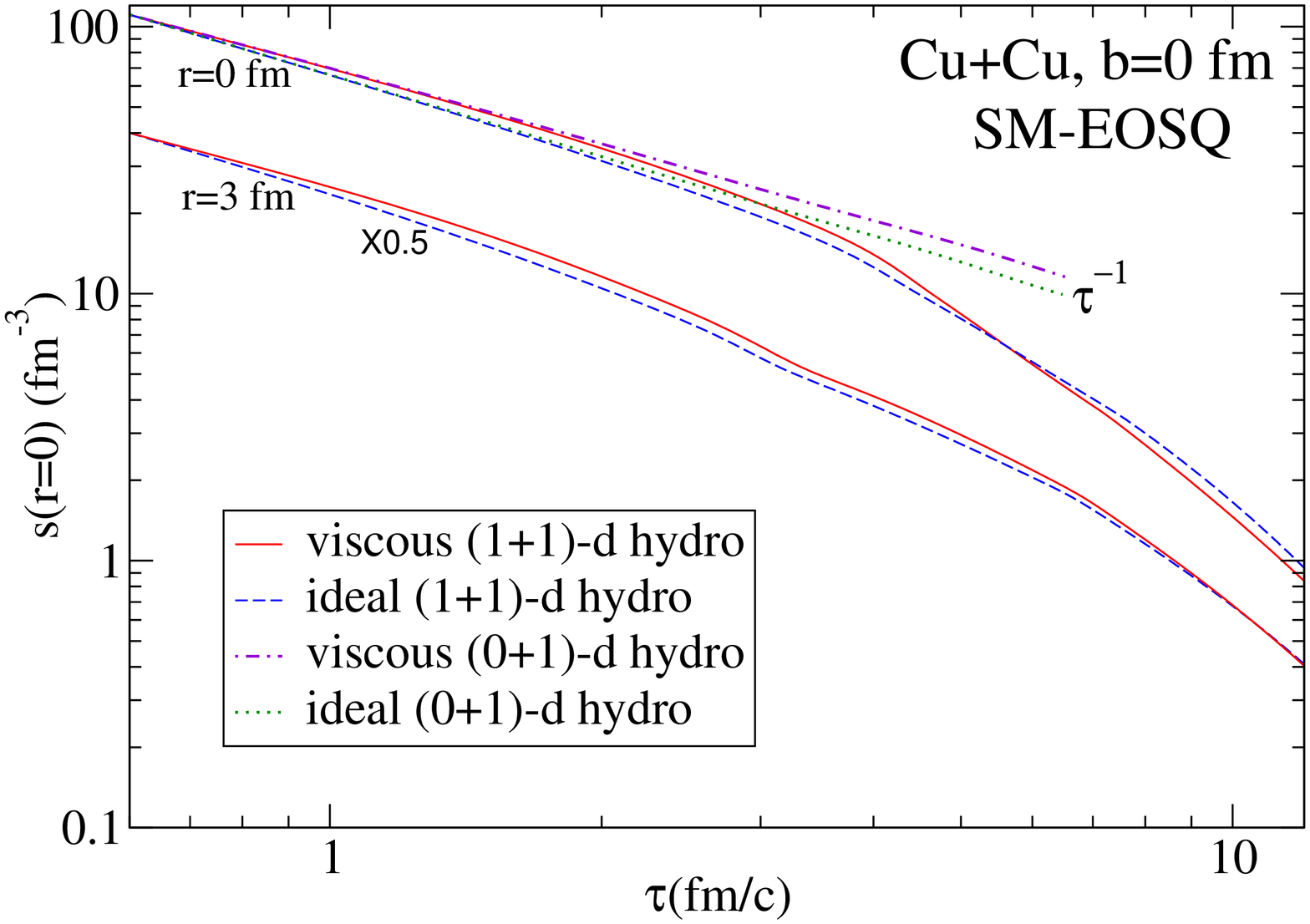}
\caption{(Color online) 
Time evolution of the local entropy density for central Cu+Cu collisions, 
calculated with EOS~I (left) and SM-EOS~Q (right), for the center of
the fireball ($r\eq0$, upper set of curves) and a point at $r\eq3$\,fm 
(lower set of curves). Same parameters and color coding as in 
Fig.~\ref{localT}. See text for discussion.
}
\label{localS}
\end{figure*}
%
Figure~\ref{source} shows what one gets numerically. Plotted are the
source terms (\ref{S00}) and (\ref{S01}), averaged over the transverse 
plane with the energy density as weight function, as a function of time, 
for evolution of central Cu+Cu collisions with two different equations 
of state, EOS~I and SM-EOS~Q. (In central collisions $\langle|{\cal 
S}^{\tau x}|\rangle\eq\langle|{\cal S}^{\tau y}|\rangle$.) One sees that 
the initially strong  viscous reduction of the (negative) source term 
${\cal S}^{\tau\tau}$, which controls the cooling by longitudinal expansion, 
quickly disappears. This is due to a combination of effects: while the 
magnitude of $\tau^2\pi^{\eta\eta}$ decreases with time, its negative 
effects are further compensated by a growing positive contribution 
$\tau\bigl(\partial_x(pv_x){+}\partial_y(pv_y)\bigl)$ arising from the 
increasing transverse flow gradients. In contrast, the viscous increase
of the (positive) transverse source term ${\cal S}^{\tau x}$ persists
much longer, until about 5\,fm/$c$. After that time, however, the viscous
correction switches sign (clearly visible in the upper inset in the right 
panel of Fig.~\ref{source}b) and turns negative, thus reducing the transverse 
acceleration at late times relative to the ideal fluid case. We can summarize 
these findings by stating that shear viscosity reduces longitudinal
cooling mostly at early times while causing initially increased but 
later reduced acceleration in the transverse direction. Due to the 
general smallness of the viscous pressure tensor components at late 
times, the last-mentioned effect (reduced acceleration) is not very strong.

The phase transition in SM-EOS~Q is seen to cause an interesting 
non-monotonic behaviour of the time evolution of the source terms (right 
panel in Fig.~\ref{source}), leading to a transient increase of the
viscous effects on the longitudinal source term while the system passes 
through the mixed phase.

The viscous slowing of the cooling process at early times and the 
increased rate of cooling at later times due to accelerated transverse 
expansion are shown in Figure~\ref{localT}. The upper set of curves shows 
what happens in the center of the fireball. For comparison we also show 
curves for boost-invariant longitudinal Bjorken expansion without 
transverse flow, labeled ``(0+1)-d hydro''. These are obtained with flat 
initial density profiles for the same value $e_0$ (no transverse 
gradients). The dotted green line in the left panel shows the well-known 
$T{\,\sim\,}\tau^{-1/3}$ behaviour of the Bjorken solution of ideal
fluid dynamics \cite{Bjorken:1982qr}, modified in the right panel by 
the quark-hadron phase transition where the temperature stays constant 
in the mixed phase. The dash-dotted purple line shows the slower 
cooling in the viscous (0+1)-dimensional case \cite{Gyulassy85}, due to 
reduced work done by the longitudinal pressure. The expansion is still 
boost-invariant a la Bjorken \cite{Bjorken:1982qr} (as it is for 
all other cases discussed in this paper), but viscous effects generate 
entropy, thereby keeping the temperature at all times higher than 
for the adiabatic case. The dashed blue (ideal) and solid red (viscous) 
lines for the azimuthally symmetric (1+1)-dimensional case show the 
additional cooling caused by transverse expansion. Again the cooling is 
initially slower in the viscous case (solid red), but at later times, due 
to faster build-up of transverse flow by the viscously increased 
transverse pressure, the viscous expansion is seen to cool the fireball 
center {\em faster} than ideal hydrodynamics. (Note also the drastic 
reduction of the lifetime of the mixed phase by transverse expansion; 
due to increased transverse flow and continued acceleration in the mixed 
phase from viscous pressure gradients, it is even more dramatic in the viscous 
than the ideal case.) The curves for $r\eq9$\,fm corroborate this, showing 
that the temperature initially increases with time due to hot matter 
being pushed from the center towards the edge, and that this temperature 
increase happens more rapidly in the viscous fluid (solid red lines), due 
to the faster outward transport of matter in this case.

%
\begin{figure*}[htb]
\includegraphics[bb=36 37 740 581,width=.49\linewidth,clip=]{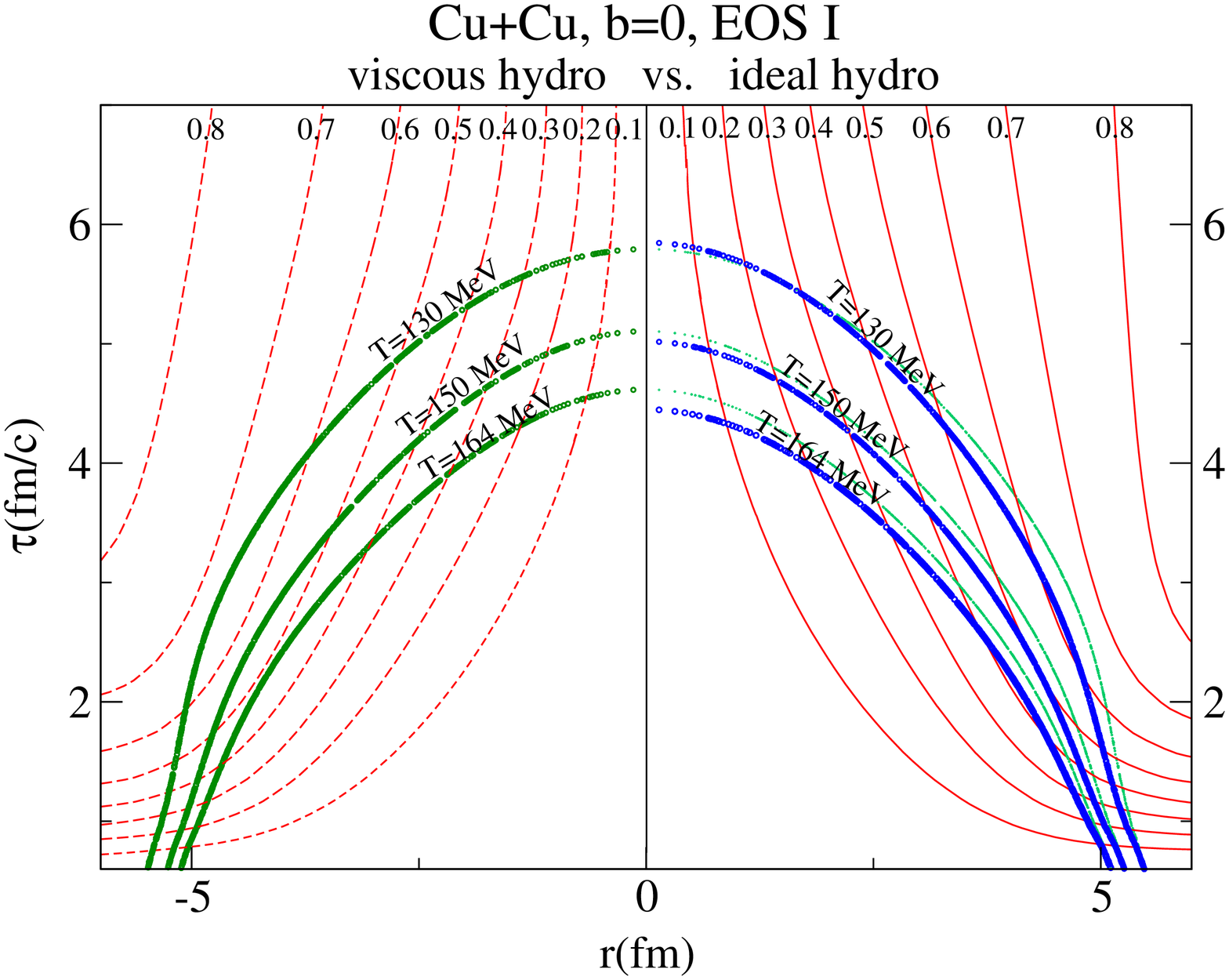}
\includegraphics[bb=36 37 740 581,width=.49\linewidth,clip=]{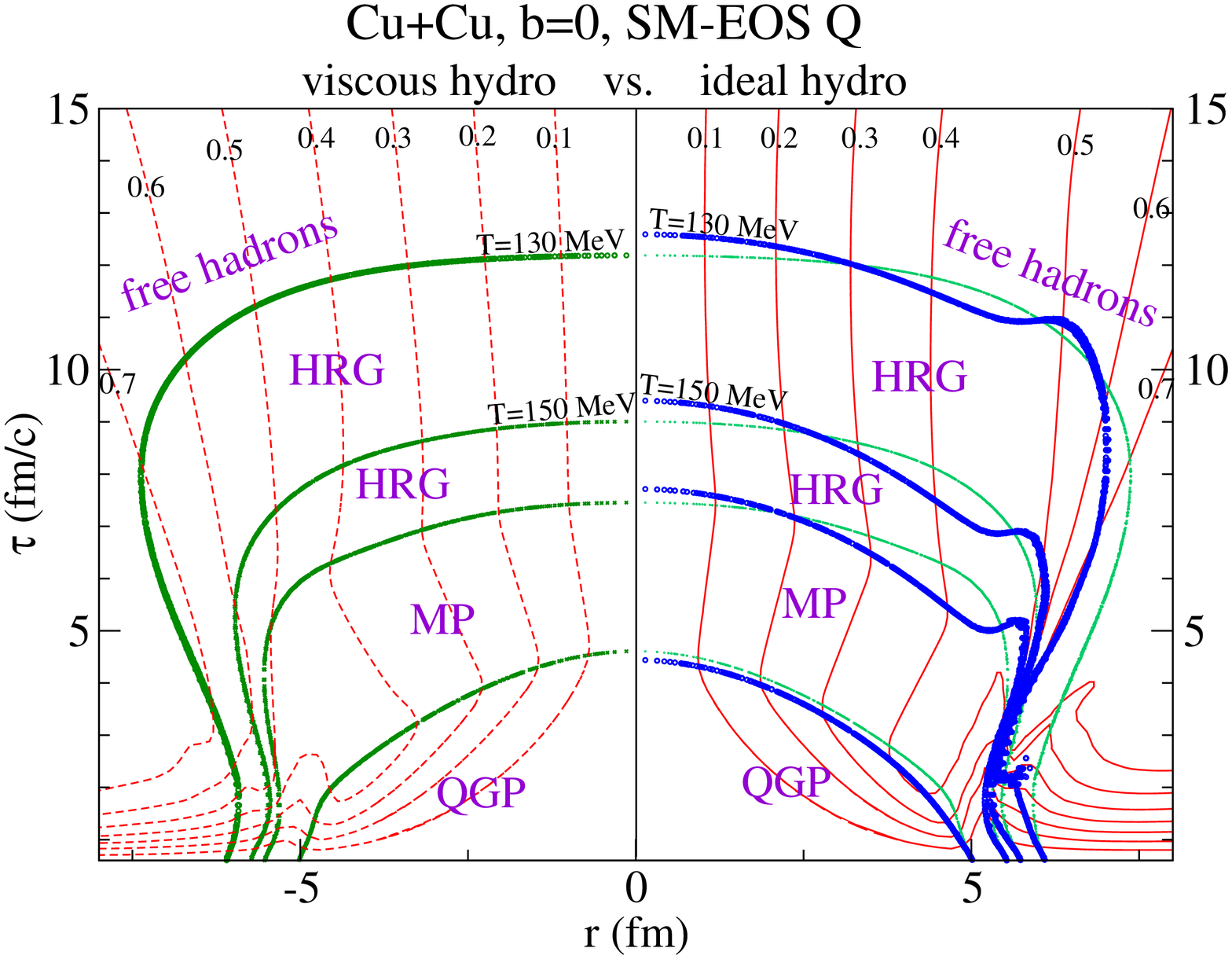}
\caption{(Color online)
Surfaces of constant temperature $T$ and constant transverse flow 
velocity $v_\perp$ for central Cu+Cu collisions, evolved with 
EOS~I (left panel) and SM-EOS~Q (right panel). In each panel, results 
from viscous hydrodynamics in the left half are directly compared with
the corresponding ideal fluid evolution in the right half. (The thin 
isotherm contours in the right halves of each panel are reflected from 
the left halves, for easier comparison.) The lines of constant $v_\perp$ 
are spaced by intervals of 0.1, from the inside outward, as indicated by 
the numbers near the top of the figures. The right panel contains two 
isotherms for $T_c\eq164$\,MeV, one separating the mixed phase (MP) from 
the QGP at energy density $e_\mathrm{Q}\eq1.6$\,GeV/fm$^3$, the other 
separating it from the hadron resonance gas (HRG) at energy density 
$e_\mathrm{H}\eq0.45$\,GeV/fm$^3$. See text for discussion.
}
\label{Contourb0}
\end{figure*}
%
%
\begin{figure*}
\includegraphics[bb=20 33 718 530,width=.49\linewidth,clip=]{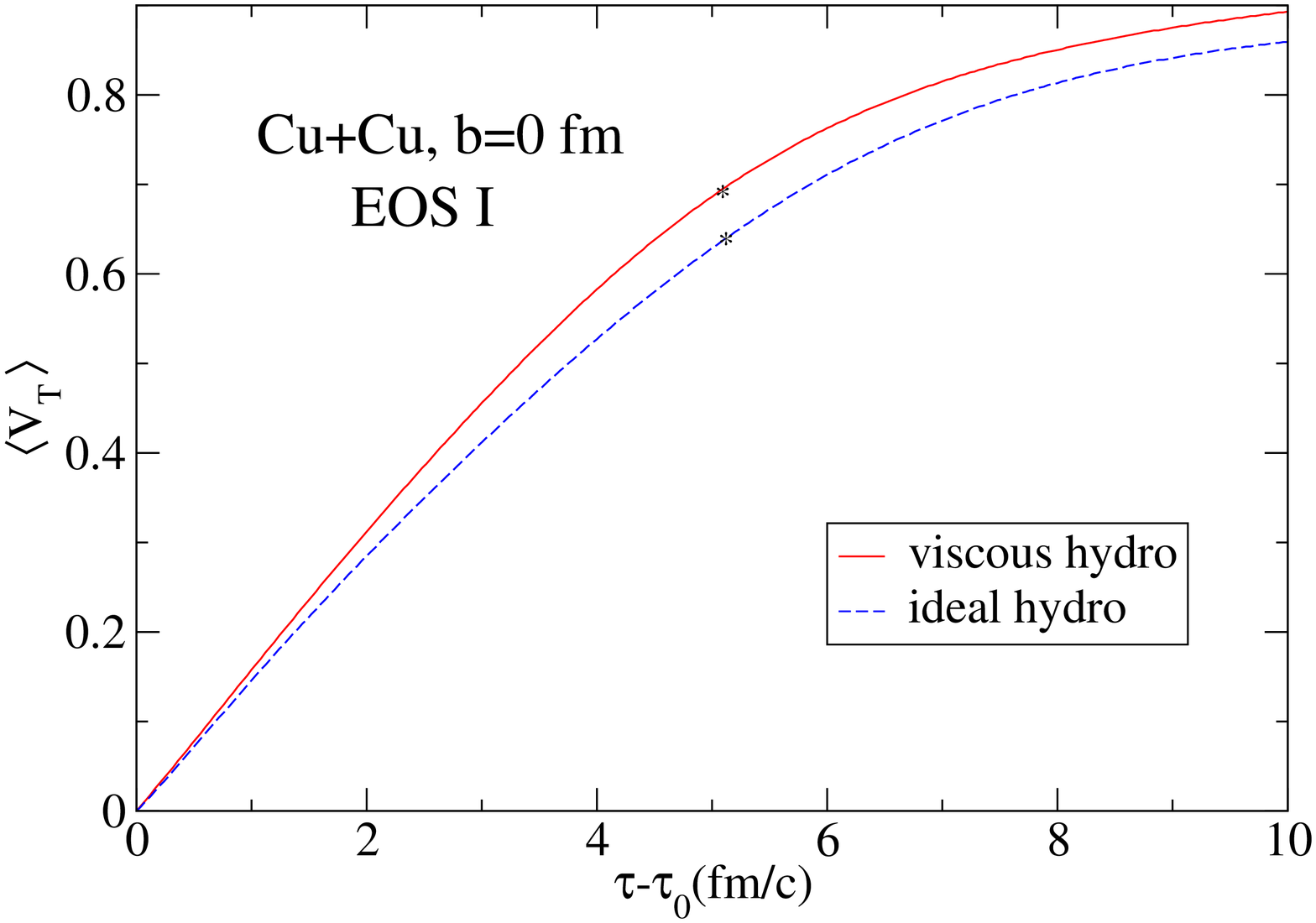}
\includegraphics[bb=20 33 718 530,width=.49\linewidth,clip=]{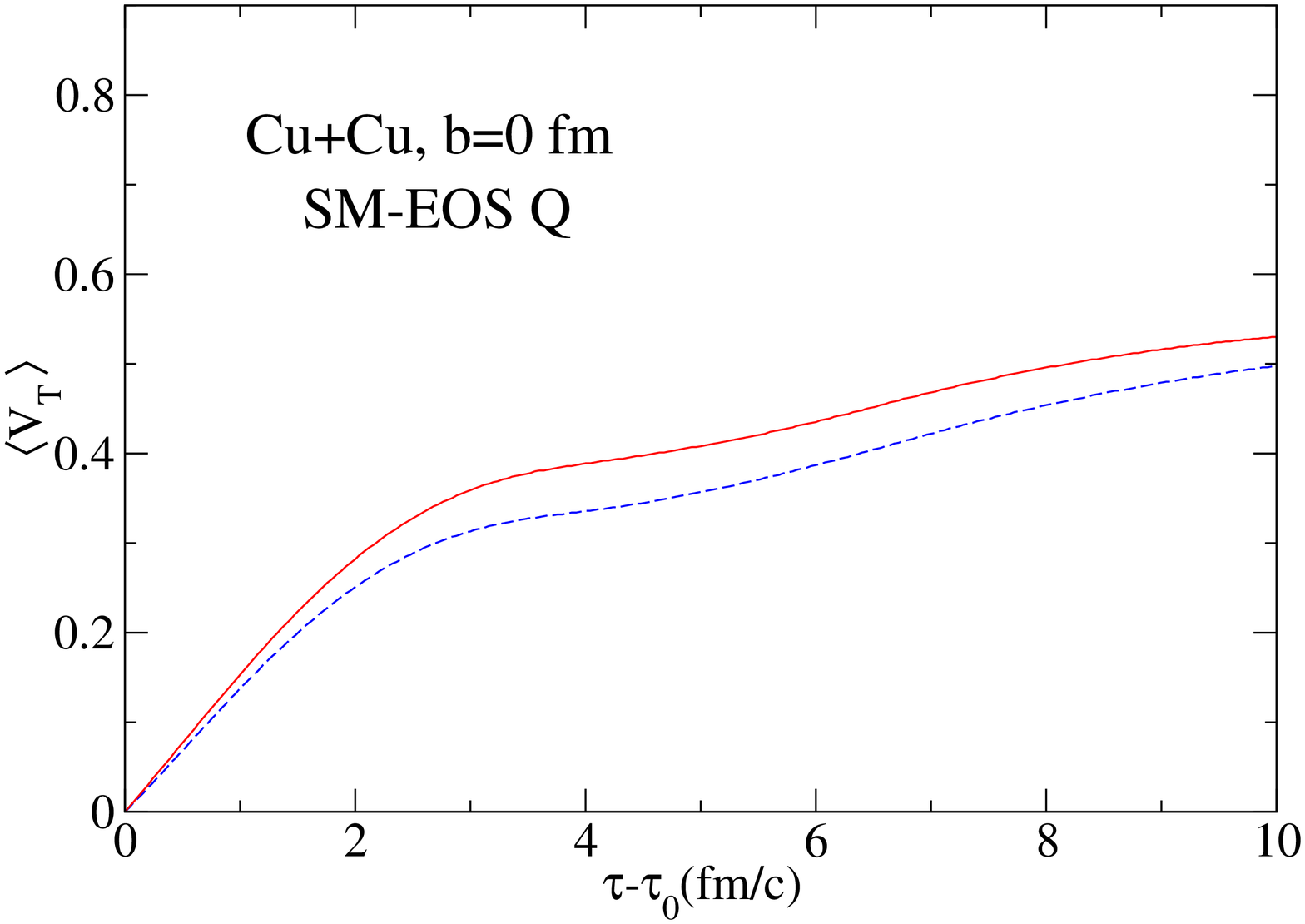}
\caption{(Color online)
Time evolution of the average radial flow velocity $\langle v_T
\rangle{\,\equiv\,}\langle v_\perp\rangle$ in central Cu+Cu 
collisions, calculated with EOS~I (left panel) and SM-EOS~Q (right panel).
Solid (dashed) lines show results from ideal (viscous) fluid dynamics.
The initially faster rate of increase reflects large positive shear
viscous pressure in the transverse direction at early times. The similar
rates of increase at late times indicate the gradual disappearance of shear
viscous effects. In the right panel, the curves exhibit a plateau
from 2 to 4\,fm/$c$, reflecting the softening of the EOS in the
mixed phase.  
}
\label{velocitySp}
\end{figure*}
%

Figure~\ref{localS} shows how the features seen in Fig.~\ref{localT} 
manifest themselves in the evolution of the entropy density. (In the 
QGP phase $s{\,\sim\,}T^3$.) The double-logarithmic presentation
emphasizes the effects of viscosity and transverse expansion on the 
power law $s(\tau){\,\sim\,}\tau^{-\alpha}$: One sees that the $\tau^{-1}$
scaling of the ideal Bjorken solution is flattened by viscous effects,
but steepened by transverse expansion. As is well-know, it takes a while 
(here about 3\,fm/$c$) until the transverse rarefaction wave reaches the 
fireball center and turns the initially 1-dimensional longitudinal 
expansion into a genuinely 3-dimensional one. When this happens, the power 
law $s(\tau){\,\sim\,}\tau^{-\alpha}$ changes from $\alpha\eq1$ in the 
ideal fluid case to $\alpha>3$ \cite{Rev-hydro}. Here 3 is the 
dimensionality of space, and the fact that $\alpha$ becomes larger than 
3 reflects relativistic Lorentz-contraction effects through the 
transverse-flow-related $\gamma_\perp$-factor that keeps increasing even 
at late times. In the viscous case, $\alpha$ changes from 1 to 3 sooner
than for the ideal fluid, due to the faster growth of transverse flow.
At late times the $s(\tau)$ curves for ideal and viscous hydrodynamics 
are almost perfectly parallel, indicating that very little entropy is 
produced during this late stage.

In Figure~\ref{Contourb0} we plot the evolution of temperature in $r{-}\tau$
space, in the form of constant-$T$ surfaces. Again the two panels compare
the evolution with EOS~I (left) to the one with SM-EOS~Q (right). In the
two halves of each panel we directly contrast viscous and ideal fluid 
evolution. (The light gray lines in the right halves are reflections of 
the viscous temperature contours in the left halves, to facilitate
comparison of viscous and ideal fluid dynamics.) Beyond the already 
noted fact that at $r\eq0$ the viscous fluid cools initially more 
slowly (thereby giving somewhat longer life to the QGP phase) but later 
more rapidly (thereby freezing out earlier), this figure also exhibits two 
other noteworthy features: (i) Moving from $r\eq0$ outward, one notes that 
contours of larger radial flow velocity are reached sooner in the viscous 
than in the ideal fluid case; this shows that radial flow builds up more 
quickly in the viscous fluid. This is illustrated more explicitly in 
Fig.~\ref{velocitySp} which shows the time evolution of the radial 
velocity $\langle v_\perp\rangle$, calculated as an average over the 
transverse plane with the Lorentz contracted energy density 
$\gamma_\perp e$ as weight function. (ii) Comparing the two sets of 
temperature contours shown in the right panel of Fig.~\ref{Contourb0}, 
one sees that viscous effects tend to smoothen any structures related 
to the (first order) phase transition in SM-EOS~Q. The reason for this 
is that, with the discontinuous change of the speed of sound at either 
end of the mixed phase, the radial flow velocity profile develops 
dramatic structures at the QGP-MP and MP-HRG interfaces \cite{Kolb:1999it}. 
This leads to large velocity gradients across these interfaces (as can be 
seen in the right panel of Fig.~\ref{Contourb0} in its lower right corner
which shows rather twisted contours of constant radial flow velocity), 
inducing large viscous pressures which drive to reduce these gradients
(as seen in lower left corner of that panel). In effect, shear viscosity 
softens the first-order phase transition into a smooth but rapid 
cross-over transition.

%
\begin{figure*}
\includegraphics[bb=10 33 711 536,width=.49\linewidth,clip=]{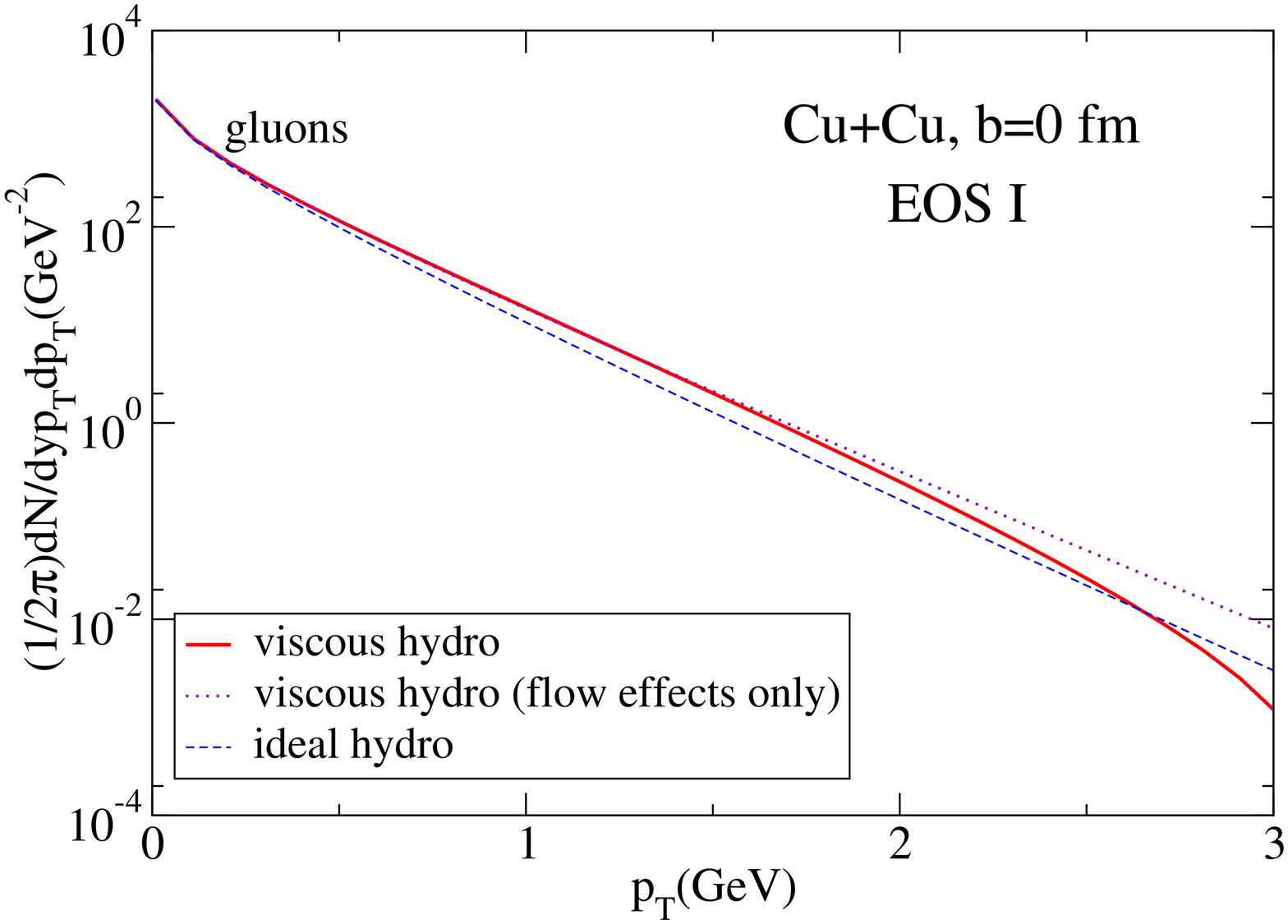}
\includegraphics[bb=10 33 711 536,width=.49\linewidth,clip=]{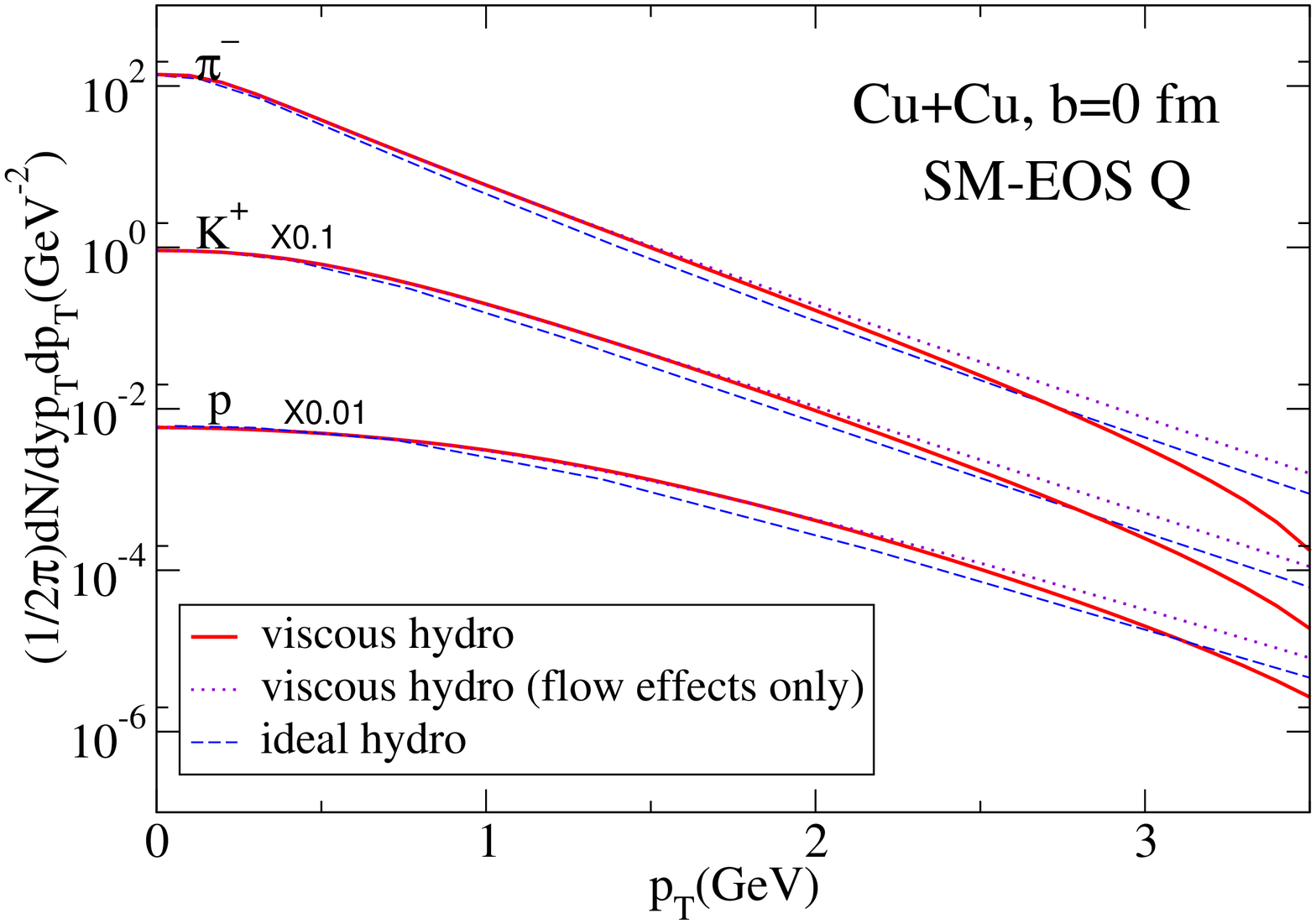}
\caption{(Color online)
Mid-rapidity particle spectra for central Cu+Cu collisions, calculated with 
EOS~I (left, gluons) and with SM-EOS~Q (right, $\pi^-$, $K^+$ and $p$). The 
solid blue (red dashed) lines are from ideal (viscous) hydrodynamics.
The purple dotted lines show viscous hydrodynamic spectra that neglect the
viscous correction $\delta f_i$ to the distribution function in 
Eq.~(\ref{Cooper}), i.e. include only the effects from the larger
radial flow generated in viscous hydrodynamics.
}
\label{Spectra}
\end{figure*}
%
These same viscous pressure gradients cause the fluid to accelerate
even in the mixed phase where all thermodynamic pressure gradients vanish
(and where the ideal fluid therefore does not generate additional flow).
As a result, the lifetime of the mixed phase is shorter in viscous 
hydrodynamics, as also seen in the right panel of Figure~\ref{Contourb0}.  

\subsection{Final particle spectra}
\label{sec3b}

After obtaining the freeze-out surface, we calculate the particle spectra 
from the generalized Cooper-Frye formula (\ref{Cooper}), using the 
AZHYDRO algorithm \cite{AZHYDRO} for the integration over the freeze-out 
surface $\Sigma$. For calculations with EOS~I which lacks the transition
from massless partons to hadrons, we cannot compute any hadron spectra. For
illustration we instead compute the spectra of hypothetical massless bosons 
(``gluons''). They can be compared with the pion spectra from SM-EOS~Q
which can also, to good approximation, be considered as massless bosons. 

The larger radial flow generated in viscous hydrodynamics, for a fixed set 
of initial conditions, leads, of course, to flatter transverse momentum 
spectra \cite{Teaney:2004qa,Chaudhuri:2005ea,Baier:2006gy} (at least at 
low $p_T$ where the viscous correction $\delta f_i$ to the distribution 
function can be neglected in (\ref{Cooper})). This is seen in 
Figure~\ref{Spectra}, by comparing the dotted and solid lines. This 
comparison also shows that the viscous spectra lie systematically above
the ideal ones, indicating larger final total multiplicity. This reflects
the creation of entropy during the viscous hydrodynamic evolution. As 
pointed out in \cite{Chaudhuri:2005ea,Baier:2006gy}, this requires a retuning 
of initial conditions (starting the hydrodynamic evolution later with
smaller initial energy density) if one desires to fit a given set of
experimental $p_T$-spectra. Since we here concentrate on investigating
the origins and detailed mechanics of viscous effects in relativistic
hydrodynamics, we will not explore any variations of initial conditions.
All comparisons between ideal and viscous hydrodynamics presented here
will use identical starting times $\tau_0$ and initial peak energy densities
$e_0$. 

The viscous correction $\delta f_i$ in Eqs.~(\ref{Cooper},\ref{deltaf})
depends on the signs and magnitudes of the various viscous pressure
tensor components along the freeze-out surface, weighted by the 
equilibrium part $f_{\mathrm{eq},i}$ of the distribution function.
Its effect on the final $p_T$-spectra (even its sign!) is not {\em a
priori} obvious. Teaney \cite{Teaney:2003kp}, using a blast-wave model to 
evaluate the velocity stress tensor $\sigma^{\mu\nu}\eq\pi^{\mu\nu}/(2\eta)$,
found that the correction is positive, growing quadratically with $p_T$.
%
\begin{figure}[bht]
\vspace*{5mm}
\includegraphics[bb=6 7 672 522,width=\linewidth,clip=]{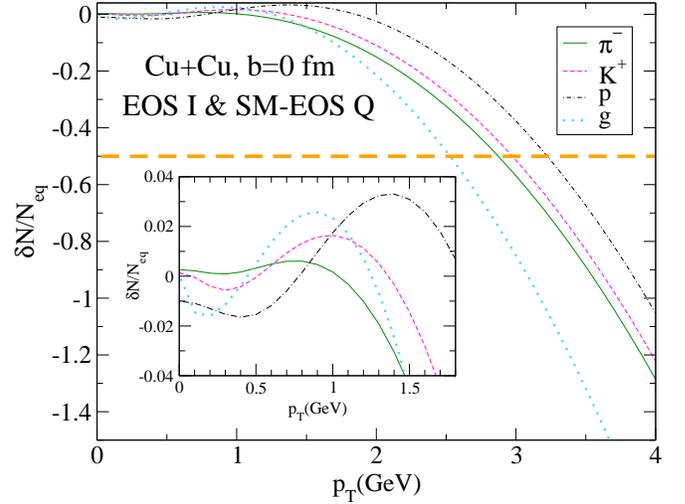}
\caption{(Color online)
Ratio of the viscous correction $\delta N$, resulting from the
non-equilibrium correction $\delta f$, Eq.~(\ref{deltaf}), to the 
distribution function at freeze-out, to the equilibrium spectrum 
$N_\mathrm{eq}{\,\equiv\,}dN_\mathrm{eq}/(dyd^2p_T)$ calculated from 
Eq.~(\ref{Cooper}) by setting $\delta f\eq0$. The gluon curves are for
evolution with EOS~I, the curves for $\pi^-$, $K^+$ and $p$ are from 
calculations with SM-EOS~Q.
}
\label{Spectra-Correction}
\end{figure}
%
Romatschke {\em et al.} \cite{Baier:2006gy,Romatschke:2007mq} did not break 
out separately the contributions from larger radial flow in 
$f_{\mathrm{eq},i}$ and from $\delta f_i$. Dusling and Teaney
\cite{Dusling:2007gi}, solving a slightly different set of viscous 
hydrodynamic equations and using a different (kinetic) freeze-out criterium
to determine their decoupling surface, found a (small) positive effect from 
$\delta f_i$ on the final pion spectra, at least up to $p_T\eq2$\,GeV/$c$, 
for freeze-out around $T_\mathrm{dec}\sim130$\,MeV, turning weakly negative 
when their effective freeze-out temperature was lowered to below 100 MeV. 
The dashed lines in Figure~\ref{Spectra} show that in our calculations for 
$p_T{\,\gtrsim\,}2$\,GeV/$c$ the effects from $\delta f_i$ have an 
overall negative sign, leading to a reduction of the $p_T$-spectra
at large $p_T$ relative to both the viscous spectra without $\delta f_i$ 
and the ideal hydrodynamic spectra. This is true for all particle species,
irrespective of the EOS used to evolve the fluid.

It turns out that, when evaluating the viscous correction $\delta f$
in Eq.~(\ref{deltaf}) with the help of Eq.~(\ref{vis-cor}), large 
cancellations occur between the first and second line in Eq.~(\ref{vis-cor}).
[After azimuthal integration, the contribution to $\delta f$ from the 
third line ${\sim\,}\Delta$ vanishes identically for central collisions.] 
These cancellations cause the final result to be quite sensitive to small 
numerical errors in the calculation of $\tau^2\pi^{\eta\eta}$ and 
$\Sigma\eq\pi^{xx}{+}\pi^{yy}$. Increased numerical stability is achieved 
by trading $\tau^2\pi^{\eta\eta}$ for 
$\pi^{\tau\tau}\eq\tau^2\pi^{\eta\eta}{+}\Sigma$ and using instead of 
Eq.~(\ref{vis-cor}) the following expression:
\begin{widetext}
\begin{eqnarray}
  p_\mu p_\nu \pi^{\mu\nu} \!\!&=&\!\! \pi^{\tau\tau}
  \left[m_T^2\bigl(2\cosh^2(y{-}\eta){-}1\bigr)
      -2 \frac{p_T}{v_\perp} m_T\cosh(y{-}\eta)
         \frac{\sin(\phi_p{+}\phi_v)}{\sin(2\phi_v)}
       +\frac{p_T^2}{v_\perp^2}\frac{\sin(2\phi_p)}{\sin(2\phi_v)}\right]
\nonumber\\
 &&\!\!\!+\,\Sigma 
  \left[-m_T^2\sinh^2(y{-}\eta)
        +p_T m_T \cosh(y{-}\eta) v_\perp
         \frac{\sin(\phi_p{-}\phi_v)}{\tan(2\phi_v)}
       +\frac{p_T^2}{2}\left(1-\frac{\sin(2\phi_p)}{\sin(2\phi_v)}\right)
  \right]
\!\!\!\!
\nonumber\\
  &&\!\!\! +\,\Delta
  \left[p_T m_T\cosh(y{-}\eta) v_\perp
        \frac{\sin(\phi_p{-}\phi_v)}{\sin(2\phi_v)}
       -\frac{p_T^2}{2}\frac{\sin(2(\phi_p{-}\phi_v))}{\sin(2\phi_v)}
  \right].
\label{vis-cor-2}
\end{eqnarray}
\end{widetext}
The first and second lines of this expression are now much smaller than 
before and closer in magnitude to the final net result for 
$p_\mu p_\nu \pi^{\mu\nu}$. This improvement carries over to non-central 
collisions as discussed in Sec.~\ref{sec4d}, where we also show the 
individual contributions from $\pi^{\tau\tau}$, $\Sigma$ and $\Delta$ 
to the particle spectra. To be able to use Eq.~(\ref{vis-cor-2}), the 
numerical code should directly evolve, in addition to $\pi^{\tau\tau}$, 
$\pi^{\tau x}$, and $\pi^{\tau y}$ which are needed for the velocity 
finding algorithm (see Appendix~\ref{appb}), the components $\pi^{xx}$
and $\pi^{yy}$. Otherwise these last two components must be computed
from the evolved $\pi^{mn}$ components using the transversality and
tracelessness constraints which necessarily involves the amplification 
of numerical errors by division with small velocity components.
 
In Figure~\ref{Spectra-Correction} we explore the non-equilibrium 
contribution to the final hadron spectra in greater detail. The figure 
shows that the non-equilibrium effects from $\delta f_i$ are largest 
for massless particles and, at high $p_T$, decrease in magnitude with 
increasing particle mass. The assumption $|\delta f|{\,\ll\,}f_\mathrm{eq}$, 
which underlies the viscous hydrodynamic formalism, is seen to break 
down at high $p_T$, but to do so later for heavier hadrons than for 
lighter ones. Once the correction exceeds ${\cal O}(50\%)$ (indicated by 
the horizontal dashed line in Fig.~\ref{Spectra-Correction}), the 
calculated spectra can no longer be trusted. 

In contrast to viscous hydrodynamics, ideal fluid dynamics has no 
intrinsic characteristic that will tell us when it starts to break down.
Comparison of the calculated elliptic flow $v_2$ from ideal fluid dynamics
with the experimental data from RHIC \cite{Rev-hydro} suggests that the 
ideal fluid picture begins to break down above $p_T{\,\simeq\,}1.5$\,GeV/$c$ 
for pions and above $p_T{\,\simeq\,}2$\,GeV/$c$ for protons. This
phenomenological hierarchy of thresholds where viscous effects appear 
to become essential is qualitatively consistent with the mass hierarchy
from viscous hydrodynamics shown in Fig.~\ref{Spectra-Correction}.

In the region $0{\,<\,}p_T{\,\lesssim\,}1.5$\,GeV/$c$, the interplay 
between $m_T$- and $p_T$-dependent terms in Eq.~(\ref{vis-cor}) 
is subtle, causing sign changes of the viscous spectral correction 
depending on hadron mass and $p_T$ (see inset in 
Fig.~\ref{Spectra-Correction}). The fragility of the sign of the
effect is also obvious from Fig.~8 in the work by Dusling and Teaney 
\cite{Dusling:2007gi} where it is shown that in this $p_T$ region 
the viscous correction changes sign from positive to negative when 
freeze-out is shifted from earlier to later times (higher to lower
freeze-out temperature). Overall, we agree with them that the viscous 
correction effects on the $p_T$-spectra are weak in this region 
\cite{Dusling:2007gi}. We will see below that a similar statement does 
not hold for the elliptic flow.

\section{Non-central collisions}
\label{sec4}
\subsection{Hydrodynamic evolution}
\label{sec4a}

%
\begin{figure*}
\includegraphics[bb= 30 30 755 573,width=.49\linewidth,clip=]{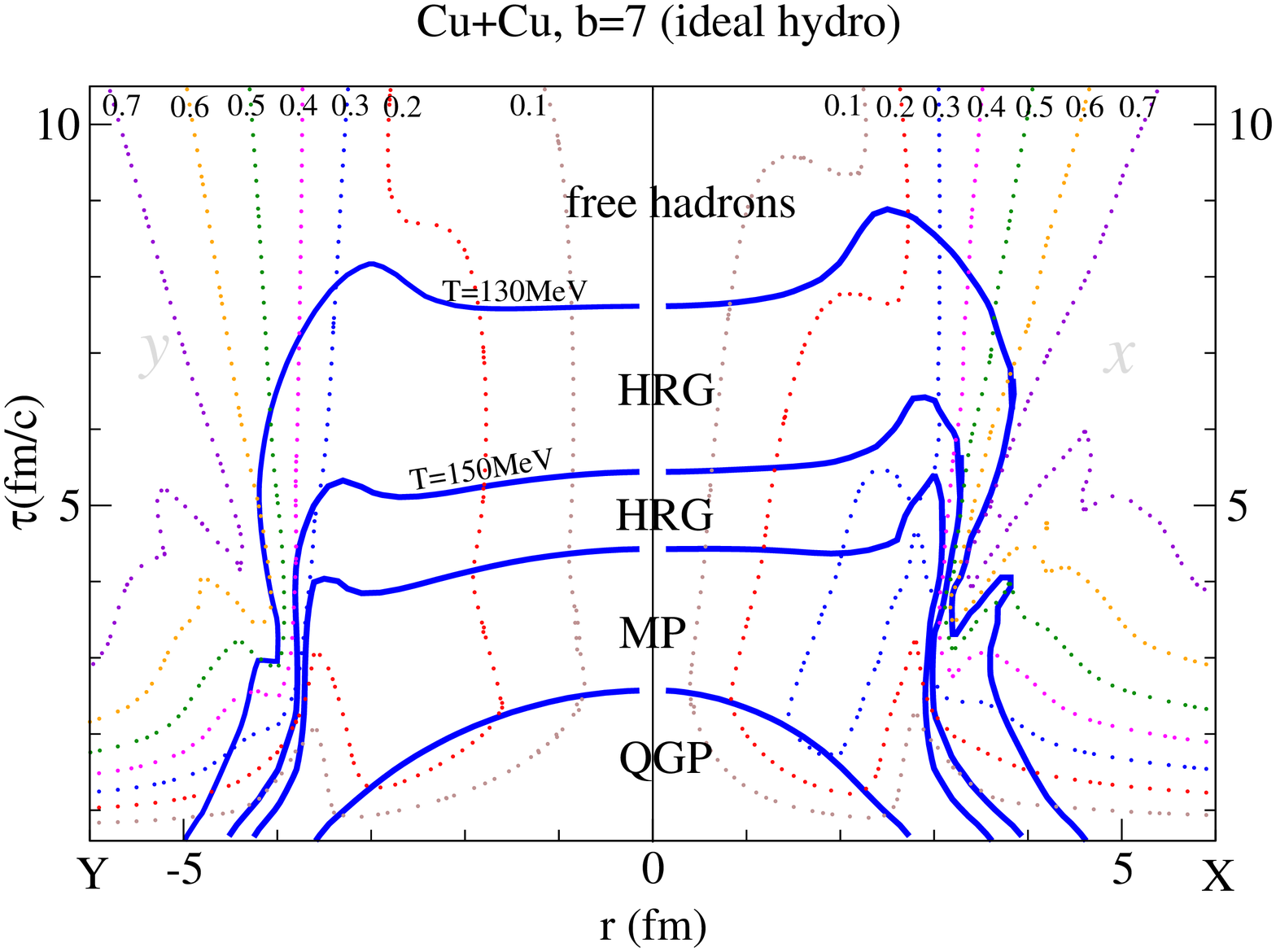}
\includegraphics[bb= 30 30 737 573,width=.49\linewidth,clip=]{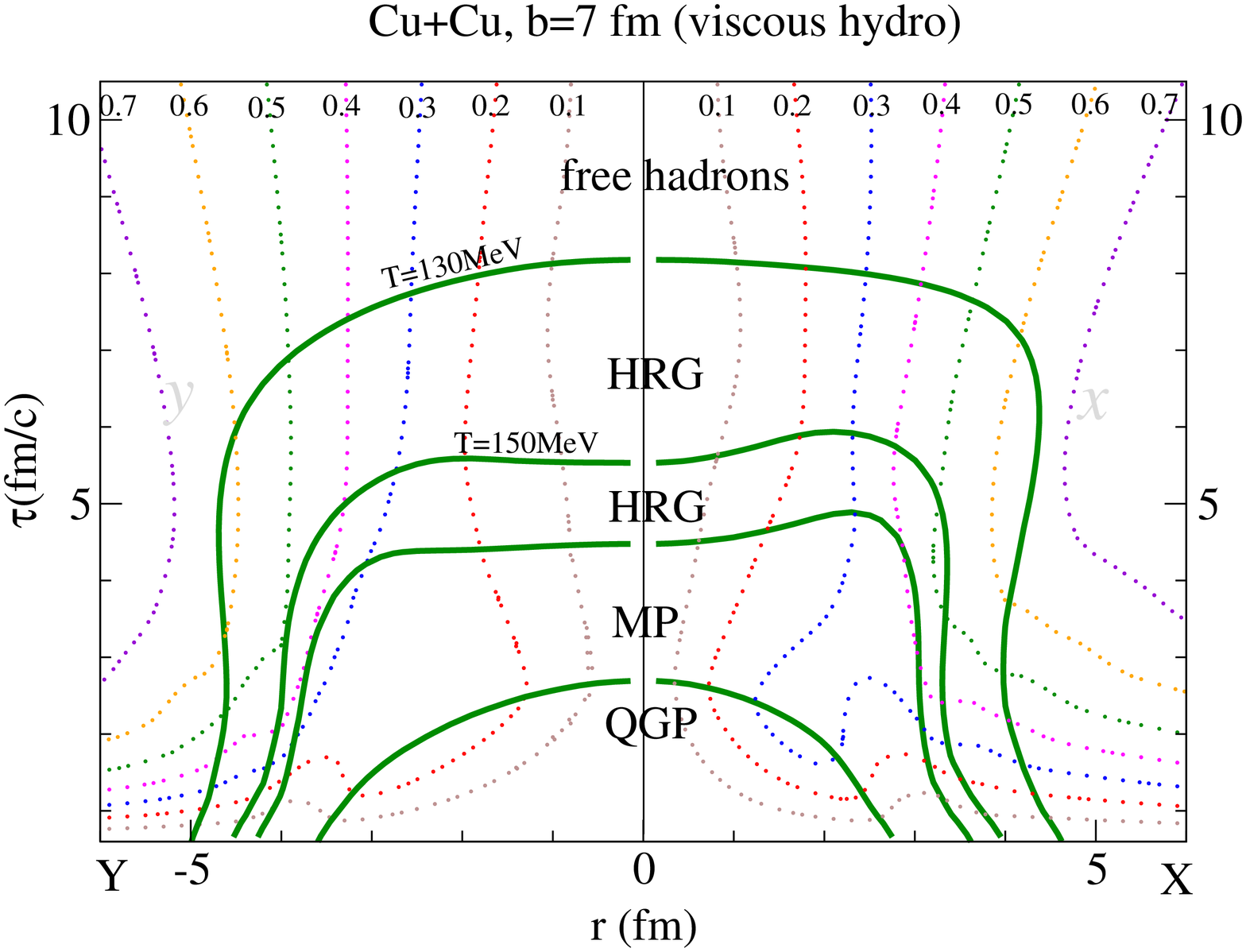}
\\
\includegraphics[bb=30 30 737 573,width=.49\linewidth,clip=]{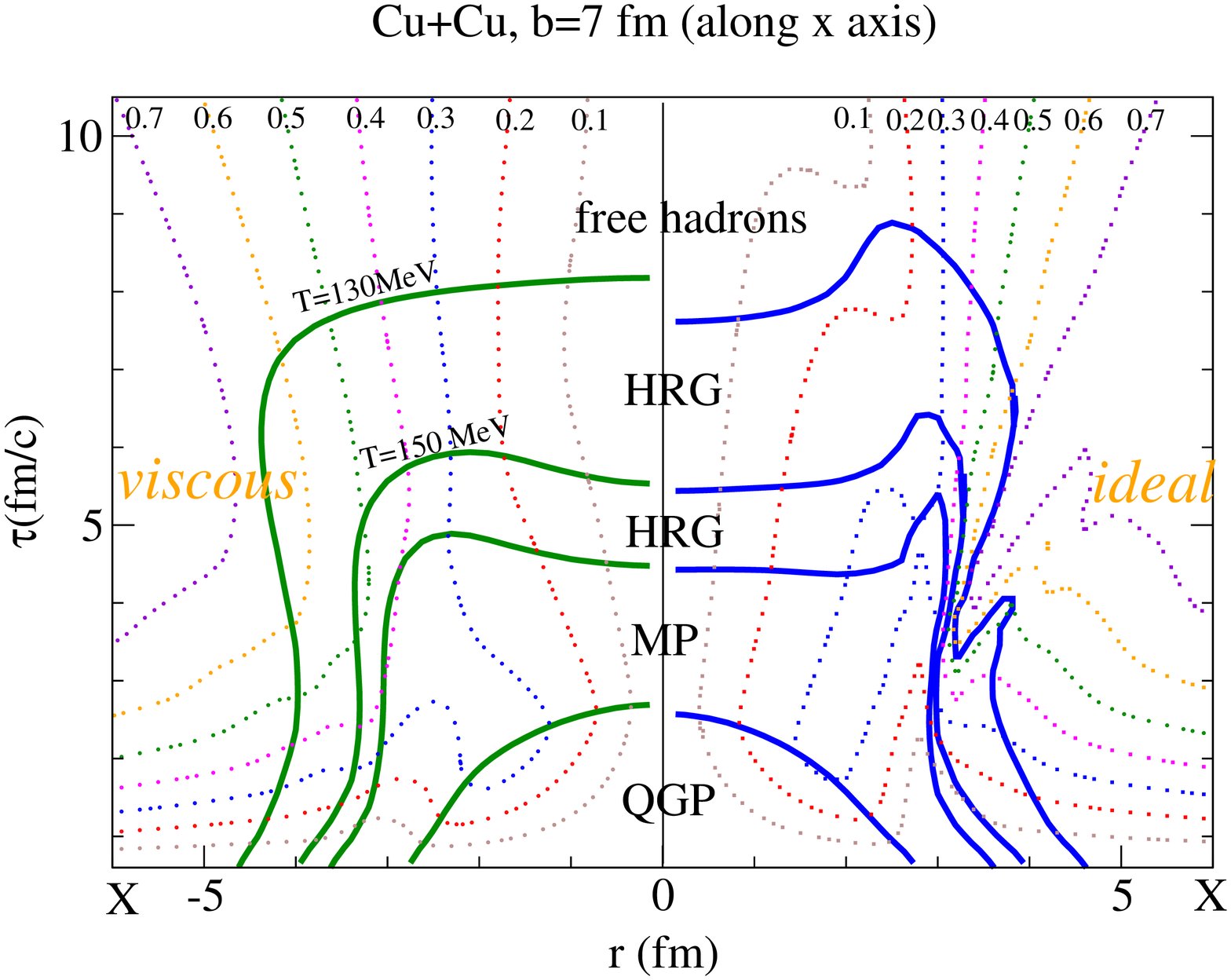}
\includegraphics[bb=30 30 737 573,width=.49\linewidth,clip=]{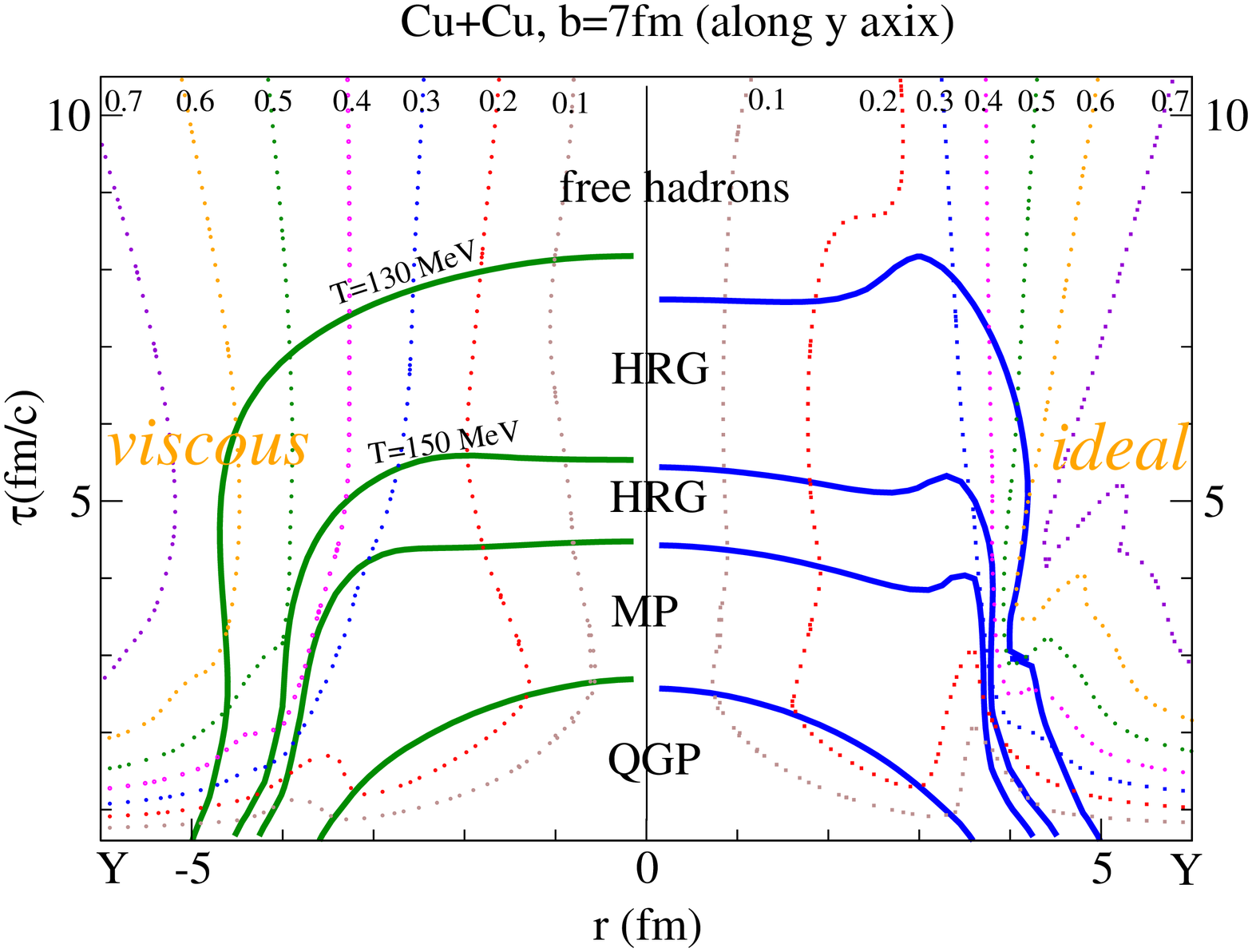}
\caption{(Color online)
Surfaces of constant temperature $T$ and constant transverse flow 
velocity $v_\perp$ for semi-peripheral Cu+Cu collisions at $b\eq7$\,fm, 
evolved with SM-EOS~Q. In the {\em top row} we contrast ideal (left panel) 
and viscous (right panel) fluid dynamics, with a cut along the $x$ axis 
(in the reaction plane) shown in the right half while the left half shows 
a cut along the $y$ axis (perpendicular to the reaction plane). In the {\em
bottom row} we compare ideal and viscous evolution in the same panel,
with cuts along the $x$ ($y$) direction shown in the left (right) panel.
See Fig.~\ref{Contourb0} for comparison with central Cu+Cu collisions.
}
\label{Contourb7}
\end{figure*}
%
%
\begin{figure*}
\includegraphics[bb=33 43 367 528,width=.4\linewidth,clip=]{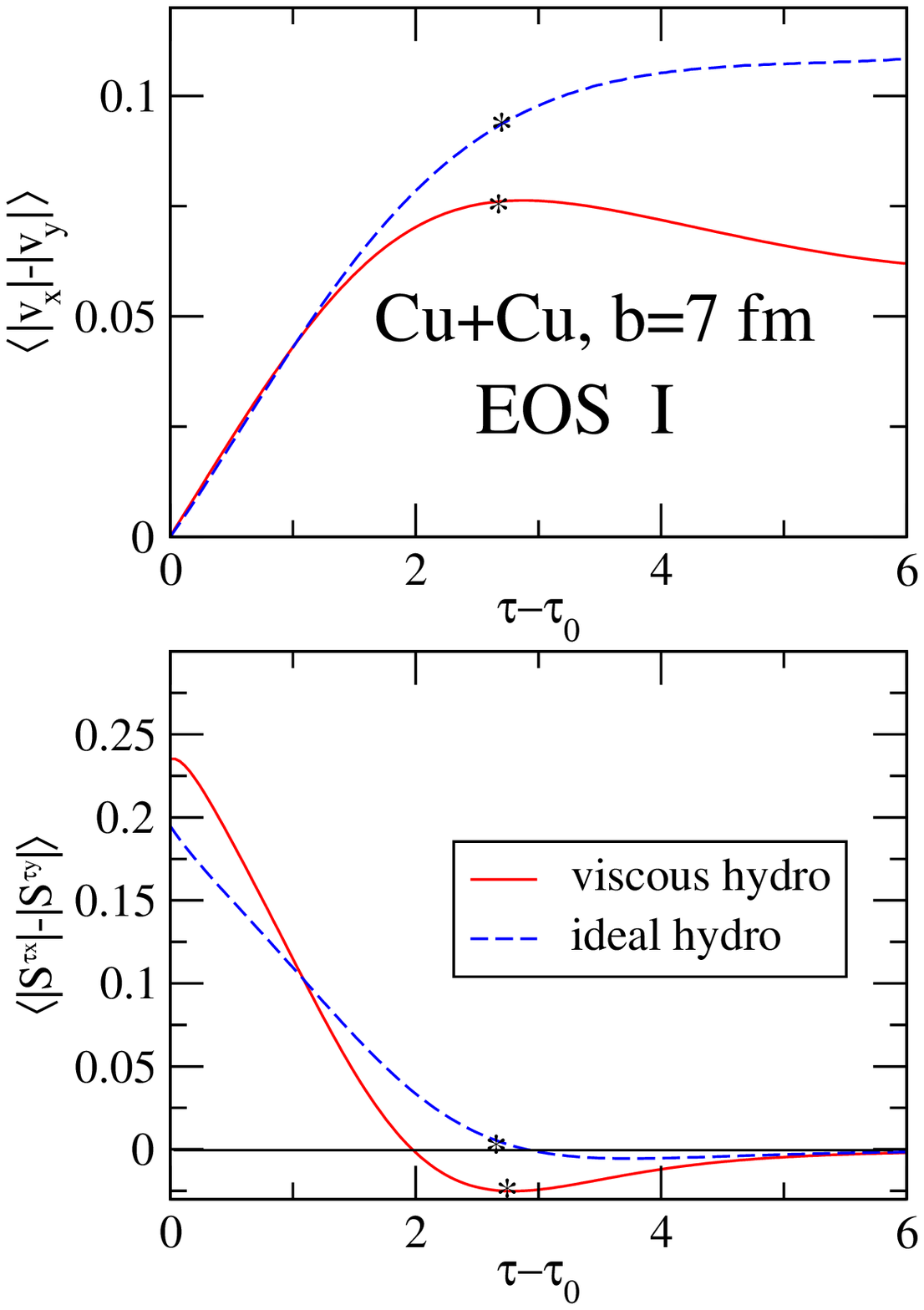}
\includegraphics[bb=33 43 367 528,width=.4\linewidth,clip=]{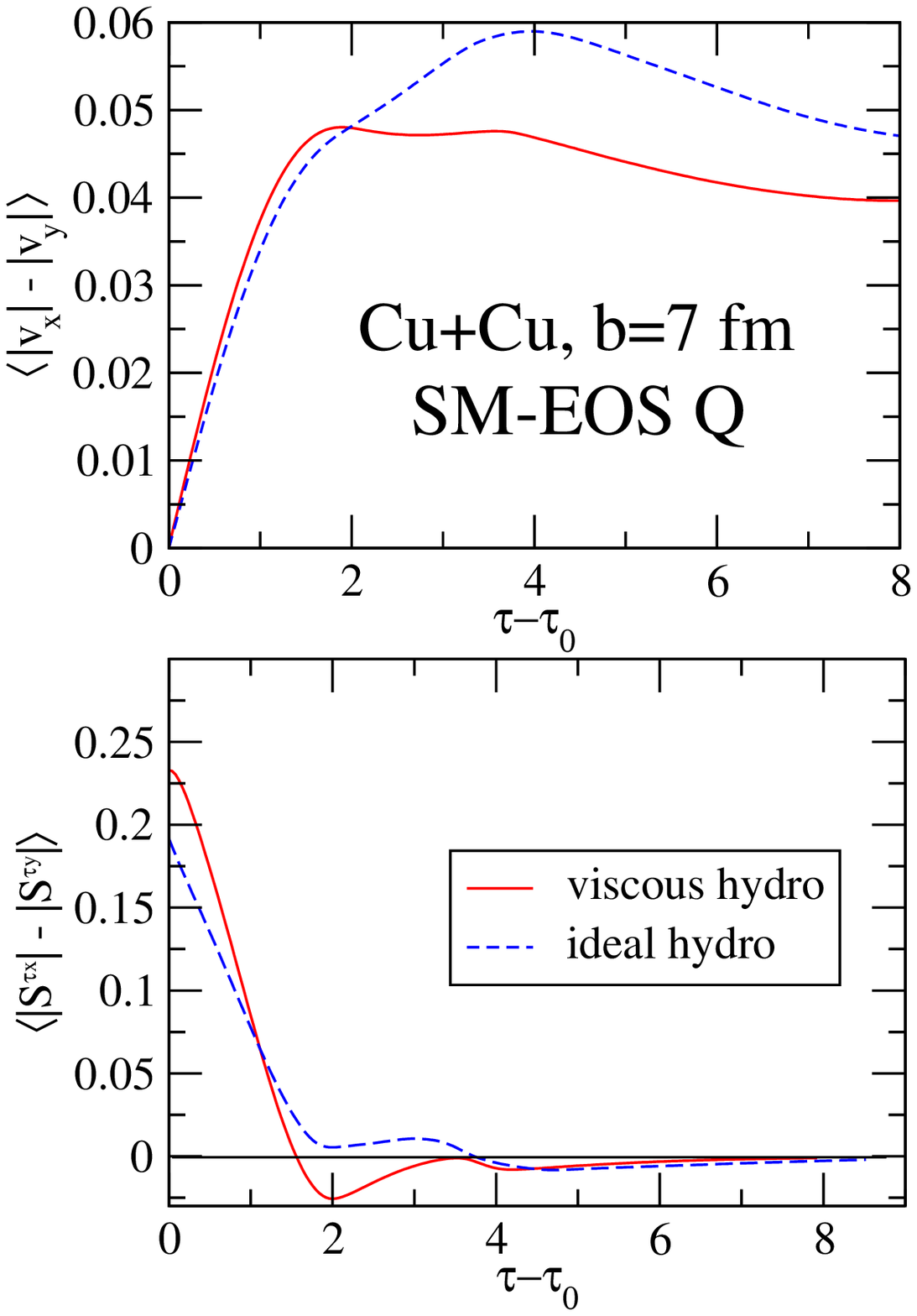}
\caption{(Color online)
Time evolution of the transverse flow anisotropy $\La|v_x|{-}|v_y|\Ra$ 
(top row) and of the anisotropy in the transverse source term 
$\La|{\cal S}^{\tau x}|{-}|{\cal S}^{\tau y}|\Ra$ (bottom row). Both quantities
are averaged over the transverse plane, with the Lorentz-contracted energy
density $\gamma_\perp$ as weight function. The left (right) column shows
results for EOS~I (SM-EOS~Q), with solid (dashed) lines representing ideal
(viscous) fluid dynamical evolution.
}
\label{VxSc}
\end{figure*}

We now take full advantage of the ability of VISH2+1 to solve the
transverse expansion in 2 spatial dimensions to explore the anisotropic
fireball evolution in non-central heavy-ion collisions. Similar to 
Fig.~\ref{Contourb0} for $b\eq0$, Figure~\ref{Contourb7} shows surfaces 
of constant temperature and radial flow for Cu+Cu collisions at $b\eq7$\,fm, 
for the equation of state SM-EOS~Q. The plots show the different evolution 
into and perpendicular to the reaction plane and compare ideal with viscous 
fluid dynamics. Again, even a minimal amount of shear viscosity 
($\frac{\eta}{s}\eq\frac{1}{4\pi}$) is seen to dramatically smoothen all 
structures related to the existence of a first-order phase transition in 
the EOS. However, in distinction to the case of central collisions, radial 
flow builds up at a weaker rate in the peripheral collisions and never 
becomes strong enough to cause faster central cooling at late times than 
seen in ideal fluid dynamics (bottom row in Fig.~\ref{Contourb7}). The 
viscous fireball cools more slowly than the ideal one at all times and 
positions, lengthening in particular the lifetime of the QGP phase, and 
it grows to larger transverse size at freeze-out. [Note that this does 
{\em not} imply larger transverse HBT radii than for ideal hydrodynamics 
(something that --in view of the ``RHIC HBT Puzzle'' \cite{Rev-hydro}-- 
would be highly desirable): the larger geometric size is counteracted by 
larger radial flow such that the geometric growth, in fact, does not lead 
to larger transverse homogeneity lengths \cite{Baier:2006gy}.]

%
\begin{figure*}
\includegraphics[bb=56 8 436 521,width=.45\linewidth,clip=]{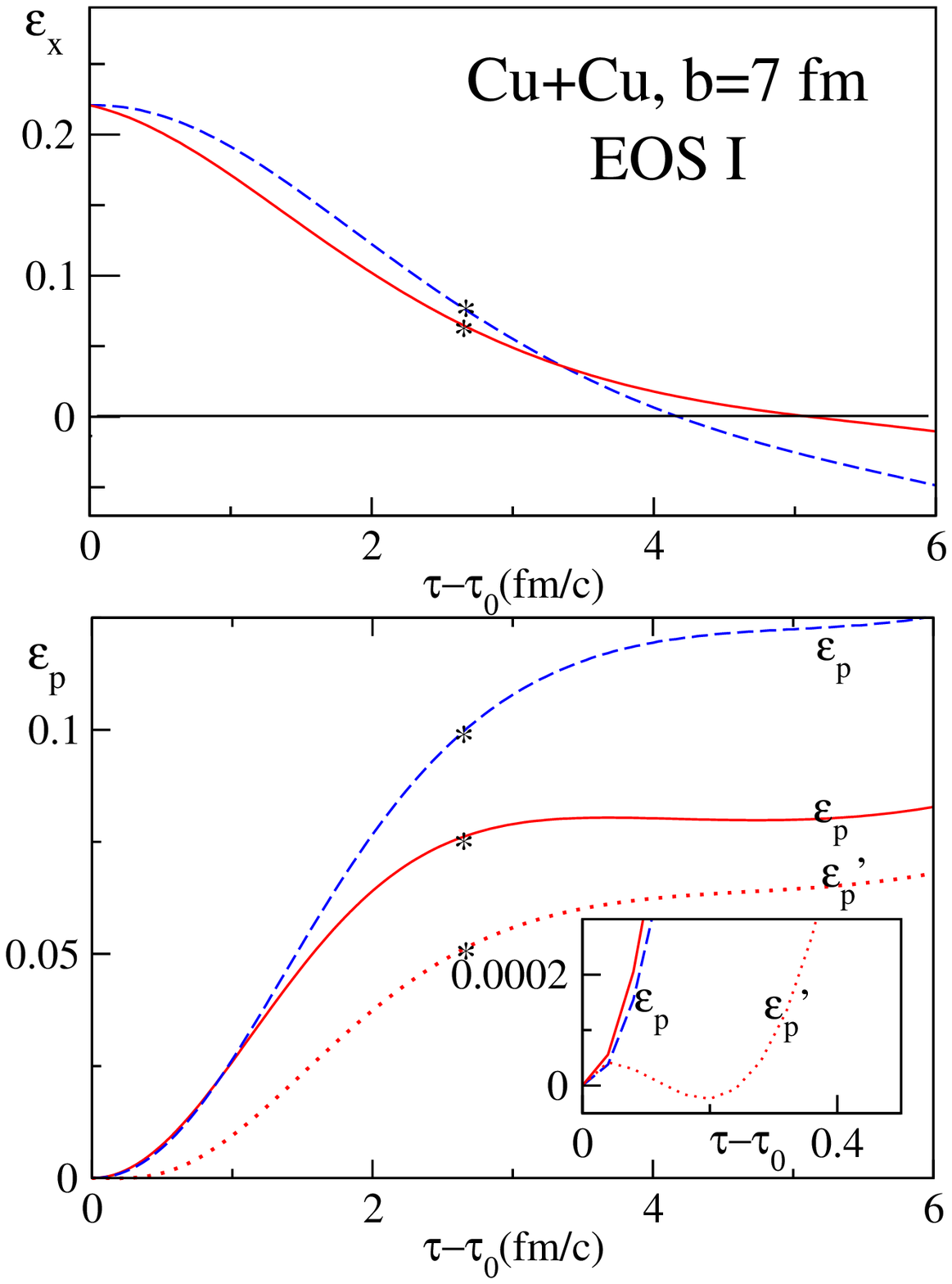}
\includegraphics[bb=55 8 436 521,width=.45\linewidth,clip=]{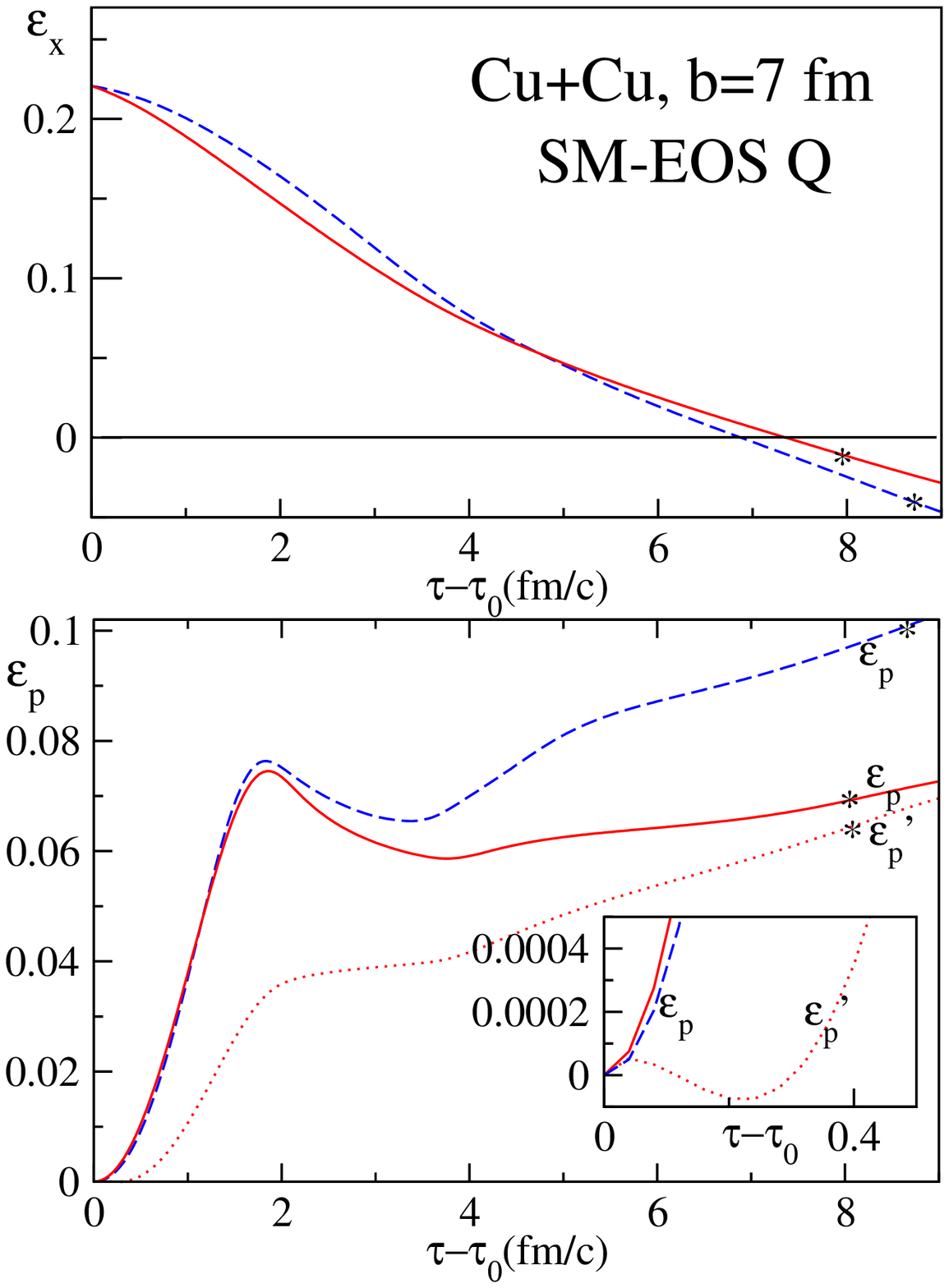}
\caption{(Color online)
Time evolution for the spatial eccentricity $\epsilon_x$, momentum 
anisotropy $\epsilon_p$ and total momentum anisotropy $\epsilon_p'$ 
(see text for definitions), calculated for $b\eq7$\,fm Cu+Cu collisions
with EOS~I (left column) and SM-EOS~Q (right column). Dashed lines are for 
ideal hydrodynamics while the solid and dotted lines show results from 
viscous hydrodynamics. See text for discussion.
}
\label{ExEp}
\end{figure*}
%

While Figure~\ref{Contourb7} gives an impression of the anisotropy 
of the fireball in coordinate space, we study now in Fig.~\ref{VxSc}
the evolution of the flow anisotropy $\La|v_x|{-}|v_y|\Ra$. In central
collisions this quantity vanishes. In ideal hydrodynamics it is driven
by the anisotropic gradients of the thermodynamic pressure. In viscous
fluid dynamics, the source terms (\ref{S01},\ref{S02}), whose difference
is shown in the bottom row of Fig.~\ref{VxSc}, receive additional 
contributions from gradients of the viscous pressure tensor which 
contribute their own anisotropies. Fig.~\ref{VxSc} demonstrates that 
these additional anisotropies {\em increase} the driving force for
anisotropic flow at very early times ($\tau{-}\tau_0{\,<\,}1$\,fm/$c$),
but {\em reduce} this driving force throughout the later evolution.
At times $\tau{-}\tau_0{\,>\,}2$\,fm/$c$ the anisotropy of the effective
transverse pressure even changes sign and turns negative, working to 
{\em decrease} the flow anisotropy. As a consequence of this, the buildup 
of the flow anisotropy stalls at $\tau{-}\tau_0{\,\approx\,}2.5$\,fm/$c$
(even earlier for SM-EOS~Q where the flow buildup stops as soon as
the fireball medium enters the mixed phase) and proceeds to slightly 
decrease therafter. This happens during the crucial period where 
ideal fluid dynamics still shows strong growth of the flow anisotropy.
By the time the fireball matter decouples, the average flow velocity
anisotropy of viscous hydro lags about 20-25\% behind the value
reached during ideal fluid dynamical evolution. 

These features are mirrored in the time evolution of the spatial 
eccentricity $\epsilon_x\eq\frac{\La x^2{-}y^2\Ra}{\La x^2{+}y^2\Ra}$
(calculated by averaging over the transverse plane with the energy 
density $e(x)$ as weight function \cite{Kolb:1999it} and shown in the top
row of Fig.~\ref{ExEp}) and of the momentum anisotropies $\epsilon_p$ and
$\epsilon'_p$ (shown in the bottom row). The momentum anisotropy 
$\epsilon_p\eq\frac{\La T^{xx}_0{-}T^{yy}_0\Ra}{\La T^{xx}_0{+}T^{yy}_0\Ra}$
\cite{Ollitrault:1992bk} measures the anisotropy of the transverse momentum 
density due to anisotropies in the collective flow pattern, as shown in 
top row of Fig.~\ref{VxSc}; it includes only the ideal fluid part of the
energy momentum tensor. The {\em total momentum anisotropy}
$\epsilon'_p\eq\frac{\La T^{xx}{-}T^{yy}\Ra}{\La T^{xx}{+}T^{yy}\Ra}$,
similarly defined in terms of the total energy momentum tensor 
$T^{\mu\nu}\eq{T}_0^{\mu\nu}{+}\pi^{\mu\nu}$, additionally counts 
anisotropic momentum contributions arising from the viscous pressure 
tensor. Since the latter quantity includes effects due to the deviation 
$\delta f$ of the local distribution function from its thermal equilibrium 
form which, according to Eq.~(\ref{Cooper}), also affects the final
hadron momentum spectrum and elliptic flow, it is this {\em total
momentum anisotropy} that should be studied in viscous hydrodynamics
if one wants to understand the evolution of hadron elliptic flow. In 
other words, in viscous hydrodynamics hadron elliptic flow is not simply
a measure for anisotropies in the collective flow velocity pattern, 
but additionally reflects anisotropies in the local rest frame momentum 
distributions, arising from deviations of the local momentum distribution
from thermal equilibrium and thus being related to the viscous pressure. 

Figure~\ref{ExEp} correlates the decrease in time of the spatial
eccentricity $\epsilon_x$ with the buildup of the momentum anisotropies
$\epsilon_p$ and $\epsilon'_p$. In viscous dynamics the spatial
eccentricity is seen to decrease initially faster than for ideal
fluids. This is less a consequence of anisotropies in the large viscous 
transverse pressure gradients at early times than due to the faster
radial expansion caused by their large overall magnitude. In fact,
it was found a while ago \cite{unpublished} that for a system of 
free-streaming partons the spatial eccentricity falls even faster
than the viscous hydrodynamic curves (solid lines) in the upper row of 
Figure~\ref{ExEp}. The effects of early pressure gradient anisotropies
is reflected in the initial growth rate of the flow-induced momentum 
anisotropy $\epsilon_p$ which is seen to slightly exceed that observed
in the ideal fluid at times up to about 1\,fm/$c$ after the beginning
of the transverse expansion (bottom panels in Fig.~\ref{ExEp}). This
parallels the slightly faster initial rise of the flow velocity
anisotropy seen in the top panels of Fig.~\ref{VxSc}. Figure~\ref{VxSc}
also shows that in the viscous fluid the flow velocity anisotropy 
stalls about 2\,fm/$c$ after start and remains about 25\% below the 
final value reached in ideal fluid dynamics. This causes the spatial
eccentricity of the viscous fireball to decrease more slowly at later
times than that of the ideal fluid (top panels in Fig.~\ref{ExEp})
which, at late times, features a significantly larger difference
between the horizontal ($x$) and vertical ($y$) expansion velocities. 
  
It is very instructive to compare the behaviour of the flow-induced
ideal-fluid contribution to the momentum anisotropy, $\epsilon_p$, with 
that of the total momentum anisotropy $\epsilon'_p$. At early times
they are very different, with $\epsilon'_p$ being much smaller than 
$\epsilon_p$ and even turning slightly {\em negative} at very early 
times (see insets in the lower panels of Fig.~\ref{ExEp}). This reflects
very large {\em negative} contributions to the anisotropy of the total 
energy momentum tensor from the shear viscous pressure whose gradients 
along the out-of-plane direction $y$ strongly exceed those within the 
reaction plane along the $x$ direction. At early times this effect 
almost compensates for the larger in-plane gradient of the thermal 
pressure. The {\em negative} viscous pressure gradient anisotropy
is responsible for reducing the growth of flow anisotropies, thereby 
causing the flow-induced momentum anisotropy $\epsilon_p$ to significantly
lag behind its ideal fluid value at later times. The negative viscous 
pressure anisotropies responsible for the difference between $\epsilon_p$
and $\epsilon'_p$ slowly disappear at later times, since all viscous
pressure components then become very small (see Fig.~\ref{AverPi} below).

The net result of this interplay is a total momentum anisotropy in Cu+Cu
collisions (i.e. a source of elliptic flow $v_2$) that for a ``minimally'' 
viscous fluid with $\frac{\eta}{s}\eq\frac{1}{4\pi}$ is 40-50\% lower than
for an ideal fluid, at all except the earliest times (where it is even 
smaller). The origin of this reduction changes with time: Initially it is
dominated by strong momentum anisotropies in the local rest frame, with
momenta pointing preferentially out-of-plane, induced by deviations from 
local thermal equilibrium and associated with large shear viscous 
pressure.
At later times, the action of these anisotropic viscous pressure gradients 
integrates to an overall reduction in collective flow anisotropy, while 
the viscous pressure itself becomes small; at this stage, the reduction 
of the total momentum anisotropy is indeed mostly due to a reduced 
anisotropy in the collective flow pattern while momentum isotropy in 
the local fluid rest frame is approximately restored.

\subsection{Elliptic flow $v_2$ of final particle spectra}
\label{sec4b}

\begin{figure*}
\includegraphics[bb=0 28 705 534,width=.48\linewidth,clip=]{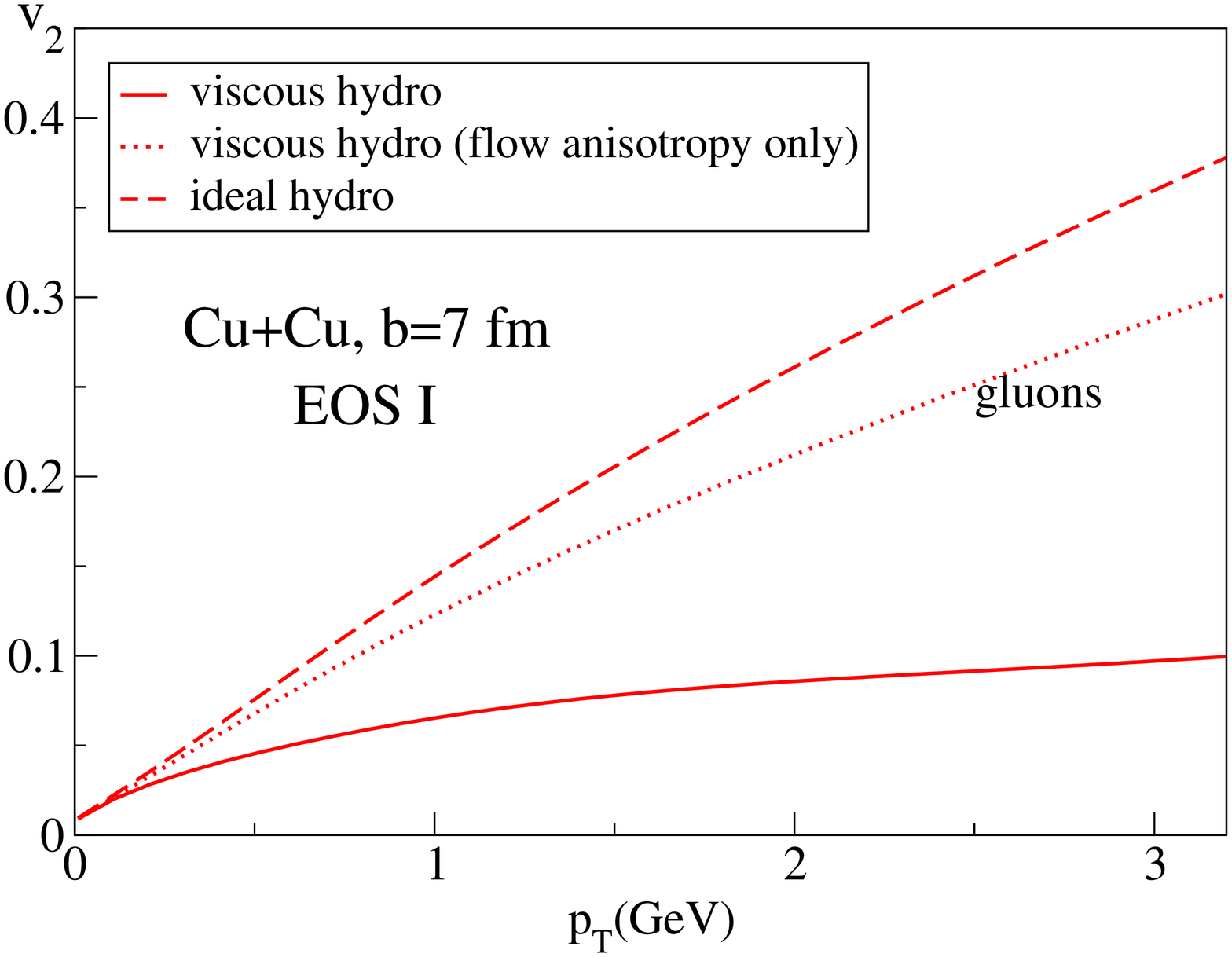}
\includegraphics[bb=0 28 705 534,width=.48\linewidth,clip=]{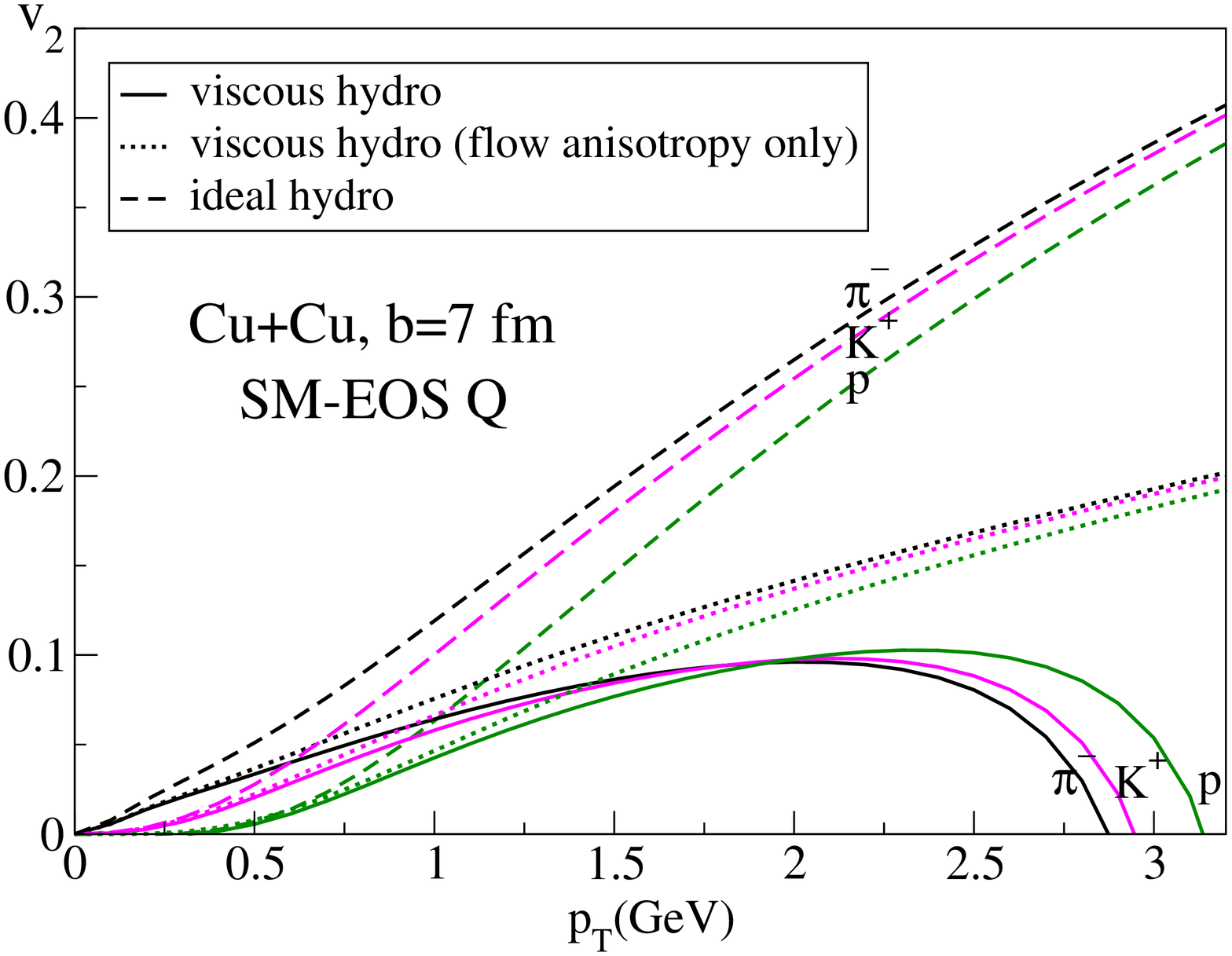}
\caption{(Color online)
Differential elliptic flow $v_2(p_T)$ for Cu+Cu collisions at $b\eq7$\,fm.
{\sl Left panel:} Gluons from evolution with EOS~I. {\sl Right panel:}
$\pi^-$, $K^+$, and $p$ from evolution with SM-EOS~Q. Dashed lines: ideal 
hydrodynamics. Solid lines: viscous hydrodynamics. Dotted lines: viscous
hydrodynamics without non-equilibrium distortion $\delta f$ of 
distribution function at freeze-out.
} 
\label{v-2}
\end{figure*}

%
\begin{figure*}
\includegraphics[bb=19 20 692 530,width=.468\linewidth,clip=]{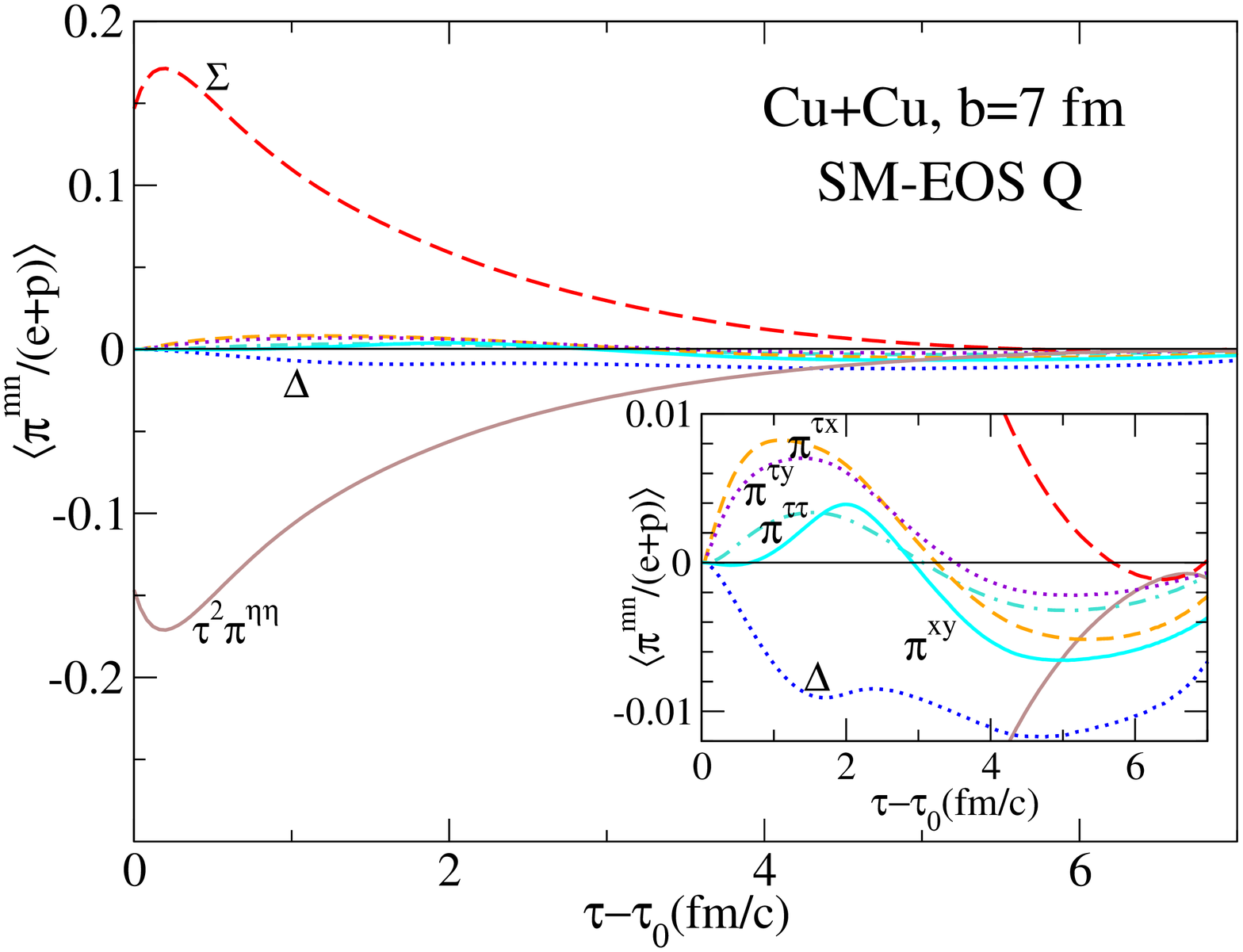}
\includegraphics[bb=25 33 711 530,width=.49\linewidth,clip=]{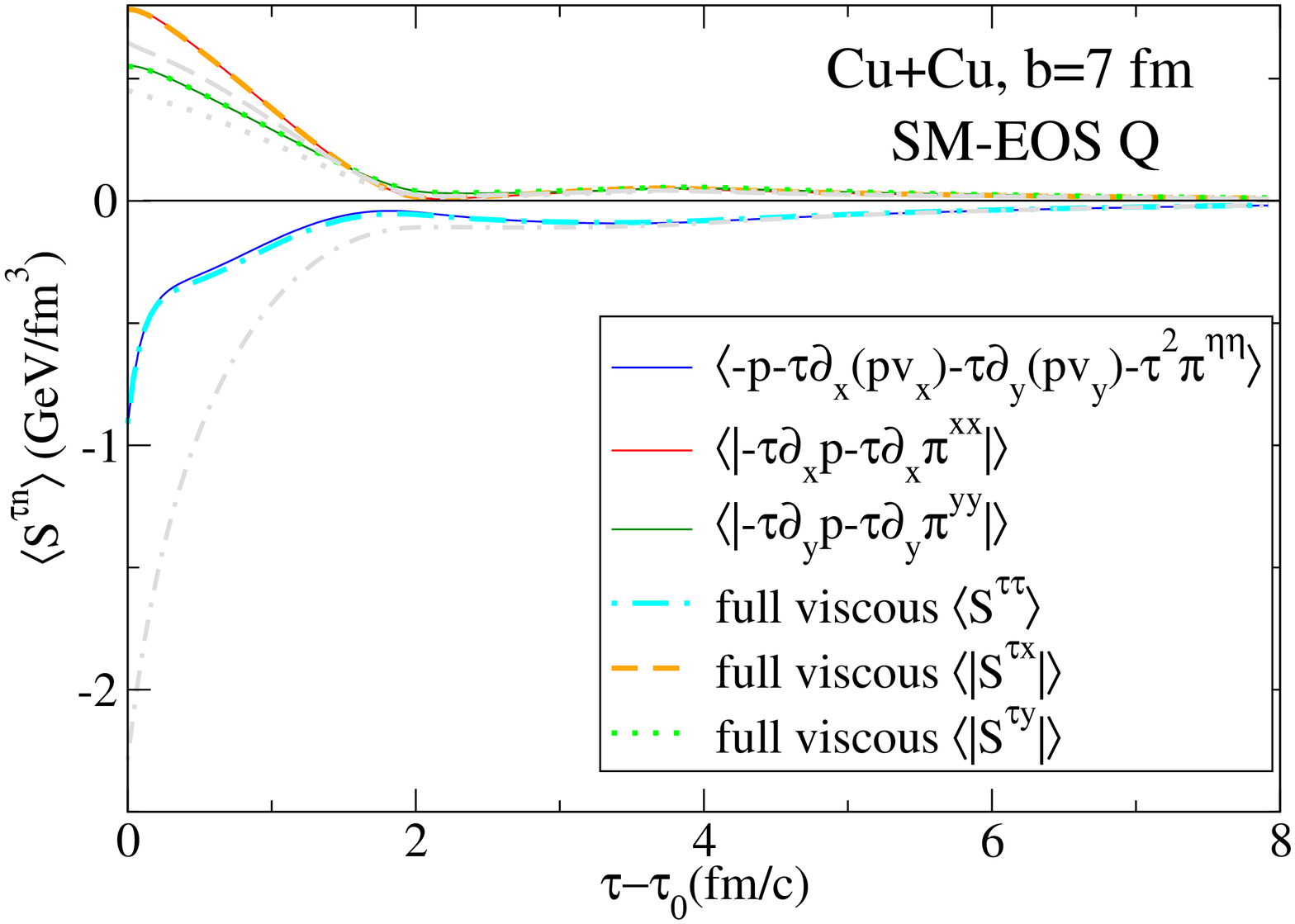}
\caption{(Color online)
{\sl Left panel:} Time evolution of the various components of the shear
viscous pressure tensor, normalized by the enthalpy and averaged in the 
transverse plane over the thermalized region inside the freeze-out surface
\cite{fn6}. Note that the normalizing factor $e{+}p{\,\sim\,}T^4$ decreases 
rapidly with time. 
{\sl Right panel:} Comparison of the full viscous hydrodynamic source 
terms ${\cal S}^{mn}$, averaged over the transverse plane, with their 
approximations given in Eqs.~(\ref{S00}-\ref{S02}), as a function of time. 
The thin gray lines indicate the corresponding source terms in ideal fluid
dynamics.
}
\label{AverPi}
\end{figure*}
%

The effect of the viscous suppression of the total momentum anisotropy
$\epsilon'_p$ on the final particle elliptic flow is shown in 
Figure~\ref{v-2}. Even for the ``minimal'' viscosity $\frac{\eta}{s}\eq\frac{
1}{4\pi}$ considered here one sees a very strong suppression of the
differential elliptic flow $v_2(p_T)$ from viscous evolution (dashed lines)
compared to the ideal fluid (solid lines). Both the viscous reduction of
the collective flow anisotropy (whose effect on $v_2$ is shown as the 
dotted lines) and the viscous contributions to the anisotropy of the 
local momentum distribution (embodied in the term $\delta f$ in 
Eq.~(\ref{Cooper})) play big parts in this reduction. The runs with
EOS~I (which is a very hard EOS) decouple more quickly than those with
SM-EOS~Q; correspondingly, the viscous pressure components are still
larger at freeze-out and the viscous corrections $\delta f$ to the 
distribution function play a bigger role. With SM-EOS~Q the fireball 
doesn't freeze out until $\pi^{mn}$ has become very small (see 
Fig.~\ref{AverPi} below), resulting in much smaller corrections from 
$\delta f$ (difference between dashed and dotted lines in Fig.~\ref{v-2})
\cite{fn4}. On the other hand, due to the longer fireball lifetime the 
negatively anisotropic viscous pressure has more time to decelerate 
the buildup of anisotropic flow, so $v_2$ is strongly reduced because of 
the much smaller flow-induced momentum anisotropy $\epsilon_p$.

The net effect of all this is that, for Cu+Cu collisions and in the 
soft momentum region $p_T{\,<\,}1.5$\,GeV/$c$, the viscous evolution 
with $\frac{\eta}{s}\eq\frac{1}{4\pi}$ leads to a suppression of $v_2$ 
by about a factor 2 \cite{fn5}, in both the slope of its $p_T$-dependence 
and its $p_T$-integrated value. (Due to the flatter $p_T$-spectra from
the viscous dynamics, the effect in the $p_T$-integrated $v_2$ is not
quite as large as for $v_2(p_T)$ at fixed $p_T$.) 

\subsection{Time evolution of the viscous pressure tensor components
and hydrodynamic source terms}
\label{sec4c}

In Figure~\ref{AverPi} we analyze the time evolution of the 
viscous pressure tensor components and the viscous hydrodynamic
source terms on the r.h.s. of Eqs.~(\ref{transport-T}). As already
mentioned, the largest components of $\pi^{mn}$ are $\tau^2\pi^{\eta\eta}$,
$\pi^{xx}$ and $\pi^{yy}$ (see Fig.~2 in \cite{Song:2007fn} and
left panel of  Fig.~\ref{AverPi} \cite{fn6}). At early times, both 
$\tau^2\pi^{\eta\eta}$ and the sum $\Sigma\eq\pi^{xx}{+}\pi^{yy}$ reach 
(with opposite signs) almost 20\% of the equilibrium enthalpy $e{+}p$.
At this stage all other components of $\pi$ are at least an order 
of magnitude smaller (see inset). The largest of these small components
is the difference $\Delta\eq\pi^{xx}{-}\pi^{yy}$ which we choose as the 
variable describing the anisotropy of the viscous pressure tensor in 
non-central collisions. At late times ($\tau{-}\tau_0{\,>\,}5$\,fm/$c$), 
when the large components of $\pi^{mn}$ have strongly decreased, $\Delta$ 
becomes comparable to them in magnitude. As a fraction of the thermal 
equilibrium enthalpy $e{+}p{\,\sim\,}T^4$ which sets the scale in ideal 
fluid dynamics and which itself decreases rapidly with time, all viscous 
pressure components are seen to decrease with time. In a fluid with a set 
ratio $\eta/s$, viscous effects thus become less important with time. 
In real life, however, the ratio $\eta/s$ depends itself on temperature 
and rises dramatically during the quark-hadron phase transition and below 
\cite{Hirano:2005wx,Csernai:2006zz}. Shear viscous effects will therefore 
be larger at late times than considered here. The consequences of this 
will be explored elsewhere.

The observation that many components of $\pi^{mn}$ are very small
throughout the fireball evolution underlies the validity of the 
approximation of the hydrodynamic source terms given in the second lines 
of Eqs.~(\ref{S00}-\ref{S02}). The excellent quality of this approximation 
is illustrated in the right panel of Fig.~\ref{AverPi}. 

\subsection{Viscous corrections to final pion spectra and elliptic flow}
\label{sec4d}

\begin{figure}[htb]
\includegraphics[bb=49 51 702 528,width=\linewidth,clip=]{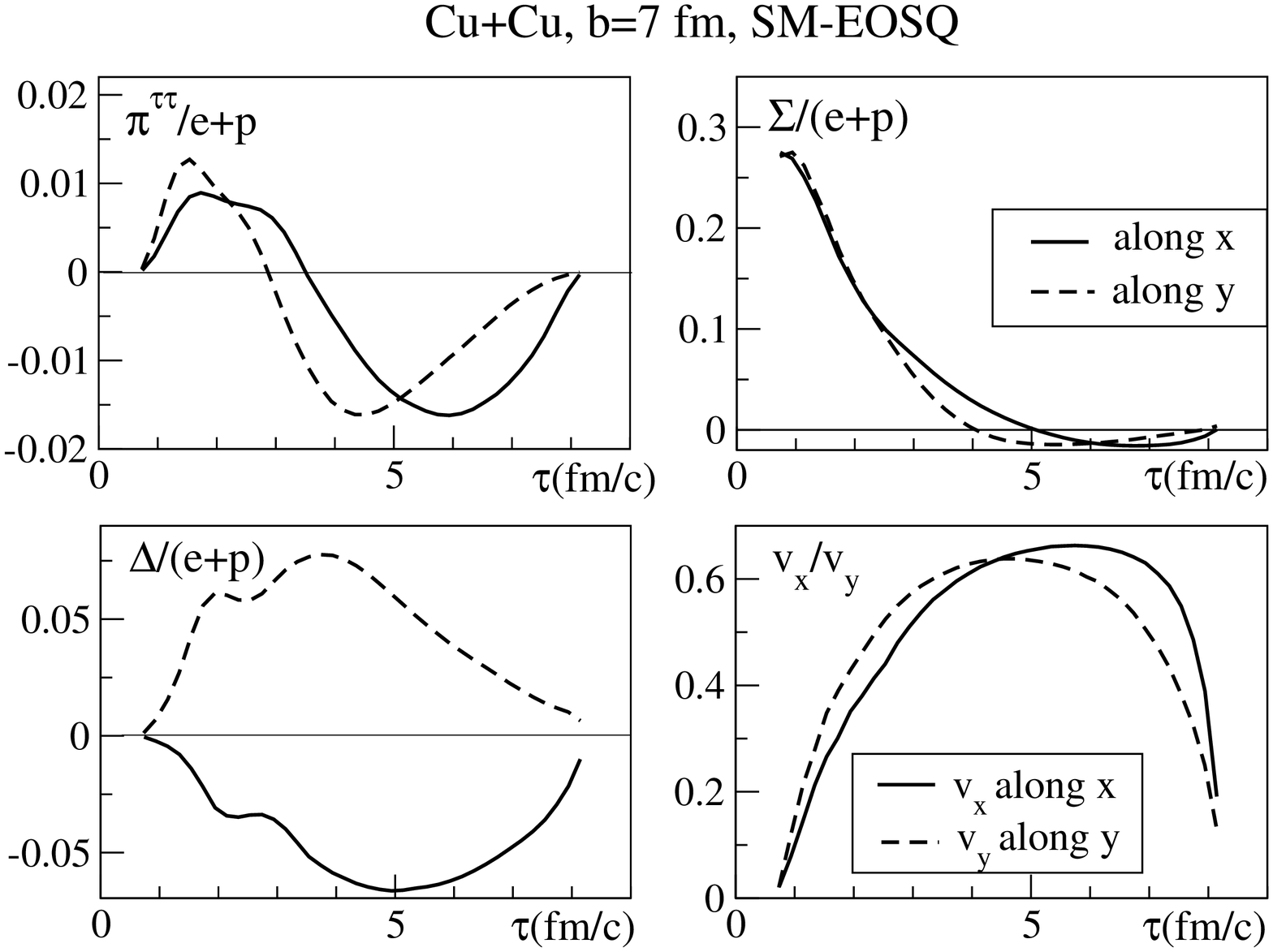}
\caption{
Time evolution of $\pi^{\tau\tau}$, $\Sigma$, and $\Delta$, as well
as that of the transverse velocity, along the decoupling surface
for $b\eq7$\,fm Cu+Cu collisions in viscous hydrodynamics with 
SM-EOS~Q, as shown in Fig.~\ref{Contourb7}. Solid (dashed) lines 
represent cuts along the $x$ ($\phi\eq0$) and $y$ ($\phi\eq\frac{\pi}{2}$) 
directions.
}
\label{Pi-dec} 
\end{figure}

The large viscous reduction of the elliptic flow seen in 
Fig.~\ref{v-2} warrants a more detailed analysis of the viscous
corrections to the particle spectra and $v_2$. In Fig.~\ref{Pi-dec}
we show, for Cu+Cu collisions at $b\eq7$\,fm evolved with SM-EOS~Q, 
%
%
the time evolution of the independent components $\pi^{\tau\tau}$,
$\Sigma$ and $\Delta$ of the viscous pressure tensor $\pi^{mn}$, 
normalized by the equilibrium enthalpy $e{+}p$, along the 
$T_\mathrm{dec}\eq130$\,MeV decoupling surface plotted in the 
upper right panel of Fig.~\ref{Contourb7}. Solid (dashed) lines show 
the behaviour along the $x$ $(y)$ direction (right (left) half of the 
upper right panel in Fig.~\ref{Contourb7}). We see that generically 
all three of these viscous pressure components are of similar magnitude,
except for $\Sigma$ which strongly dominates over the other two
during the first 2\,fm/$c$ after the beginning of the expansion stage.
However, since most particle production, especially that of low-$p_T$ 
particles, occurs at late times ($\tau{\,>\,}4$\,fm/$c$ for 
$b\eq7$\,fm/$c$ Cu+Cu, see Fig.~\ref{Contourb7} and the discussion 
around Fig.~27 in Ref.~\cite{Rev-hydro}), the regions where $\Sigma$ 
is large do not contribute much. As far as the non-equilibrium 
contribution to the spectra is concerned, we can thus say that the 
viscous pressure at freeze-out is of the order of a few percent of $e{+}p$.
The anisotropy term $\Delta$ is even smaller, due to cancellations 
between the in-plane ($x$) and out-of-plane ($y$) contributions when
integrating over the azimuthal angle in Eq.~(\ref{Cooper}).  

\begin{figure}[htb]
\includegraphics[bb=4 31 705 560,width=\linewidth,clip=]{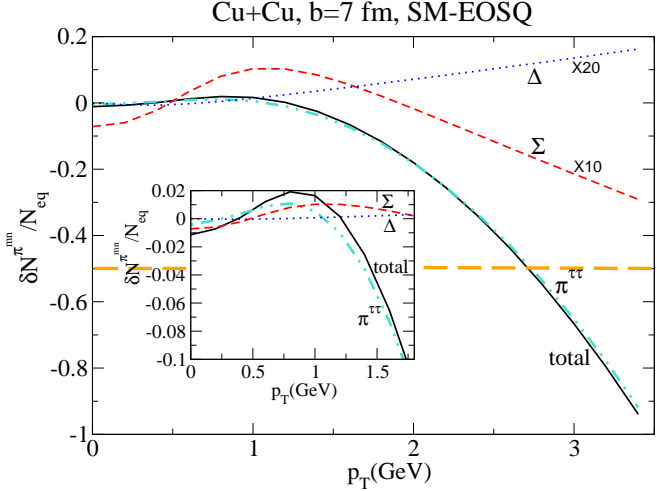}
\caption{(Color online)
Viscous corrections to the azimuthally averaged pion spectrum resulting 
from individual components of the viscous pressure tensor $\pi^{mn}$
as indicated, as well as the total correction $\delta N/N_\mathrm{eq}$,
for Cu+Cu collisions at $b\eq7$\,fm with SM-EOS~Q. The horizontal dashed
line at -50\% indicates the limit of validity.
}
\label{dPidN}
\end{figure}

These viscous pressure components generate the non-equilibrium 
contribution $\delta f$ to the distribution function on the freeze-out
surface according to Eqs.~(\ref{deltaf}) and (\ref{vis-cor-2}), resulting
in a corresponding viscous correction to the azimuthally integrated 
particle spectrum 
$\delta N{\,\equiv\,}\int d\phi_p \,\delta\left(E\frac{d^3N}{d^3p}\right)$. 
Figure~\ref{dPidN} shows these non-equilibrium contributions for pions, 
normalized by the azimuthally averaged equilibrium part 
$N_\mathrm{eq}{\,\equiv\,}\int d\phi_p \frac{d^3N_\mathrm{eq}}{d^3p}$.
We show both the total viscous correction and the individual 
contributions arising from the three independent pressure tensor 
components used in Eq.~(\ref{vis-cor-2}) and shown in Fig.~\ref{Pi-dec}. 

In the viscous correction, the term (\ref{vis-cor-2}) (normalized by 
$T^2(e{+}p)$) is weighted by particle production via the equilibrium 
distribution function $f_\mathrm{eq}(x,p)$. It is well-known (see Fig.~27 
in Ref.~\cite{Rev-hydro}) that for low-$p_T$ particles this weight is 
concentrated along the relatively flat top part of the decoupling
surface in Fig.~\ref{Contourb7}, corresponding to 
$\tau{\,\gtrsim\,}5{-}6$\,fm/$c$ in Fig.~\ref{Pi-dec}. In this momentum
range, the contributions from $\pi^{\tau\tau}$, $\Sigma$ and $\Delta$
to $\delta N/N_\mathrm{eq}$ are of similar magnitude and alternating 
signs (see Fig.~\ref{dPidN}), making the sign of the overall viscous
correction to the spectra hard to predict.  

%
%
High-$p_T$ particles, on the other hand, come from those regions in the 
fireball which feature the largest transverse flow velocity at freeze-out. 
Fig.~\ref{Contourb7} shows that this restricts their emission mostly to 
the time interval $3{\,<\,}\tau{\,<\,}6$\,fm/$c$. In this region 
$\pi^{\tau\tau}$ is negative, see Fig.~\ref{Pi-dec}. A detailed study 
of the different terms in Eq.~(\ref{vis-cor-2}) reveals that (after 
azimuthal integration) the expression multiplying $\pi^{\tau\tau}$ is 
positive, hence the negative sign of $\pi^{\tau\tau}$ explains its 
negative contribution to $\delta N/N_\mathrm{eq}$ at high $p_T$, as 
seen in Fig.~\ref{dPidN}. Figure~\ref{dPidN} also shows that in the 
region $p_T{\,\gtrsim\,}1$\,GeV/$c$ the first line ${\sim\,}\pi^{\tau\tau}$ 
in Eq.~(\ref{vis-cor-2}) completely dominates the viscous correction to 
the spectra. We found that this involves additional cancellations between 
terms of opposite sign (after azimuthal integration) inside the square 
brackets multiplying $\Sigma$ and $\Delta$ in the second and third line 
of Eq.~(\ref{vis-cor-2}). Furthermore, the term ${\sim\,}\pi^{\tau\tau}$ 
is the only contribution whose magnitude grows quadratically with $p_T$. 
For the contributions involving $\Sigma$ and $\Delta$, the apparent 
quadratic momentum dependence seen in Eq.~(\ref{vis-cor-2}) is tempered 
by the integrations over space-time rapidity $\eta$ and azimuthal angle 
$\phi$ in (\ref{Cooper}), resulting in only linear growth at large $p_T$.

In the absence of higher-order momentum anisotropies $v_n$, $n{\,>\,}2$,
the elliptic flow $v_2(p_T)$ can be easily computed from the momentum
spectra in $x$ ($\phi_p{=}0$) and $y$ ($\phi_p{=}\frac{\pi}{2}$) 
directions: 
\begin{eqnarray}
\label{v2-eqn2}
  2 v_2(p_T) &=& \frac{N_x-N_y}{N} 
\nonumber\\
  &=& \frac{(N_{x,\mathrm{eq}}{-}N_{y,\mathrm{eq}})
          + (\delta N_x{-}\delta N_y)}
           {N_\mathrm{eq} + \delta N},
\end{eqnarray}
where $N\eq{N}_\mathrm{eq}{+}\delta N$ is shorthand for the 
%
\begin{figure}[htb]
\includegraphics[bb=44 34 711 524,width=\linewidth,clip=]{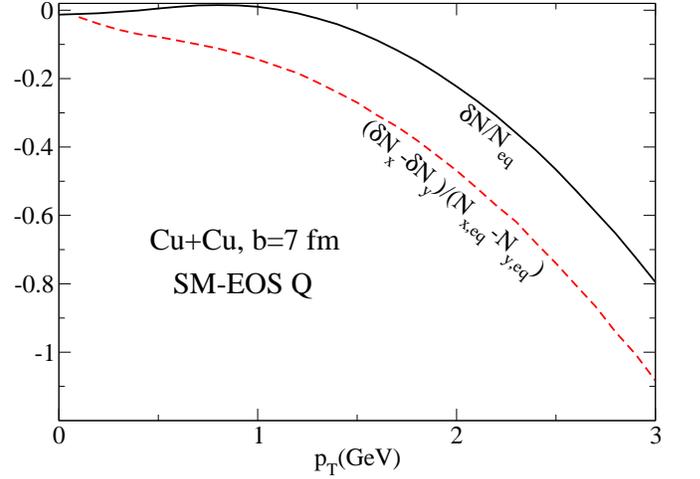}
\caption{(Color online)
Ratio of non-equilibrium and equilibrium contributions to particle 
production (solid line) and to its momentum anisotropy
(dashed line), as a function of $p_T$ for pions from Cu+Cu collisions 
at $b\eq7$\,fm with SM-EOS~Q.
}
\label{NxNy-V2}
\end{figure}
%
azimuthally averaged spectrum $dN/(2\pi\,dy\,p_T dp_T)$, and
$N_{x,y}$ denote the $p_T$ spectra along the $x$ and $y$
directions, respectively:
$N_{x}{\,\equiv\,}N_{x,\mathrm{eq}}{+}\delta N_x{\,\equiv\,}\frac{
d^3N}{dy\,p_T dp_T\,d\phi_p}(\phi_p{=}0)$, and similarly for $N_y$
with $\phi_p{=}\frac{\pi}{2}$.
Equation~(\ref{v2-eqn2}) shows that $v_2$ receives contributions 
from anisotropies in the equilibrium part of the distribution
function $f_\mathrm{eq}$, which reflect the hydrodynamic flow 
anisotropy along the freeze-out surface, and from the viscous 
correction $\delta f$, which reflects non-equilibrium momentum 
anisotropies in the local fluid rest frame. The dashed line in 
Figure~\ref{NxNy-V2} shows the relative magnitude of these two
anisotropy contributions, $\frac{\delta N_x{-}\delta N_y}{N_{x,
\mathrm{eq}}{-}N_{y,\mathrm{eq}}}$, and compares it with the 
relative magnitude $\frac{\delta N}{N_\mathrm{eq}}$ of the 
non-equilibrium and equilibrium contributions to the total, 
$\phi_p$-integrated pion spectrum for Cu+Cu at $b\eq7$\,fm. We
see that the non-equilibrium contribution to the momentum
anisotropy $v_2$ is always negative and larger in relative 
magnitude than the non-equilibrium contribution to the azimuthally
averaged spectrum. Since $v_2$ is a small quantity reflecting the 
anisotropic distortion of the single-particle spectrum, it reacts 
more sensitively than the spectrum itself to the (anisotropic) 
non-equilibrium contributions caused by the small viscous pressure
$\pi^{mn}$ on the decoupling surface. Furthermore, the viscous 
corrections to the $\phi_p$-integrated spectrum change sign as a 
function of $p_T$, the corrections to $v_2$ are negative everywhere, 
decreasing $v_2(p_T)$ at all values of $p_T$, but especially at large 
transverse momenta.
 
\section{Sensitivity to input parameters and limits of applicability}
\label{sec5}
\subsection {Initialization of $\pi^{mn}$}
\label{sec5a}

Lacking input from a microscopic model of the pre-equilibrium stage
preceding the (viscous) hydrodynamic one, one must supply initial
conditions for the energy momentum tensor, including the viscous 
pressure $\pi^{mn}$. The most popular choice has been to initialize
$\pi^{mn}$ with its Navier-Stokes value, i.e. to set initially
$\pi^{mn}\eq2\eta\sigma^{mn}$. Up to this point, this has also been 
our choice in the present paper. Ref.~\cite{Romatschke:2007mq} advocated
the choice $\pi^{mn}\eq0$ at time $\tau_0$ in order to minimize
viscous effects and thus obtain an upper limit on  $\eta/s$ by
comparison with experimental data. In the present subsection we 
explore the sensitivity of the final spectra and elliptic flow
to these different choices of initialization, keeping all other
model parameters unchanged.  

%
\begin{figure*}[htb]
\includegraphics[bb=19 20 692 530,width=.468\linewidth,clip=]{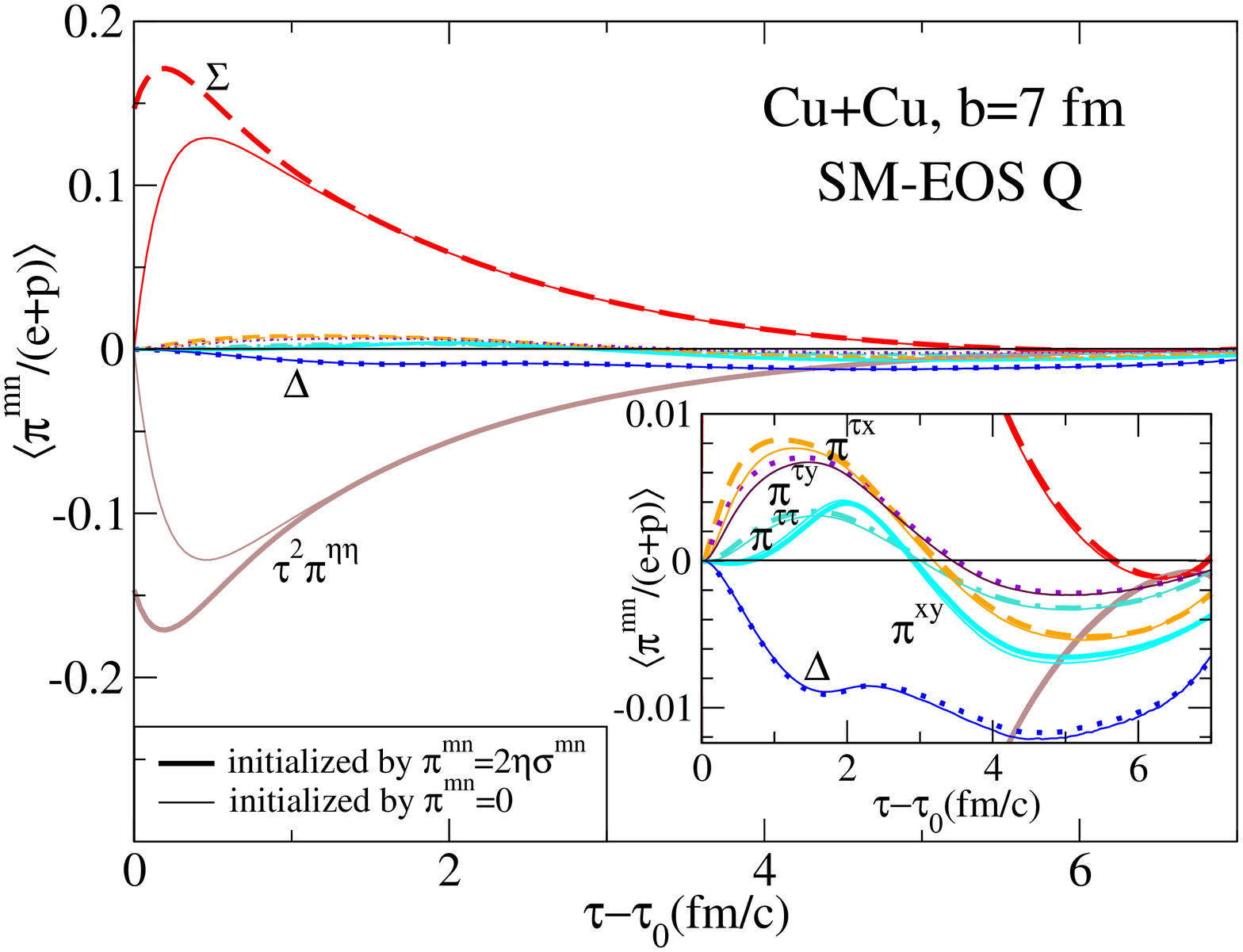}
\includegraphics[bb=25 33 711 530,width=.49\linewidth,clip=]{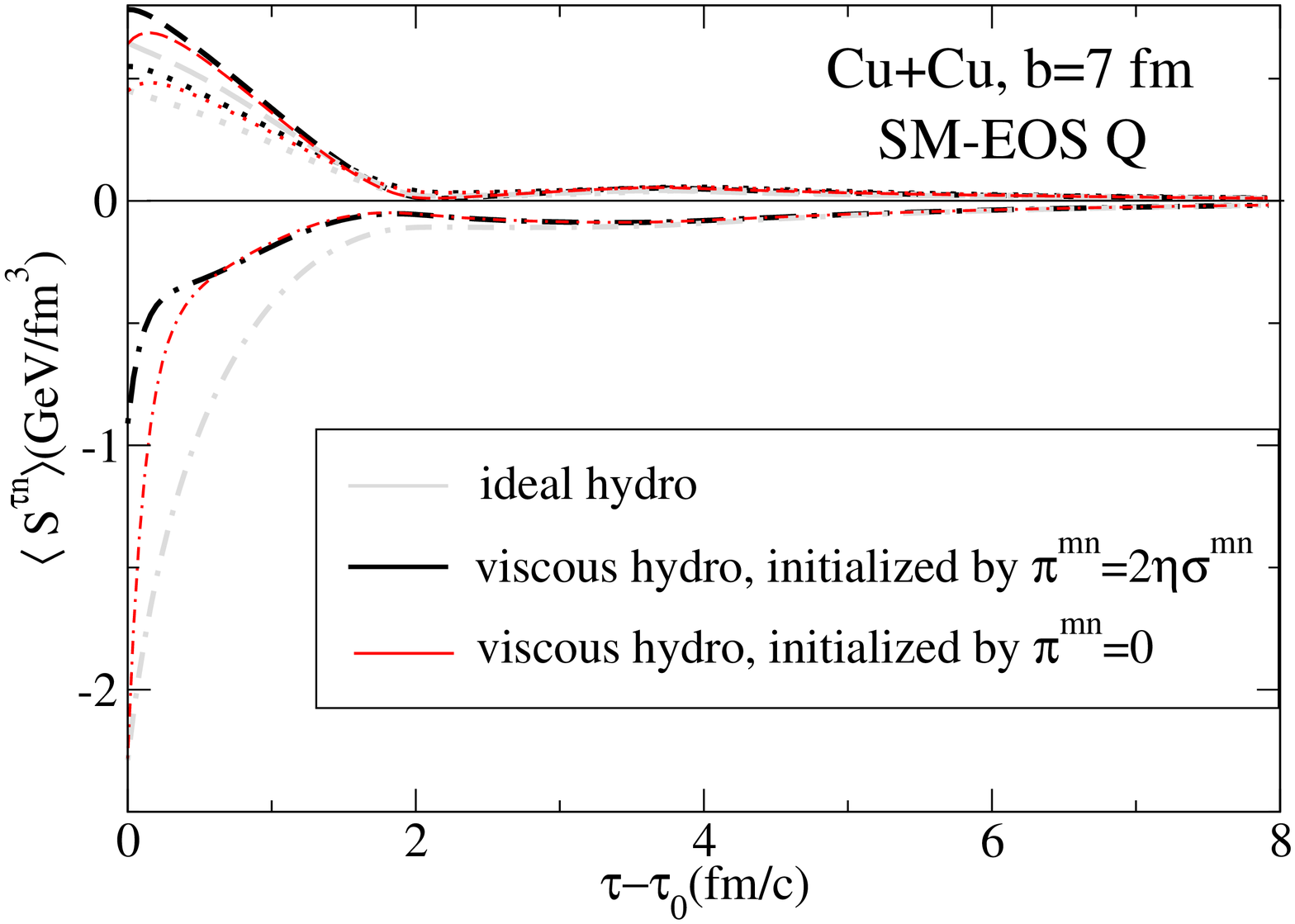}
\caption{(Color online)
Similar to Figure~\ref{AverPi}, but now comparing runs with different
initial conditions. The thick lines reproduce the results from 
Figure~\ref{AverPi}, obtained with $\pi^{mn}\eq2\eta
\sigma^{mn}$at initial time $\tau_0$, while thin lines of the 
same type show the corresponding results obtained by setting initially
$\pi^{mn}\eq0$. The right panel shows the full viscous source 
terms, without approximation: $\langle|{\cal S}^{\tau x}|\rangle$ (dashed),
$\langle|{\cal S}^{\tau y}|\rangle$ (dotted), and
$\langle{\cal S}^{\tau\tau}\rangle$ (dash-dotted).
}
\label{ScT-2}
\end{figure*}
%
Figure~\ref{ScT-2} shows the time evolution of the viscous pressure
tensor and viscous hydrodynamic source terms for the two different
initializations. Differences with respect to the results shown 
Fig.~\ref{AverPi} (which are reproduced in Fig.~\ref{ScT-2} for 
comparison) are visible only at early times $\tau{-}\tau_0{\,\lesssim\,}5
\tau_\pi{\,\approx\,}1$\,fm/$c$. After $\tau_\pi{\,\sim\,}0.2$\,fm/$c$,
the initial difference $\pi^{mn}{-}2\eta\sigma^{mn}$ has decreased by
roughly a factor $1/e$, and after several kinetic scattering times 
$\tau_\pi$ the hydrodynamic evolution has apparently lost all
memory how the viscous terms were initialized. 

\begin{figure}[htb]
\includegraphics[bb=52 28 705 563,width=\linewidth,clip=]{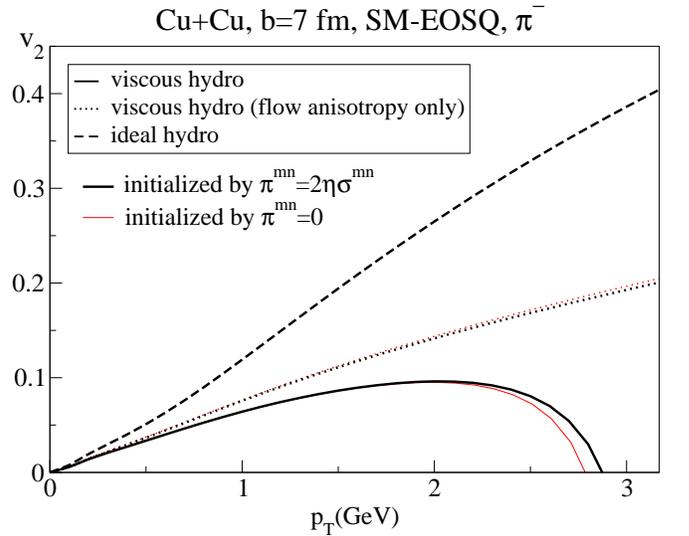}
\caption{(Color online)
Differential elliptic flow $v_2(p_T)$ for pions from $b\eq7$\,fm Cu+Cu 
collisions with SM-EOS~Q. Thick lines reproduce the pion curves from 
Figure~\ref{v-2}, obtained with $\pi^{mn}\eq2\eta\sigma^{mn}$at initial 
time $\tau_0$, while thin lines of the same type show the corresponding 
results obtained by setting initially $\pi^{mn}\eq0$. 
\vspace*{-8mm}
} 
\label{v2-comp}
\end{figure}

Correspondingly, the final spectra and elliptic flow show very little 
sensitivity to the initialization of $\pi^{mn}$, as seen in
Fig.~\ref{v2-comp}. With vanishing initial viscous pressure, 
viscous effects on the final flow anisotropy are a little weaker
(dotted lines in Fig.~\ref{v2-comp}), but this difference is 
overcompensated in the total elliptic flow by slightly stronger 
anisotropies of the local rest frame momentum distributions at 
freeze-out (dashed lines in Fig.~\ref{v2-comp}). For shorter
kinetic relaxation times $\tau_\pi$, the differences resulting from 
different initializations of $\pi^{mn}$ would be smaller still.
%
%

\subsection{Kinetic relaxation time $\tau_\pi$}
\label{sec5b}

While the finite relaxation time $\tau_\pi$ for the viscous pressure 
tensor in the Israel-Stewart formalism eliminates problems with 
superluminal signal propagation in the relativistic Navier-Stokes theory,
it also keeps the viscous pressure from ever fully approaching its
Navier-Stokes limit $\pi^{mn}\eq2\eta\sigma^{mn}$. In this subsection
we explore how far, on average, the viscous pressure evolved by
VISH2+1 deviates from its Navier-Stokes limit, and how this changes if 
we reduce the relaxation time $\tau_\pi$ by a factor 2. 

\begin{figure}[htb]
\includegraphics[bb=40 50 369 522,width=\linewidth,clip=]{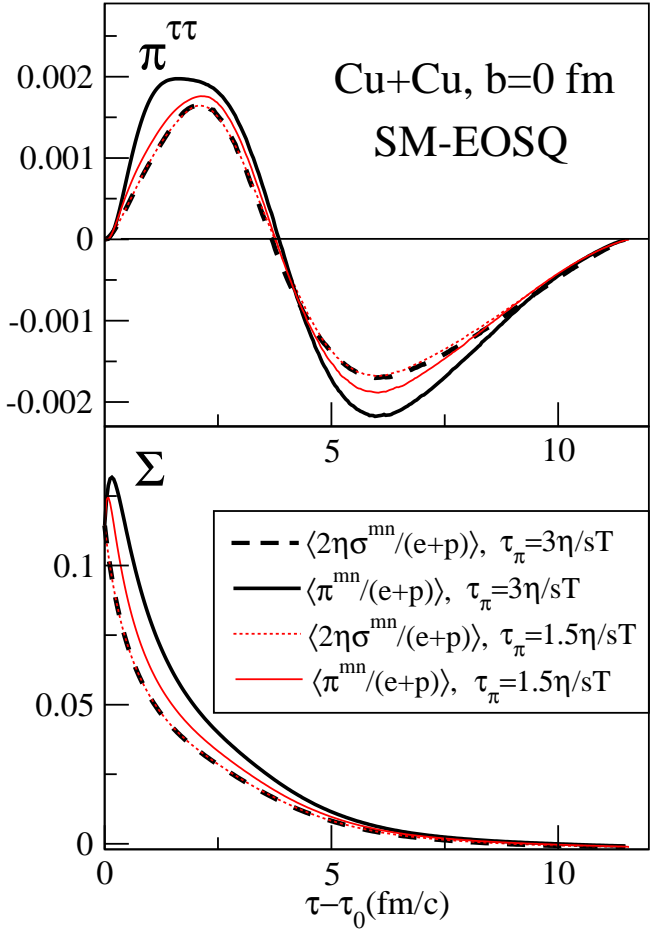}
\caption{(Color online)
Time evolution of the two independent viscous pressure tensor
components $\pi^{\tau\tau}$ and $\Sigma\eq\pi^{xx}{+}\pi^{yy}$
for central Cu+Cu collisions (solid lines), compared with their 
Navier-Stokes limits $2\eta\sigma^{\tau\tau}$ and 
$2\eta(\sigma^{xx}{+}\sigma^{yy})$ (dashed lines), for two values 
of the relaxation time, $\tau_\pi\eq3\eta/sT$ (thick lines) and 
$\tau_\pi\eq1.5\eta/sT$ (thin lines). All quantities are scaled
by the thermal equilibrium enthalpy $e{+}p$ and transversally 
averaged over the thermalized region inside the decoupling surface.
} 
\label{1st2nd}
\end{figure}

In Figure~\ref{1st2nd} we compare, for central Cu+Cu collisions, the 
%
%
time evolution of the scaled viscous pressure tensor, averaged in the 
transverse plane over the thermalized region inside the freeeze-out 
surface, with its Navier-Stokes limit, for two values of $\tau_\pi$, 
$\tau_\pi\eq3\eta/sT\eq\tau_\pi^\mathrm{class}/2$ and 
$\tau_\pi\eq\tau_\pi^\mathrm{class}/4$. For the larger relaxation time,
the deviations from the Navier-Stokes limit reach 25-30\% at early
times, but this fraction gradually decreases at later times. For the 
twice shorter relaxation time, the fractional deviation from 
Navier-Stokes decreases by somewhat more than a factor 2 and never
exceeds a value of about 10\%.

%
\begin{figure}[htb]
\includegraphics[bb= 52 28 705 565,width=\linewidth,clip=]{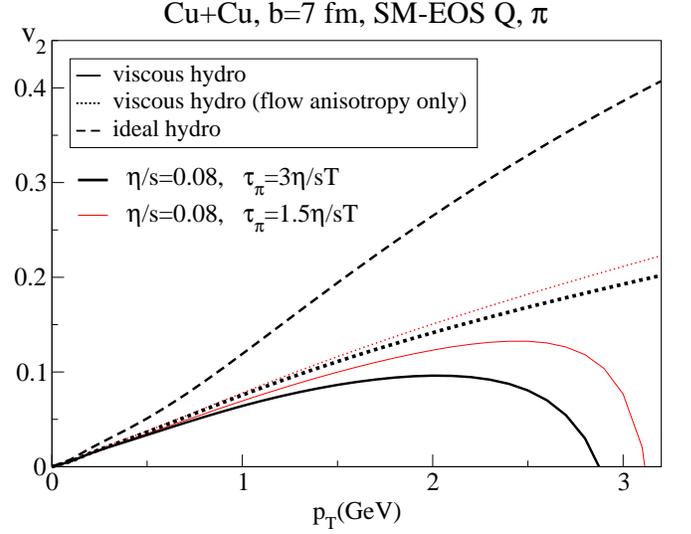}
\caption{(Color online)
Differential elliptic flow $v_2(p_T)$ for $\pi^-$ from $b\eq7$\,fm Cu+Cu 
collisions with SM-EOS~Q, calculated from viscous hydrodynamics with two 
different values for the relaxation time $\tau_\pi$. Thick lines reproduce 
the pion curves from Figure~\ref{v-2}, thin lines show results obtained
with a twice shorter relaxation time. For the standard (twice larger) 
classical relaxation time value $\tau_\pi\eq6\eta/sT$ 
\cite{Israel:1976tn,Baier:2006um} deviations from ideal hydrodynamics 
would exceed those seen in the thick lines.
}  
\label{v-2-Beta}
\end{figure}
%

Figure~\ref{v-2-Beta} shows that, small as they may appear, these 
deviations of $\pi^{mn}$ from its Navier-Stokes limit $2\eta\sigma^{mn}$ 
(especially on the part of the decoupling surface corresponding to early 
times $\tau{-}\tau_0$) still play an important role for the viscous 
reduction of elliptic flow observed in our calculations. While a 
decrease of the relaxation time by a factor 2 leads to only a small 
reduction of the viscous suppression of flow anisotropies (dotted lines 
in Fig.~\ref{v-2-Beta}), the contribution to $v_2(p_T)$ resulting from 
the viscous correction ${\sim\,}p_m p_n \pi^{mn}$ to the final particle 
spectra is reduced by about a factor 2, too, leading to a significant 
overall increase of $v_2(p_T)$ in the region $p_T{\,>\,}1$\,GeV/$c$. To 
avoid strong sensitivity to the presently unknown value of the relaxation 
time $\tau_\pi$ in the QGP, future extractions of the specific shear 
viscosity $\eta/s$ from a comparison between experimental data and 
viscous hydrodynamic simulations should therefore be performed at 
{\em low transverse momenta}, $p_T{\,<\,}1$\,GeV/$c$, where our results 
appear to be reasonably robust against variations of $\tau_\pi$.
 
\subsection{Breakdown of viscous hydrodynamics at high $p_T$}
\label{sec5c}

As indicated by the horizontal dashed lines in 
Figs.~\ref{Spectra-Correction} and \ref{dPidN}, the assumption
$|\delta f|{\,\ll\,}|f_\mathrm{eq}|$ under which the viscous 
hydrodynamic framework is valid breaks down at sufficiently
large transverse momenta. For a quantitative assessment we assume 
that viscous hydrodynamic predictions become unreliable when the 
viscous corrections to the particle spectra exceed 50\%. 
Fig.~\ref{Spectra-Correction} shows that the characteristic transverse
momentum $p_T^*$ where this occurs depends on the particle species
and increases with particle mass. To be specific, we here consider 
$p_T^*$ for pions --- the values for protons would be about 15\% 
higher. The discussion in the preceding subsection of the $
\tau_\pi$-dependence of viscous corrections to the final spectra makes
it clear that reducing $\tau_\pi$ will also push $p_T^*$ to larger 
values. Since we do not know $\tau_\pi$ we refrain from a 
quantitative estimate of this effect. 

%
\begin{figure}[htb]
\includegraphics[bb=40 22 724 531,width=\linewidth,clip=]{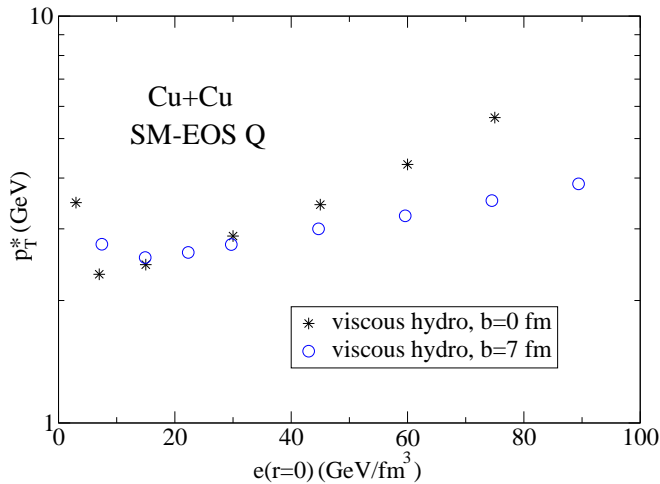}
\caption{(Color online)
Characteristic transverse momentum $p_T^*$ where the viscous corrections
to the final pion spectrum become so large (${>\,}50\%$) that the
spectrum becomes unreliable, as a function of the initial energy density
in the center of the fireball. Stars are for central Cu+Cu collisions,
open circles for peripheral Cu+Cu collisions at $b\eq7$\,fm. Note that
identical $e(r{=}0)$ values correspond to higher collision energies in
peripheral than in central collisions. $p_T^*$ values are higher for
more massive hadrons (see Fig.~\ref{Spectra-Correction}), and they also 
increase for smaller relaxation times $\tau_\pi$ (see discussion of 
Fig.~\ref{v-2-Beta}).
}
\label{V2-dnN}
\end{figure}
%
In Fig.~\ref{V2-dnN} we show the breakdown momentum $p_T^*$ for
pions as a function of the peak initial energy density in the fireball
center (i.e. indirectly as a function of collision energy), for both 
central and peripheral Cu+Cu collisions. (The initial time was held 
fixed at $\tau_0\eq0.6$\,fm.) Generically, $p_T^*$ rises with collision 
energy. The anomaly at low values of $e(r{=}0)$ results, as far as we 
could ascertain, from effects connected with the phase transition in 
SM-EOS~Q. The rise of $p_T^*$ with increasing $e(r{=}0)$ reflects the 
growing fireball lifetime which leads to smaller viscous pressure 
components at freeze-out. This lifetime effect is obviously stronger 
for central than for peripheral collisions, leading to the faster rise 
of the stars than the open circles in Fig.~\ref{V2-dnN}. Taking further 
into account that a given beam energy leads to higher $e(r{=}0)$ values 
in central than in peripheral collisions such that, for a given 
experiment, the peripheral collision points are located farther to the 
left in the figure than the central collision points, we conclude that 
in central collisions the validity of viscous hydrodynamics extends to 
{\em significantly larger} values of $p_T$ than in peripheral 
collisions: Viscous effects are more serious in peripheral than in 
central collisions.

\section{\bf Summary and conclusions}
\label{sec6}

In this paper, we numerically studied the shear viscous effects to
the hydrodynamic evolution, final hadron spectra, and elliptic flow 
$v_2$, using a (2+1)-dimensional causal viscous hydrodynamic code, 
VISH2+1, based on the 2nd order Israel-Stewart formalism. Using a 
fixed set of initial and final conditions, we explored the effects 
of shear viscosity for a ``minimally'' \cite{son} viscous fluid with 
$\frac{\eta}{s}\eq\frac{1}{4\pi}$ in central and peripheral Cu+Cu 
collisions, comparing the evolution with two different equations of 
state, an ideal massless parton gas (EOS~I) and an EOS with a 
semirealistic parametrization of the quark-hadron phase transition 
(SM-EOS~Q). Final hadron spectra and their elliptic flow were calculated 
from the hydrodynamic output using the Cooper-Frye prescription. 

We found that shear viscosity decelerates longitudinal expansion, but 
accelerates the build-up of transverse flow. This slows the cooling 
process initially, leading to a longer lifetime for the QGP phase,
but causes accelerated cooling at later stages by faster transverse
expansion. Viscous pressure gradients during the mixed phase increase
the acceleration during this stage and slightly reduce its lifetime.
They counteract large gradients of the radial velocity profile that 
appear in ideal fluid dynamics as a result of the softness of the EOS 
in the mixed phase, thereby {\it de facto} smoothing the assumed 
first-order phase transition into a rapid cross-over transition. In the
end the larger radial flow developing in viscous hydrodynamics leads to 
flatter transverse momentum spectra of the finally emitted particles, 
while their azimuthal anisotropy in non-central heavy-ion collisions is 
found to be strongly reduced.

Although the viscous hardening of the hadron $p_T$-spectra can be largely
absorbed by retuning the initial conditions, starting the transverse 
expansion later and with lower initial entropy density 
\cite{Baier:2006um,Baier:2006gy}, this only acerbates the viscous effects
on the elliptic flow $v_2$ which in this case is further reduced by the 
decreased fireball lifetime. The reduction of the elliptic flow
$v_2$ by shear viscous effects is therefore a sensitive and robust 
diagnostic tool for shear viscosity in the fluid \cite{Heinz:2002rs}.

Our results indicate that in semiperipheral Cu+Cu collisions even
a ``minimal'' amount of shear viscosity \cite{son} causes a reduction
of $v_2$ by almost 50\% relative to ideal fluid dynamical simulations.
In the present paper we explored the origin of this reduction in great
detail. The effects observed by us for Cu+Cu collisions \cite{Song:2007fn} 
are larger than those recently reported in 
Refs.~\cite{Romatschke:2007mq,Dusling:2007gi}
for Au+Au collisions. While some of these differences can be attributed
to an increased importance of viscous effects in smaller systems
\cite{fn5}, the bulk of the difference appears to arise from the fact 
that the different groups solve somewhat different sets of viscous 
hydrodynamic equations \cite{next,Paul_private}. (See also the recent
interesting suggestion by Pratt \cite{Pratt:2007gj} for a phenomenological
modification of the Isreal-Stewart equations for systems with large 
velocity gradients.) This raises serious questions: if theoretical 
ambiguities in the derivation of the viscous hydrodynamic equations 
reflect themselves in large variations of the predicted elliptic flow, 
any value of the QGP shear viscosity extracted from relativistic 
heavy-ion data will strongly depend on the specific hydrodynamic model 
used in the comparison. A reliable quantitative extraction of $\eta/s$ 
from experimental data will thus only be possible if these ambiguities 
can be resolved.

Our studies show that shear viscous effects are strongest during the
early stage of the expansion phase when the longitudinal expansion rate
is largest. At later times the viscous corrections become small, although
not negligible. Small non-zero viscous pressure components along the
hadronic decoupling surface have significant effects on the final
hadron spectra that grow with transverse momentum and thus limit the 
applicability of the viscous hydrodynamic calculation to transverse
momenta below 2-3\,GeV/$c$, depending on impact parameter, collision 
energy and particle mass. Viscous effects are more important in 
peripheral than in central collisions, and larger for light than for 
heavy particles. They increase with the kinetic relaxation time for
the viscous pressure tensor. Since the breakdown of viscous hydrodynamics 
is signalled by the theory itself, through the relative magnitude of the 
viscous pressure, the applicability of the theory can be checked 
quantitatively case by case and during each stage of the expansion. 

For the kinetic relaxation times $\tau_\pi$ considered in the present 
work, sensitivities to the initial value of the viscous pressure tensor 
were found to be small and practically negligible. Sensitivity to 
the value of $\tau_\pi$ was found for the hadron spectra, especially 
the elliptic flow, at large transverse momenta. This leads us to suggest
to restrict any comparison between theory and experiment with the goal of 
extracting the shear viscosity $\eta/s$ to the region
$p_T{\,\lesssim\,}1$\,GeV/$c$ where the sensitivity to $\tau_\pi$ 
is sufficiently weak.

The dynamical analysis of shear viscous effects on the momentum 
anisotropy and elliptic flow in non-central collisions reveals
an interesting feature: The total momentum anisotropy receives two
types of contributions, the first resulting from the anisotropy of the
collective flow pattern and the second arising from a local momentum
anisotropy of the phase-space distribution function in the local
fluid rest frame, reflecting viscous corrections to its local thermal
equilibrium form. During the early expansion stage the latter effect
(i.e. the fact that large viscous pressure effects generate momentum
anisotropies in the local fluid rest frame) dominate the viscous effects
on elliptic flow. At later times, these local momentum anisotropies get 
transferred to the collective flow profile, manifesting themselves as a 
viscous reduction of the collective flow anisotropy. The time scale for 
transferring the viscous correction to $v_2$ from the local rest frame 
momentum distribution to the collective flow pattern appears to be of 
the same order as that for the evolution of the total momentum 
anisotropy itself.  

Several additional steps are necessary before the work presented here
can be used as a basis for a quantitative interpretation of relativistic
heavy-ion data. First, the abovementioned ambiguity of the detailed form 
of the kinetic evolution equations for the viscous pressure must be 
resolved. Second, the equation of state must be fine-tuned to lattice 
QCD data and other available information, to make it as realistic as 
presently possible. The hydrodynamic scaling of the final elliptic flow 
$v_2$ with the initial source eccentricity $\epsilon_x$ \cite{scaling} 
and its possible violation by viscous effects need to be explored 
\cite{next}, in order to assess the sensitivity of the scaled elliptic 
flow $v_2/\epsilon_x$ to details of the model used for initializing the 
hydrodynamic evolution \cite{Hirano:2005xf}. The temperature dependence 
of the specific shear viscosity $\eta/s$, especially across the 
quark-hadron phase transition \cite{Hirano:2005wx,Csernai:2006zz}, must 
be taken into account, and bulk viscous effects, again particularly near 
$T_c$, must be included. To properly account for the highly viscous nature 
of the hadron resonance gas during the last collision stage it may be 
necessary to match the viscous hydrodynamic formalism to a microscopic 
hadronic cascade to describe the last part of the expansion until hadronic 
decoupling \cite{Hirano:2005wx}. We expect to report soon on progress
along some of these fronts. 

{\it Note added:} Just before submitting this work for publication we 
became aware of Ref.~\cite{Baier:2007ix} where the form of the kinetic 
evolution equations for the viscous pressure is revisited and it is 
argued that Eqs.~(\ref{Pi-transport},\ref{pi-transport}) must be amended 
by additional terms which reduce the strong viscous suppression of the 
elliptic flow observed by us \cite{Paul_private}. While details of the 
numerical results will obviously change if these terms are included 
({\it cf.} Refs.~\cite{Romatschke:2007mq,Dusling:2007gi}), our 
discussion of the driving forces behind the finally observed viscous 
corrections to ideal fluid results and of the evolution of these 
corrections with time is generic, and the insights gained in the present 
study are expected to hold, at least qualitatively, also for future
improved versions of VISH2+1 that properly take into account the new 
findings reported in \cite{Baier:2007ix}.

\acknowledgments
The viscous hydrodynamic code VISH2+1 employs several subroutines from 
AZHYDRO, the (2+1)-d ideal hydrodynamic code developed by P. Kolb 
\cite{AZHYDRO,Kolb:1999it}, especially the flux-corrected SHASTA 
transport algorithm \cite{SHASTA} for evolving the hydrodynamic 
equations. We thank R.~Baier, E. Frodermann, 
P. Kolb, S. Pratt, P. Romatschke, D.~Teaney, and U.~Wiedemann for 
fruitful discussions. This work was supported by the U.S. Department of 
Energy under contract DE-FG02-01ER41190.

\appendix

\section{Expressions for $\tilde{\pi}^{mn}$ and
$\tilde{\sigma}^{mn}$}
\label{appa}

The expressions for $\tilde{\pi}^{mn}$ and $\tilde{\sigma}^{mn}$ in
Eq.~(\ref{transport-pi}) are
\begin{eqnarray}
\label{tilde-pi}
  \tilde{\pi}^{mn} = 
  \begin{pmatrix} \pi^{\tau\tau}&\pi^{\tau x}&\pi^{\tau y}&0\\ 
                  \pi^{\tau x}&\pi^{xx}&\pi^{x y}&0\\
                  \pi^{\tau y}&\pi^{xy}&\pi^{y y}&0\\
                  0 & 0& 0& \tau^2\pi^{\eta\eta}
  \end{pmatrix},
\end{eqnarray}
\begin{eqnarray}
\label{tilde-sigma}
  \tilde{\sigma}^{mn} &=&
  \begin{pmatrix} \partial_\tau u^\tau & 
                  \frac{\partial_\tau u^x{-}\partial_x u^\tau}{2} & 
                  \frac{\partial_\tau u^y{-}\partial_y u^\tau}{2} & 0\\[0.5ex]
                  \frac{\partial_\tau u^x{-}\partial_x u^\tau}{2} & 
                  -\partial_x u^x & 
                  -\frac{\partial_x u^y{+}\partial_y u^x}{2} & 0\\[0.5ex]
                  \frac{\partial_\tau u^y{-}\partial_y u^\tau}{2} &
                  -\frac{\partial_x u^y{+}\partial_y u^x}{2} & 
                  -\partial_y u^y & 0\\[0.5ex]
                  0 & 0 & 0 & -\frac{u^\tau}{\tau}
  \end{pmatrix}
\nonumber\\
  && -\frac{1}{2}
  \begin{pmatrix} 
        D\left((u^\tau)^2\right) & D(u^\tau u^x) & D(u^\tau u^y) & 0\\[0.5ex]
        D(u^\tau u^x)    & D\left((u^x)^2\right) & D(u^xu^y)     & 0\\[0.5ex]
        D(u^\tau u^y)    & D(u^x u^y)    & D\left((u^y)^2\right) & 0\\[0.5ex]
        0                & 0             & 0                     & 0
  \end{pmatrix} 
\\
  && +\frac{1}{3}(\partial\cdot u)
  \begin{pmatrix} (u^\tau)^2{-}1 & u^\tau u^x  & u^\tau u^y  & 0\\[0.5ex]
                  u^\tau u^x     & (u^x)^2{+}1 & u^x u^y     & 0\\[0.5ex]
                  u^\tau u^y     & u^x u^y     & (u^y)^2{+}1 & 0\\[0.5ex]
                  0              & 0           & 0           & 1
  \end{pmatrix}.
\nonumber 
\end{eqnarray}
Here $D\eq{u^\tau}\partial_\tau{\,+\,}u^x\partial_x{\,+\,}u^y\partial_y$ and
$\partial\cdot u\eq\partial_\tau u^\tau + \partial_x u^x + \partial_y u^y
+\frac{u^\tau}{\tau}$. 

\section{Velocity finding} \label{appb}

As shown in \cite{Heinz:2005bw}, since we evolve all three components
$\pi^{\tau\tau}$, $\pi^{\tau x}$, and $\pi^{\tau y}$ (one of which is
redundant due to the constraint $\pi^{\tau m}u_m\eq0$), the flow velocity
and energy density can be found from the energy-momentum tensor components 
with the same efficient one-dimensional zero-search algorithm employed in 
ideal hydrodynamics \cite{Rischke95}. This is important since this step 
has to be performed after each time step at all spatial grid points in
order to evaluate the EOS $p(e)$. 
 
Using the output from the numerical transport algorithm, one defines the 
two-dimensional vector $\bm{M}=(M_x, M_y){\,\equiv\,}(T^{\tau x}{-}\pi^{\tau
 x}, T^{\tau y}{-}\pi^{\tau y})$. This is the ideal fluid part of the 
transverse momentum density vector; as such it is parallel to the 
tranverse flow velocity $\bm{v}_\bot\eq(v_x,v_y)$. Introducing 
further $M_0{\,\equiv\,}T^{\tau\tau}{-}\pi^{\tau\tau}$, one can write
the energy density as
\begin{eqnarray}
\label{edensity}
  e=M_0-\bm{v}_\bot \cdot \bm{M} = M_0 - v_\bot M,
\end{eqnarray}
where $v_\perp\eq\sqrt{v_x^2{+}v_y^2}$ is the transverse flow speed and 
$M\equiv\sqrt{M_x^2{+}M_y^2}$. One sees that solving for $e$ requires only
the magnitude of $\bm{v}_\perp$ which is obtained by solving the implicit 
relation \cite{Rischke95,Heinz:2005bw}
\begin{eqnarray}
\label{vperp}
  v_\perp = \frac{M}{M_0+p(e{=}M_0{-}v_\perp M)}.
\end{eqnarray}
by a one-dimensional zero-search. The flow velocity components are 
then reconstructed using
\begin{eqnarray}
\label{vi}
  v_x=v_\perp\frac{M_x}{M}, \qquad  v_y=v_\bot\frac{M_y}{M}.
\end{eqnarray}

Note that this requires direct numerical propagation of all three
components ($\pi^{\tau \tau}$, $\pi^{\tau x}$ and $\pi^{\tau y}$)
since the flow velocity is not known until after the velocity 
finding step has been completed. Hence the transversality constraint
$\pi^{\tau m}u_m\eq0$ cannot be used to determine, say, $\pi^{\tau\tau}$
from $\pi^{\tau x}$ and $\pi^{\tau y}$. However, it can be used after
the fact to test the numerical accuracy of the transport code.

\section{$\pi^{mn}$ in transverse polar coordinates} \label{appc}

%
\begin{figure*}[htb]
\includegraphics[bb=27 20 700 531,width=.49\linewidth,clip=]{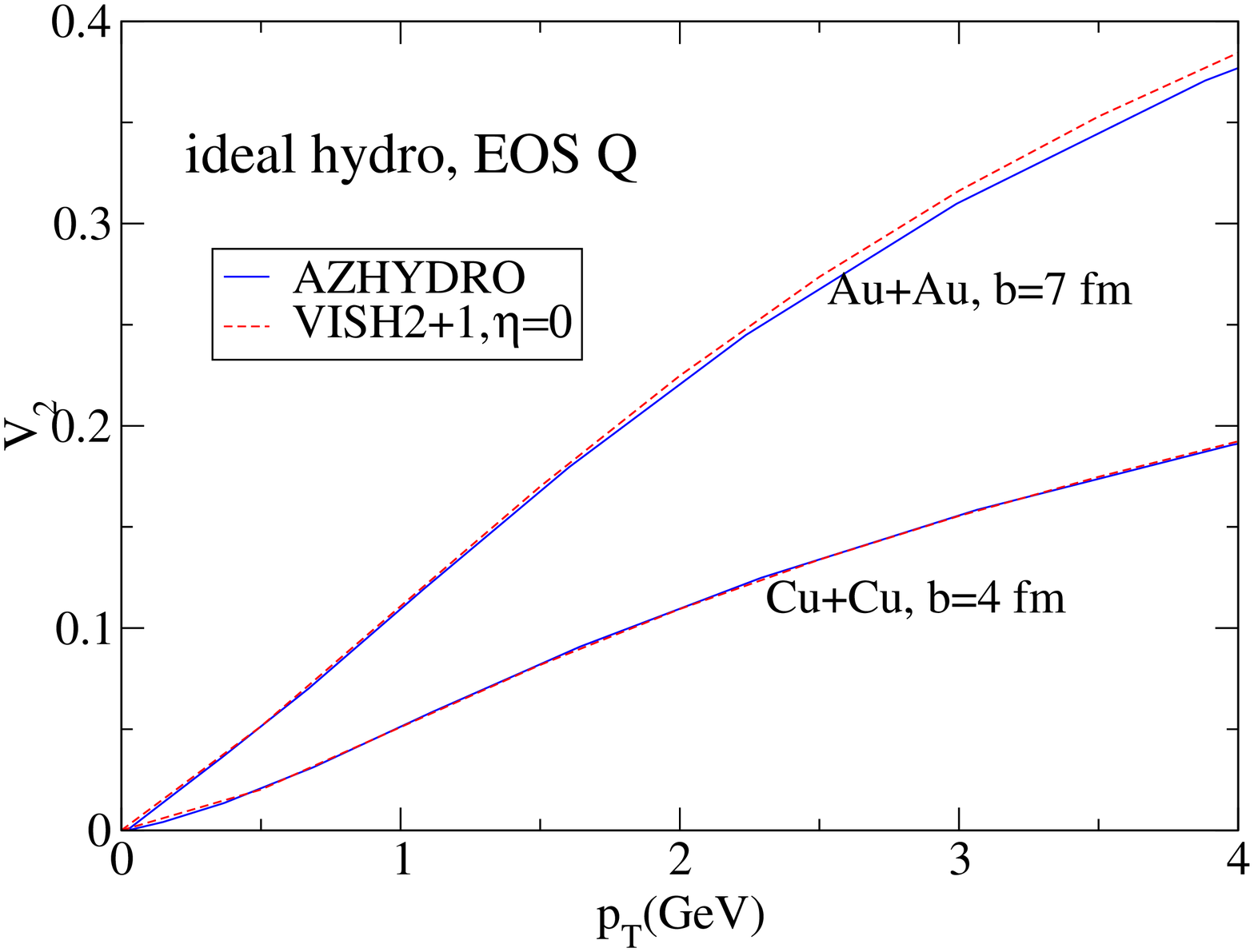}
\includegraphics[bb=27 15 714 522,width=.49\linewidth,clip=]{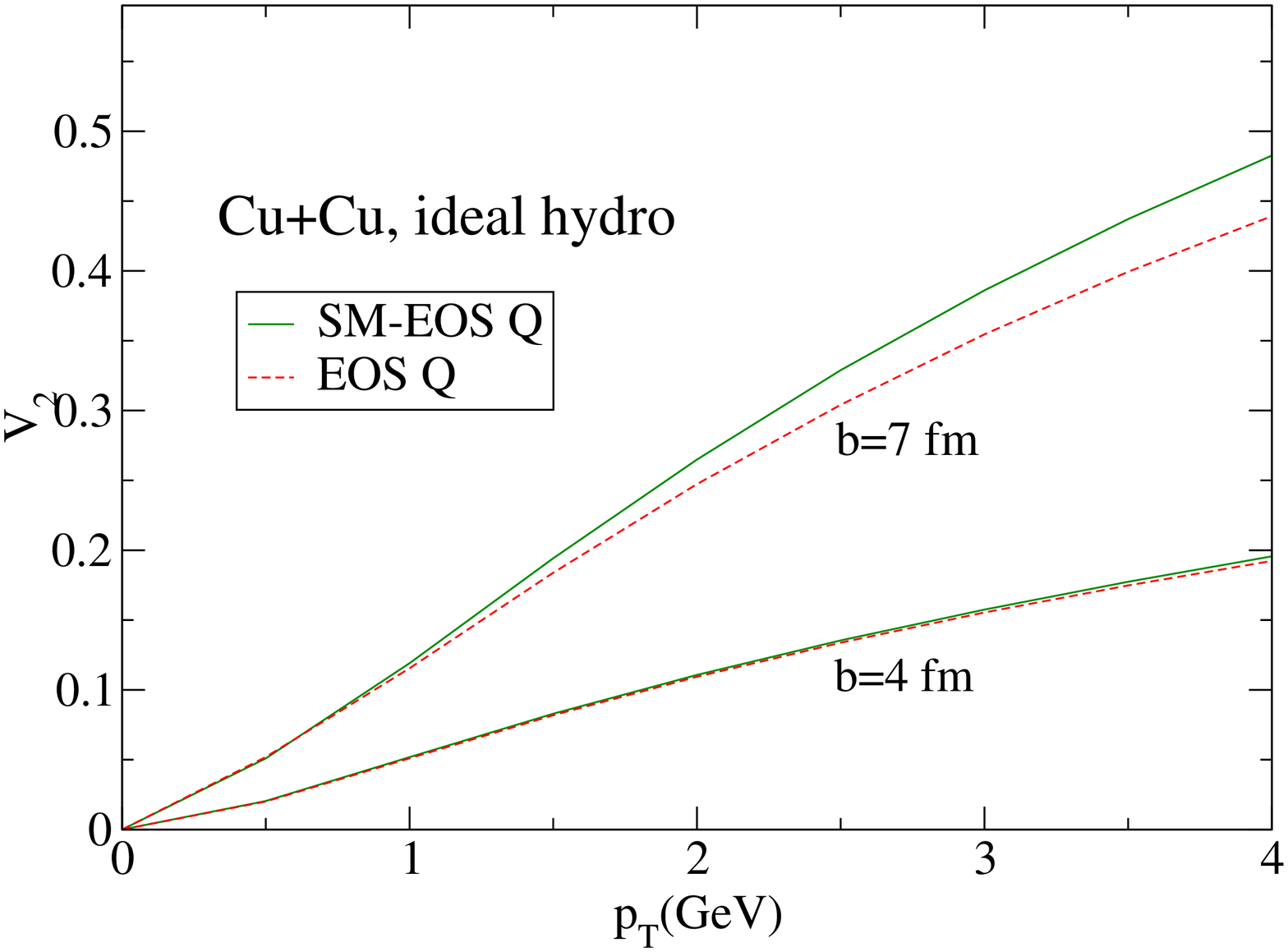}
\caption{(Color online)
{\sl Left:} Differential elliptic flow $v_2(p_T)$ for $\pi^-$ from 
$b\eq4$\,fm Cu+Cu collisions and $b\eq7$\,fm Au+Au collisions, using EOS~Q.
Results from VISH2+1 for $\eta\eq0$ and $\pi^{mn}\eq0$ (dashed lines)
are compared with the ideal fluid code AZHYDRO (solid lines).
{\sl Right:} $v_2(p_T)$ for $\pi^-$ from Cu+Cu collisions at impact 
parameters $b\eq4$ and 7\,fm, comparing VISH2+1 evolution with EOS~Q 
(dashed) and SM-EOS~Q (solid) in the ideal fluid limit $\eta\eq0$,
$\pi^{mn}\eq0$.
}
\label{v2-CompEOS4}
\end{figure*}
%

Although VISH2+1 uses Cartesian $(x,y)$ coordinates in the transverse
plane, polar $(r,\phi)$ coordinates may be convenient to understand
some of the results in the limit of zero impact parameter where azimuthal
symmetry is restored. In $(\tau,r,\phi,\eta)$ coordinates the flow velocity 
takes the form $u^m=\gamma_\perp(1,v_r,v_\phi,0)$, with 
$\gamma_\perp\eq1/\sqrt{1{-}v_\perp^2}\eq1/\sqrt{1{-}v_r^2{-}r^2v_\phi^2}$.
The polar coordinate components of the shear pressure tensor components 
$\pi^{mn}$ are obtained from those in $(\tau, x, y, \eta)$
coordinates by the transformations
\begin{eqnarray}
\label{C1}
  \pi^{\tau r}&=&\ \ \pi^{\tau x} \cos \phi+\pi^{\tau y}\sin\phi, 
\nonumber\\
  r \pi^{\tau \phi}&=&-\pi^{\tau x} \sin \phi+\pi^{\tau y}\cos\phi, 
\\
  \pi^{r r}&=&\pi^{x x}\cos^2\phi + 2\pi^{x y}\sin\phi\cos\phi
                                  + \pi^{yy}\sin^2\phi, 
\nonumber\\
  r^2 \pi^{\phi\phi}&=&\pi^{x x}\sin^2\phi - 2\pi^{xy}\sin\phi\cos\phi
                                           + \pi^{y y}\cos^2\phi, 
\nonumber\\
  r\pi^{r\phi}&=&(\pi^{yy}{-}\pi^{xx})\sin\phi\cos\phi
               + \pi^{xy}(\cos^2\phi{-}\sin^2\phi),
\nonumber
\end{eqnarray}
with $\cos\phi\eq{x/r}$ and $\sin\phi\eq{y/r}$. In terms of these the 
independent components $\Sigma$ and $\Delta$ of 
Eqs.~(\ref{indep},\ref{vis-cor}) are given as
\begin{eqnarray}
\label{C2}
  \Sigma &=& \pi^{rr} + r^2\pi^{\phi\phi},
\nonumber\\
  \Delta &=& \cos(2\phi)\bigl(\pi^{rr}{-}r^2\pi^{\phi\phi}\bigr)
             -2\sin(2\phi)\,r\pi^{r\phi},
\end{eqnarray}
from which we easily get
\begin{eqnarray}
\label{C3}
  2\pi^{xx} &=& \pi^{rr}(\bigl(1+\cos(2\phi)\bigr) 
              + r^2\pi^{\phi\phi}(\bigl(1-\cos(2\phi)\bigr),
\nonumber\\
  2\pi^{yy} &=& \pi^{rr}(\bigl(1-\cos(2\phi)\bigr) 
              + r^2\pi^{\phi\phi}(\bigl(1+\cos(2\phi)\bigr).
\end{eqnarray}
Note that azimuthal symmetry at $b\eq0$ implies $\pi^{r\phi}\eq0$ and
a vanishing azimuthal average for $\Delta$: $\langle\Delta\rangle_\phi\eq0$
or $\langle\pi^{xx}\rangle_\phi\eq\langle\pi^{yy}\rangle_\phi$.

\section{Tests of the viscous hydro code VISH2+1}\label{appd}
\subsection{Testing the ideal hydro part of VISH2+1}\label{appd1}

When one sets $\pi^{mn}\eq0$ initially and takes the limit $\eta\eq0$, 
VISH2+1 simulates the evolution of an ideal fluid, and its results
should agree with those of the well-tested and publicly available 
(2+1)-dimensional ideal fluid code AZHYDRO \cite{AZHYDRO}. Since
VISH2+1 was written independently, using only the flux-corrected 
SHASTA transport algorithm from the AZHYDRO package \cite{AZHYDRO,SHASTA} 
in its evolution part, this is a useful test of the code. The left
panel in Fig.~\ref{v2-CompEOS4} shows that, for identical initial and
final conditions as described in Sec.~\ref{sec2}, the two codes indeed
produce almost identical results. The small difference in the Au+Au 
system at $b\eq7$\,fm is likely due to the slightly better accuracy
of AZHYDRO which, in contrast to VISH2+1, invokes an additional 
timesplitting step in its evolution algorithm.
 
When comparing our VISH2+1 results with AZHYDRO we initially found 
somewhat larger discrepancies which, however, could be traced back
to different versions of the EOS used in the codes: EOS~Q in AZHYDRO,
the smoothed version SM-EOS~Q in VISH2+1. In the left panel of  
Fig.~\ref{v2-CompEOS4} this difference has been removed, by running
also VISH2+1 with EOS~Q. In the right panel we compare VISH2+1 results
for EOS~Q and for SM-EOS~Q, showing that even the tiny rounding effects
resulting from the smoothing procedure used in SM-EOS~Q (which renders
the EOS slightly stiffer in the mixed phase) lead to differences in the 
elliptic flow for peripheral collisions of small nuclei which exceed 
the numerical error of the code.

\subsection{Comparison with analytical results for (0+1)-d boost-invariant 
viscous hydro}\label{appd2}

For boost-invariant longitudinal expansion without transverse flow,
the relativistic Navier-Stokes equations read \cite{Gyulassy85} 
\begin{eqnarray}
 &&\frac{\partial e}{\partial \tau} +\frac{e+p+\tau^2 \pi^{\eta\eta}}{\tau}=0,
\\
 &&\tau^2 \pi^{\eta\eta}=-\frac{4}{3} \frac{\eta}{\tau}.
\end{eqnarray}
For an ideal gas EOS $p\eq\frac{1}{3}e{\,\sim\,}T^4$ this leads 
%
\begin{figure}[htb]
\includegraphics[bb=27 30 705 522,width=\linewidth,clip=]{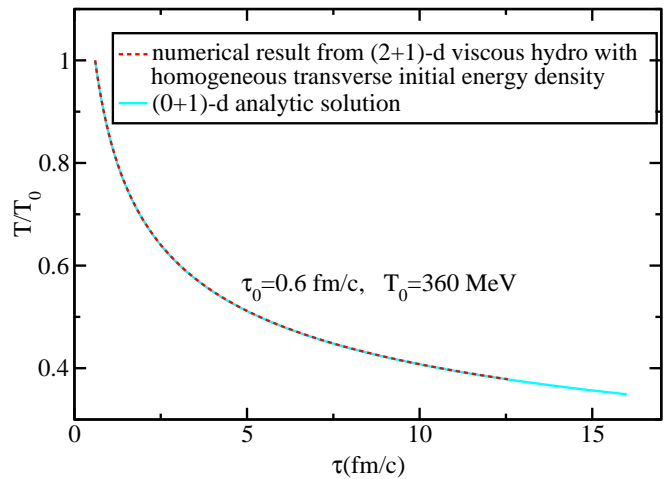}
\caption{(Color online)
Comparison between the analytical temperature evolution for (0+1)-d 
boost-invariant Navier-Stokes viscous hydrodynamics (solid line) and 
numerical results from VISH2+1 with homogeneous transverse initial energy
density profiles (dashed line).
} 
\label{CompAnaly}
\end{figure}
%
to the following analytic solution for the temperature evolution 
\cite{Gyulassy85}:
\begin{eqnarray}
\label{anal}
  \frac{T(\tau)}{T_0}=\Bigl(\frac{\tau_0}{\tau}\Bigr)^{1/3}
  \Bigl[1+\frac{2\eta}{3 s \tau_0 T_0}
        \Bigl(1-\Bigl(\frac{\tau_0}{\tau}\Bigr)^{2/3}\Bigr)\Bigr].
\end{eqnarray}
To test our code against this analytical result we initialize
VISH2+1 with homogeneous tranverse density distributions (not
transverse pressure gradients and flow) and use the Navier-Stokes
identification $\pi^{mn}\eq2\eta\sigma^{mn}$ in the hydrodynamic
part of the evolution algorithm, sidestepping the part of the code
that evolves $\pi^{mn}$ kinetically. It turns out that in this case
the relativistic Navier-Stokes evolution is numerically stable.
Fig.~\ref{CompAnaly} compares the numerically computed temperature 
evolution from VISH2+1 with the analytic formula (\ref{anal}), for 
$\eta/s\eq0.08$ and $T_0\eq360$\,MeV at $\tau_0\eq0.6$\,fm/$c$.
They agree perfectly.

\subsection{Reduction of VISH2+1 to relativistic Navier-Stokes theory
for small $\eta$ and $\tau_\pi$} 
\label{appd3}

Having tested the hydrodynamic part of the evolution algorithm in 
Appendix~\ref{appd1}, we would like to demonstrate also the accuracy
of the kinetic evolution algorithm that evolves the viscous pressure
tensor components. A straightforward approach would be to take VISH2+1,
set the relaxation time $\tau_\pi$ as close to zero as possible, and 
compare the result with a similar calculation as in Appendix~\ref{appd1}
where we sidestep the kinetic evolution algorithm and instead insert
into the hydrodynamic evolution code directly the Navier-Stokes 
%
\begin{figure}[htb]
\includegraphics[bb=52 28 711 534,width=\linewidth,clip=]{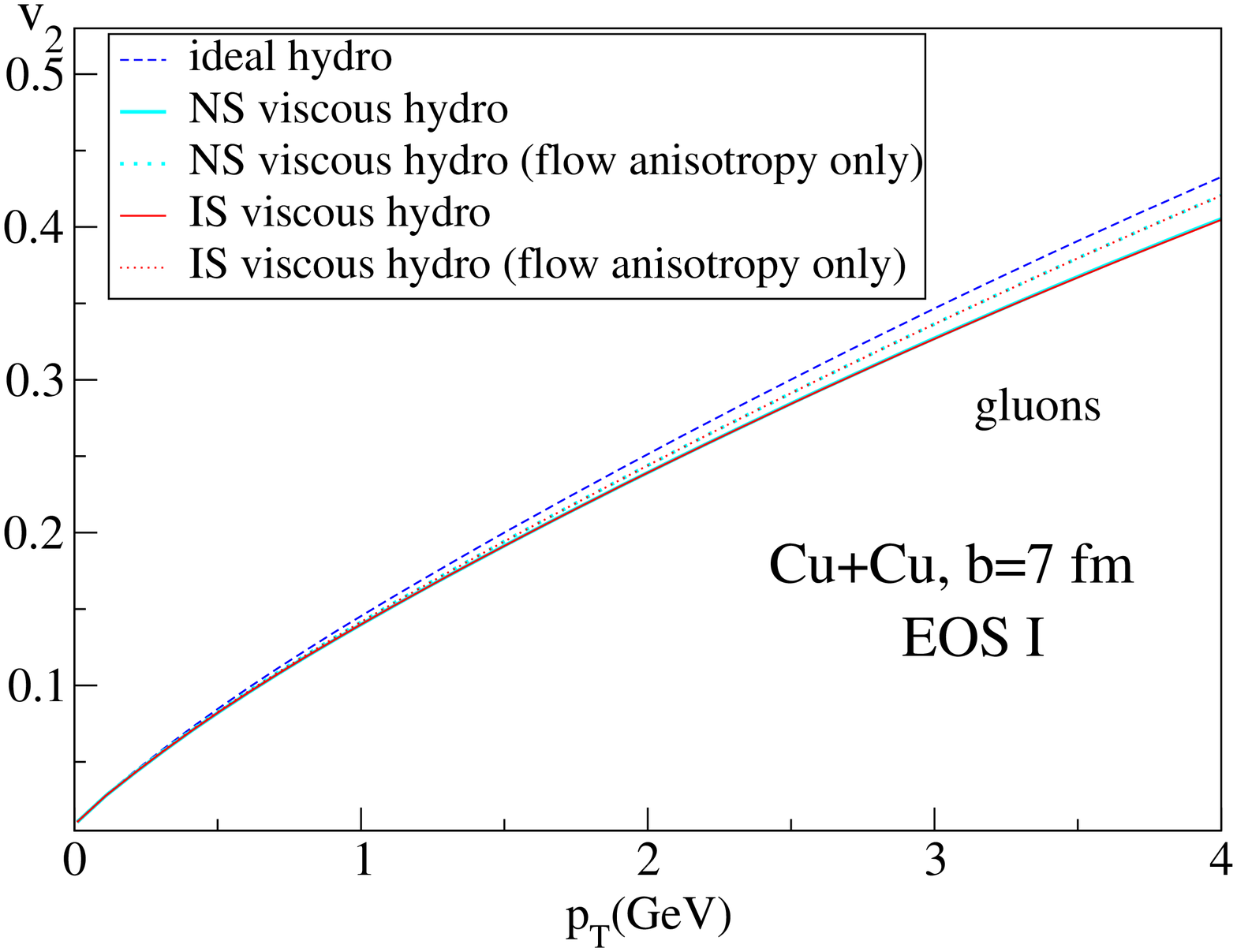}
\caption{(Color online)
Differential elliptic flow $v_2(p_T)$ for gluons from $b\eq7$\,fm
Cu+Cu collisions, calculated with ideal hydrodynamics (blue dashed line),
relativistic Navier-Stokes (NS) hydrodynamics (light blue lines), and
Israel-Stewart (IS) viscous hydrodynamics with $\frac{\eta}{s}\eq\frac{T}{2
\,\mathrm{GeV}}$ and $\tau_\pi\eq0.03$\,fm/$c$ (red lines), using EOS~I. 
The lines for NS and IS viscous hydrodynamics are almost indistinguishable.
Solid lines show the full results from viscous hydrodynamics, dotted lines
neglect viscous corrections to the spectra and take only the flow 
anisotropy effect into account.
}
\vspace*{-5mm}
\label{v2-2}
\end{figure}
%
%
\begin{figure*}[htb]
\includegraphics[bb=47 67 710 530,width=.62\linewidth,clip=]{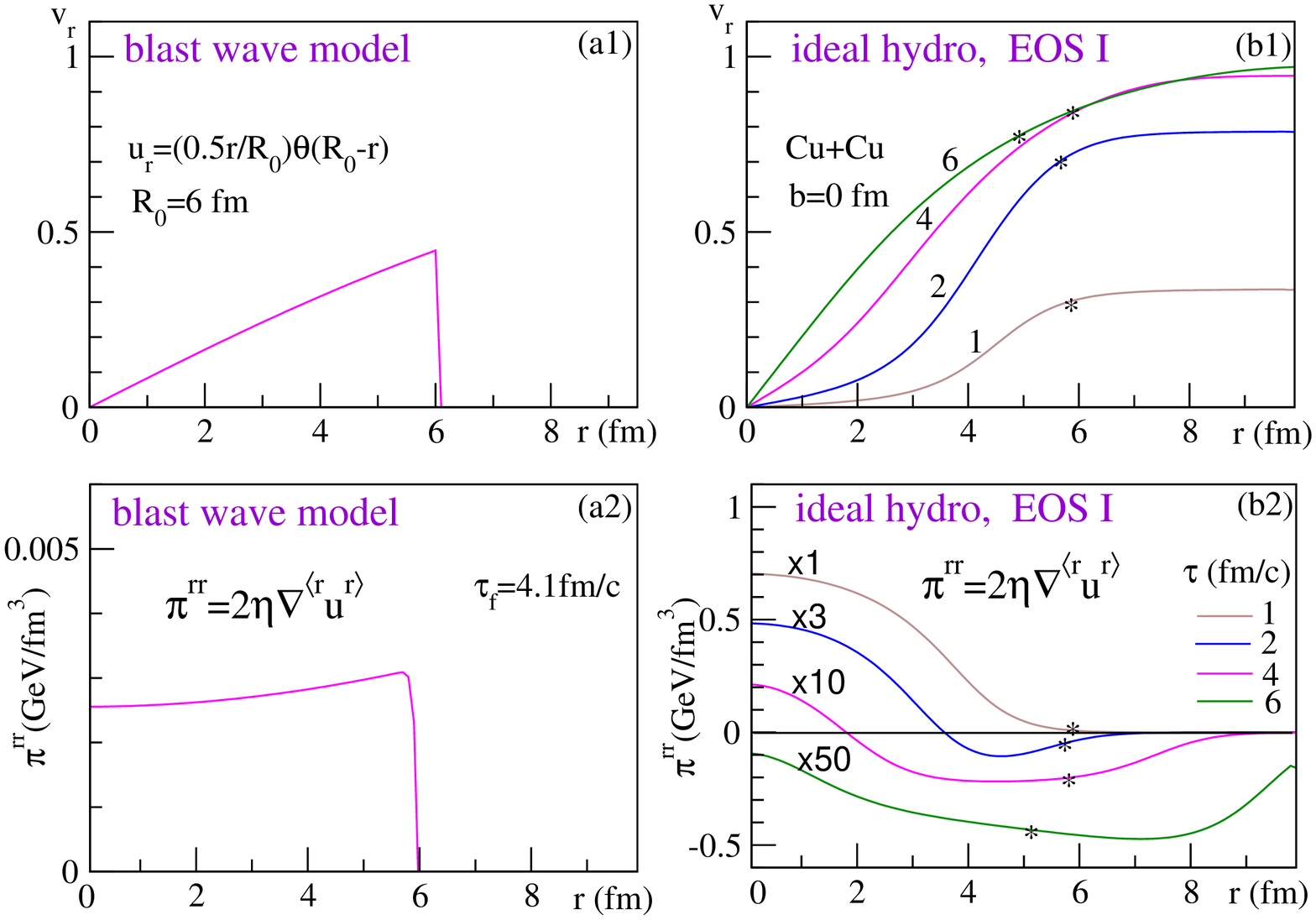}
\includegraphics[bb=31 69 402 551,width=.35\linewidth,clip=]{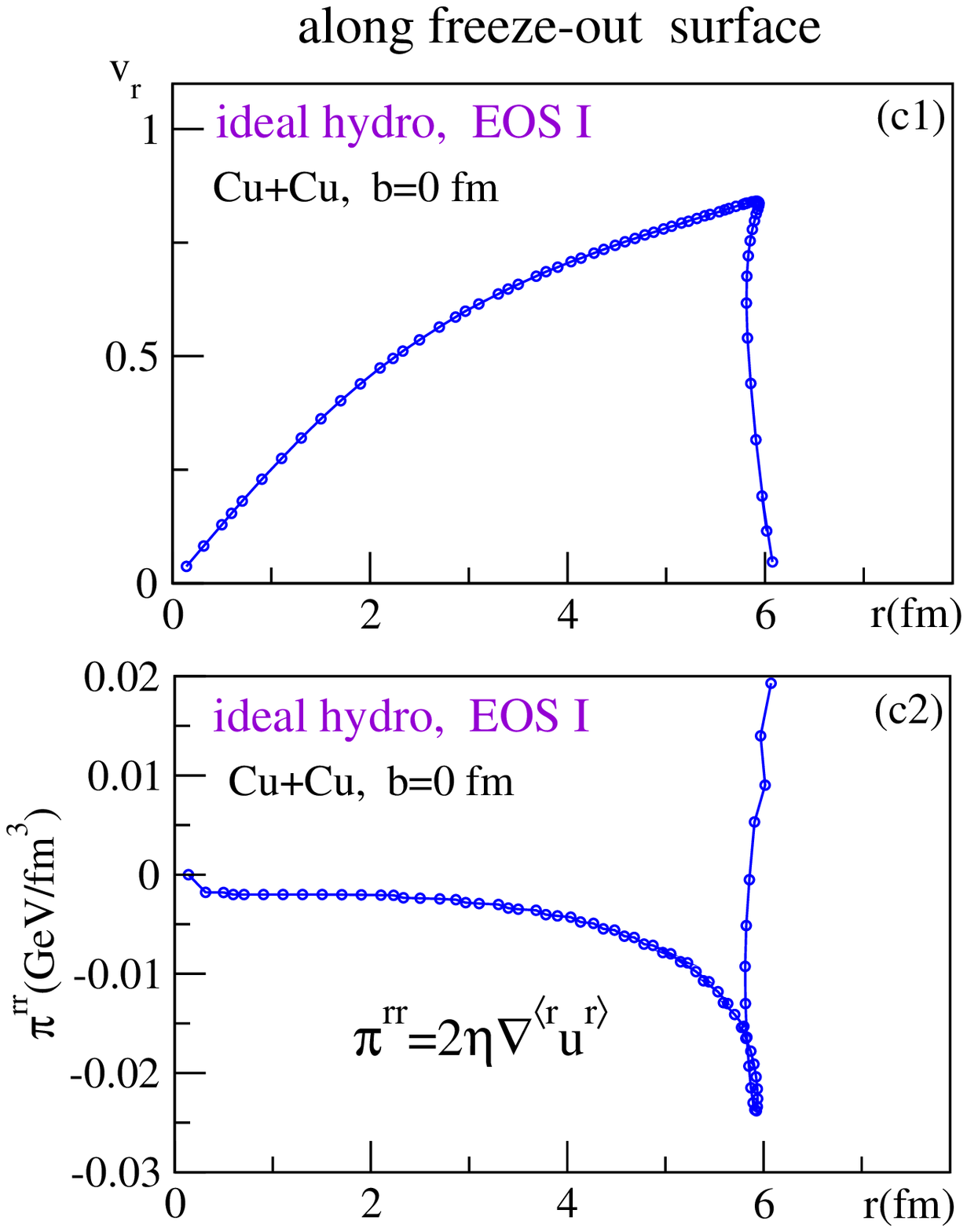}
\caption{(Color online)
{\sl Top row:} Velocity profiles from the blast wave model (left) and
from the hydrodynamic model with EOS~I at fixed times (middle) and along 
the decoupling surface (right).
{\sl Bottom row:} The corresponding profiles for the transverse shear 
viscous pressure $\pi^{rr}$ in the Navier-Stokes limit, 
$\pi^{rr}\eq2\eta\nabla^{\left\langle\mu\right.}u^{\left.\nu\right\rangle}$. 
Calculations are for central Cu+Cu collisions, and the curves in the 
middle panels correspond to the times $\tau\eq1$, 2, 4, and 6\,fm/$c$.
See text for discussion.
}
\label{BW-dyn}
\end{figure*}
%
identity $\pi^{mn}\eq2\eta\sigma^{mn}$. Unfortunately, this naive 
procedure exposes us to the well-known instability and acausality 
problems of the relativistic Navier-Stokes equations. The suggested
procedure only works if a set of initial conditions and transport
coefficients can be found where these instabilities don't kick in
before the freeze-out surface has been reached.

We found that sufficiently stable evolution of the relativistic 
Navier-Stokes algorithm (i.e. of VISH2+1 with the identification 
$\pi^{mn}\eq2\eta\sigma^{mn}$) can be achieved for standard initial 
density profiles in Cu+Cu collisions and the simple ideal gas equation 
of state EOS~I by choosing a very small and temperature dependent 
specific shear viscosity 
$\frac{\eta}{s}\eq0.01\,\frac{T}{200\,\mathrm{MeV}}\eq\frac{T}
{2\,\mathrm{GeV}}$. For the Israel-Stewart evolution we use 
a relaxation time which is correspondingly short: 
$\tau_\pi\eq\frac{3\eta}{sT}\eq0.03$\,fm/$c$. 

Figure~\ref{v2-2} shows the differential elliptic flow $v_2(p_T)$ for 
gluons in $b\eq7$\,fm Cu+Cu collisions evolved with these parameters. 
The dashed line gives the ideal fluid result. The solid and dotted lines 
show the total elliptic flow and the anisotropic flow contribution to 
$v_2(p_T)$, respectively, similar to the left panel Fig.~\ref{v-2}.
There are two solid and dotted lines with different colors, corresponding
to Israel-Stewart and Navier-Stokes evolution; they are indistinguishable,
but clearly different from the ideal fluid result. We conclude that, for 
small shear viscosity $\eta/s$ and in the limit $\tau_\pi{\,\to\,0}$,
the second-order Israel-Stewart algorithm reproduces the Navier-Stokes
limit and that, therefore, VISH2+1 evolves the kinetic equations for 
$\pi^{mn}$ accurately.

\section{Hydrodynamics vs. blast wave model}
\label{appe}

As discussed in Sec.~\ref{sec3b}, the viscous corrections to the final
pion spectra from the hydrodynamic model have a different sign
(at least in the region $p_T{\,>\,}1$\,GeV) than those originally
obtained by Teaney \cite{Teaney:2003kp}. In this Appendix we try to 
explore the origins of this discrepancy. We will see that the sign and
magnitude of viscous corrections to the (azimuthally averaged) particle 
spectra are fragile and depend on details of the dynamical evolution 
and hydrodynamic properties on the freeze-out surface. Fortunately, they 
same caveat does not seem to apply to the viscous corrections to elliptic 
flow where hydrodynamic and blast wave model calculations give 
qualitatively similar answers.

Following Teaney's procedure, we calculate $\pi^{mn}$ in the Navier-Stokes
limit $\pi^{mn}\eq2\eta\sigma^{mn}$. We do this both in the blast wave
model and using the results for $\sigma^{mn}$ from VISH2+1. 
For the blast wave model we assume like Teaney freeze-out at constant
$\tau$ with a box-like density profile $e(r)\eq{e}_\mathrm{dec}
\theta(R_0{-}r)$, where $e_\mathrm{dec}\eq0.085$\,GeV/fm$^3$ is the 
same freeze-out energy density as in the hydrodynamic model for EOS~I,
and $R_0\eq6$\,fm. The velocity profile in the blast wave model
is taken to be linear, $u_r(r)\eq{a}_0\frac{r}{R_0}\theta(R_0{-}r)$, with 
$a_0=0.5$; freeze-out is assumed to occur at $\tau_\mathrm{dec}\eq4.1
$\,fm/$c$. $R_0$, $a_0$ and $\tau_\mathrm{dec}$ are somewhat smaller than 
in Ref.~\cite{Teaney:2003kp} since we study Cu+Cu instead of Au+Au 
collisions. We concentrate here on a discussion of $\pi^{rr}$ for
illustration; the expression for $\sigma^{rr}$ is found in 
Ref.~\cite{Heinz:2005bw}, Eq.~(A11c). While $\pi^{rr}$ from VISH2+1 
differs from $2\eta\sigma^{rr}$ due to the finite relaxation time 
$\tau_\pi$ (see Sec.~\ref{sec5c}), we have checked that the signs of 
these two quantities are the same on the freeze-out surface so that 
our discussion provides at least a qualitatively correct analysis
of the viscous spectra corrections in the two models.

In Fig.~\ref{BW-dyn} we compare the freeze-out profiles for the
radial flow velocity and $2\eta\sigma^{rr}$ from the blast wave model.
In spite of qualitative similarity of the velocity profiles, the
freeze-out profiles of $2\eta\sigma^{rr}$ are entirely different and 
even have the opposite sign in the region where most of the hydrodynamic
particle production occurs (left and right columns in Fig.~\ref{BW-dyn}).
The middle column shows that at fixed times $\tau$, the hydrodynamic profile
for $2\eta\sigma^{rr}$ shows some similarity with the blast wave model in
that $2\eta\sigma^{rr}$ is positive throughout most of the
interior of the fireball. What matters for the calculation of the 
spectra via Eq.~(\ref{Cooper}), however, are the values of $2\eta\sigma^{rr}$
on the freeze-out surface $\Sigma$ where they are negative, mostly
due to radial velocity derivatives. This explains the opposite sign of 
the viscous correction to the spectra in the hydrodynamic model and shows 
that, as far as an estimate of these viscous corrections goes, the blast 
wave model has serious limitations.

%


\begin{thebibliography}{99}
\bibitem{fn5}
  Recently completed first simulations of Au+Au collisions indicate that
  the viscous suppression effects or not quite as big in this larger
  collision system than for the smaller Cu+Cu fireball studied here
  \cite{next}.
\end{thebibliography}
\end{document}